# Simplifying the Configuration of
# 802.11 Wireless Networks with Effective SNR

Daniel Chaim Halperin

A dissertation
submitted in partial fulfillment of the
requirements for the degree of

Doctor of Philosophy

University of Washington

2012

Reading Committee:

David J. Wetherall, Chair

Thomas E. Anderson, Chair

Jitendra D. Padhye

Program Authorized to Offer Degree:
Computer Science and Engineering

University of Washington

**Abstract**

Simplifying the Configuration of 802.11 Wireless Networks with Effective SNR

Daniel Chaim Halperin

Co-Chairs of the Supervisory Committee:
Professor David J. Wetherall
Computer Science and Engineering

Professor Thomas E. Anderson
Computer Science and Engineering

Advances in the price, performance, and power consumption of Wi-Fi (IEEE 802.11) technology have led to the adoption of wireless functionality in diverse consumer electronics. These trends have enabled an exciting vision of rich wireless applications that combine the unique features of different devices for a better user experience. To meet the needs of these applications, a wireless network must be configured well to provide good performance at the physical layer. But because of wireless technology and usage trends, finding these configurations is an increasingly challenging problem.

Wireless configuration objectives range from simply choosing the fastest way to encode data on a single wireless link to the global optimization of many interacting parameters over multiple sets of communicating devices. As more links are involved, as technology advances (e.g., the adoption of OFDM and MIMO techniques in Wi-Fi), and as devices are used in changing wireless channels, the size of the configuration space grows. Thus algorithms must find good operating points among a growing number of options.

The heart of every configuration algorithm is evaluating of the performance of a wireless link in a particular operating point. For example, if we know the performance of all three links between a source, a destination, and a potential relay, we can easily determine whether or not using the relay will improve aggregate throughput. Unfortunately, the two standard approaches to this task fall short. One approach uses aggregate signal strength statistics to estimate performance, but these do not yield accurate predictions of performance. Instead, the approach used in practice measures performance by actually trying the possible configurations. This procedure takes a long time to converge and hence is ill-suited to large configuration spaces, multiple devices, or changing channels, all of which are trends today.

As a result, the complexity of practical configuration algorithms is dominated by optimizing this performance estimation step.

In this thesis, I develop a comprehensive way to rapidly and accurately predict the performance of every operating point in a large configuration space. I devise a simple but powerful model that uses a single low-level channel measurement and extrapolates over a wide configuration space. My work makes the most complex step of today's configuration algorithms—estimating the effectiveness of a particular configuration—trivial, achieving better performance in practice and enabling the practical solution of larger problems.

# TABLE OF CONTENTS

ii

iv

# LIST OF FIGURES

vi

vii

# LIST OF TABLES

x

# ACKNOWLEDGMENTS

Writing acknowledgments is always tricky business, because one invariably forgets someone. To circumvent this pitfall, I chose someone very important and left them out deliberately. You know who you are.

I feel extremely privileged to have worked as part of an amazing community of people during my graduate studies. First and foremost, I am immensely grateful to David Wetherall and Tom Anderson, my amazing and inspiring advisors. They have guided me through the research process, taught me how to frame problems and analyze research, and have supported me no matter what over the past six years. Beyond this, David's unfailingly positive attitude and his excellent advice renewed my spirits when I was low on energy; I always walked out of our meetings with a newfound vigor and enthusiasm for my work. And Tom's ability to get to the heart of an issue and distill the key points has helped me focus my work. These are only a few ways in which I have benefited tremendously from both relationships.

I have enjoyed collaborating with many other professors, students, and researchers over the years. In particular, Wenjun Hu and Anmol Sheth were great mentors and key contributors to my thesis work. I was fortunate to work with Victor Bahl, Srikanth Kandula, Jitu Padhye, Yoshi Kohno, Arvind Krishnamurthy, Ben Greenstein, Kevin Fu, Bill Maisel, Tom Heydt-Benjamin, George Nychis, Ben Ransford, Vincent Liu, Josie Ammer, Shane Clark, Benessa Defend, Dongsu Han, Tom Kenney, Will Morgan, Eldad Perahia, Srini Seshan, Robert Stacey, and Peter Steenkiste over the years, and I wish we could have worked together more.

I thank all the inhabitants of the networking lab over the years for making life fun inside and outside the lab, creating a work environment I wanted to visit seven days a week (and often did). I also thank all my friends and colleagues at UW CSE—the kind of place where those words are synonymous—for creating an atmosphere of learning, respect, and friendship where it is easy to succeed. I would especially like to thank Neva Cherniavsky, Tomas Isdal, and Ratul Mahajan for always being there with advice and emotional support and serving as great examples of success.

None of us, especially me, would have gotten anything done in graduate school were it not for the tireless and heroic efforts of Melody Kadenko, our program manager, and Lindsay Michimoto, our graduate program advisor. They kept our *i*s dotted and our *t*s crossed, and generally made sure we stayed in line. They gracefully and graciously handled every request no matter how last-minute (or late). I also appreciate the support of Karla Danson and Julie Svendsen.

I would like to thank all my roommates, Saleema Amershi, Sal Guarnieri, Alex Jaffe, Nodira Khoussainova, Ian McDonald, and Dustin Shilling. Living with them was always a pleasure and was especially important in making it possible to get through deadlines. Along with Neva, Tomas, Didi, Mike, Melissa, John, Kayur, and many more, they always reminded me that it is important to play as well as work.

My graduate work was supported by the National Science Foundation, the UW Clairmont L. Egtvedt Fellowship, an Intel Foundation Ph.D. Fellowship, and through internships at Intel Labs Seattle and Microsoft Research.

Most importantly, I'd like to thank my family. They have been there for me through every stage of my life, and their encouragement and support have made all the difference.

# DEDICATION

*This dissertation is dedicated to my grandparents, for teaching me the joy of solving puzzles, the value of education, and the importance of improving the world.*

Chapter 1

# INTRODUCTION

Wireless local area networks are used today to connect devices wirelessly at high rates in locations such as cafés, shopping malls, corporate offices, and homes. The dominant technology for these networks is Wi-Fi (IEEE 802.11 [44]), which emerged in 1997 as a way to connect computers wirelessly to a nearby (within 100 m) Internet "access point" at rates up to 2 Mbps.

The past fifteen years have seen Wi-Fi technology improve dramatically. Today's commercial Wi-Fi devices come at low cost, have a small physical footprint, and offer dramatically increased speeds of up to 600 Mbps in IEEE 802.11n [45]. Wi-Fi is no longer limited to traditional computing devices such as laptop and desktop computers, but is also being adopted by consumer electronics such as smartphones, printers, speakers, video cameras, televisions, and DVD players. An ABI Research report [1] forecast that more than half of the 1 billion Wi-Fi chipsets shipped in 2011 would be used in consumer electronics.

Because of its rapid adoption in a diverse set of devices, Wi-Fi is poised at the heart of the next networking revolution: The combining of these diverse consumer devices to build rich applications that leverage each device's unique features. This stands in sharp contrast with today's access point model, in which devices only use wireless connectivity to interact with the Internet at large and hence the access point, which provides the only point of contact with the Internet, is a natural point of centralization. To support this shift away from the access point model, a new protocol called Wi-Fi Direct [122] was standardized in late 2010 that enables Wi-Fi devices to form networks that better match their applications. Wi-Fi Direct has seen great uptake: A second ABI Research study [2], conducted in late 2011, forecast a 50% annual growth rate for Wi-Fi Direct support and predicted that there will be 2 billion Wi-Fi Direct-enabled devices by 2016.

Despite these technology, standardization, and adoption trends aligning to enable future rich wireless applications, there is one major challenge: *The underlying Wi-Fi technologies and network architectures have become rather complex, and how to configure and control them has become a significant decision problem that presently lacks a simple, comprehensive solution*.

What does it mean to configure a network? In this thesis, I use the term *configuration* to describe an assignment of values to the physical-layer parameters of one or more wireless devices. This includes the choice of operating frequency, transmit power level, transmit rate, how many and which antennas are used, and more. A configuration problem is the task of configuring all of these parameters for one device, for two devices sending together as a link, or for many devices operating at the same time in a wireless network. Configuration problems can therefore be defined as search problems with the goal of finding a good operating point among the parameter space.

The heart of every configuration algorithm is evaluating the performance of a wireless link in a particular operating point. Consider rate selection, the problem of picking the fastest way to transmit data on a wireless link. In the first version of 802.11, released in 1997, the rate selection task consisted of choosing between two modulations to transmit data. Early algorithms just tried both rates and picked the better one. This worked well through 802.11b and 802.11a/g, because with up to 12 different rates to choose from, "try-it-and-see" algorithms that probed all options, though not perfect, generally sufficed. But trends in Wi-Fi make this probe-based approach much less effective.

First, the configuration space is growing much larger. One reason is technology trends: Modern 802.11n devices achieve their fast rates by relying on the ability to send with multiple antennas. This adds another dimension to the search space—how many antennas are used—and expands the number of rates into the hundreds. The other reason is usage trends: The device-to-device nature of new networks like Wi-Fi Direct means that coordination within a network is no longer limited a client and its access point. Instead, configuring the network requires extensive coordination between pairs and sets of devices in a network, growing the search space exponentially. Finally, wireless devices are increasingly used in changing environments. For instance, wireless devices are increasingly used while mobile, both while walking indoors and in vehicles. This combination of factors means that algorithms to configure the network need to respond faster to match changing channels, while simultaneously choosing from among more possibilities.

This issue affects all configuration problems, not just rate selection. An example configuration problem for a device-to-device network is choosing a multi-hop path between a source and destination device, possibly using intermediate devices as relays. One step in solving this problem involves assessing the performance achievable on each potential link. This should include taking into account the effect of using different rates, number and sets of antennas, and even the quality of using the best among multiple operating channels for each link. This set of parameters results in a very large configuration space, which proves impractical for probe-based solutions to search because the search will not converge if the channel conditions change. Instead, past solutions to this subproblem tend to assume away most dimensions of the configuration space—e.g., by assuming homogeneous single-antenna nodes and fixing the entire network to a single bitrate, frequency, and transmit power, so that the system need only probe packet delivery for a single rate. This simplification narrows the space to something feasible to probe, but by eliminating much of the configuration space from consideration it also likely results in relatively poor performance.

An alternative approach to probing uses measurements of the wireless channel to attempt to predict how well an operating point will work. For instance, 802.11 receivers can measure the total amount of power received from the transmitter. Since using slower rates generally requires less power than using faster rates, this signal strength measurement might serve as a useful indicator of whether a particular rate works. However, in practice this approach does not provide accurate predictions for Wi-Fi links, for fundamental reasons that I discuss below. Some proposals [55] attempt to train a mapping between signal strength and rate, but suffer from the same convergence problems as with probing because this mapping must be updated independently for each configuration point.

To get good performance in future wireless networks, configuration algorithms need to be able to take into account the entire configuration space. Fundamentally, this requires the ability to rapidly assess how well each operating point will work. Past solutions indicate that the probe-based algorithms used until now will not scale to handle these future systems. (My study in Chapter 7 reinforces this point.) Neither will the approaches based on aggregate signal strength information that must be trained for each operating point. Instead, Wi-Fi systems need a way to predict how well an operating point will work without trying it and without requiring online, per-wireless-link calibration.

In this dissertation, I provide a practical, effective solution to this problem. In particular, I develop a comprehensive way to inform these complex decision problems using low-level wireless channel measurements. I devise a simple but powerful model that can predict the performance of every operating point in the entire configuration space, using only a small set of measurements and without online training. Because it only uses one or a few measurements, my model can rapidly update its predictions as the channel changes. My approach enables algorithms to adjust *all* parameters and to adapt quickly to changing conditions, thus enabling the type of configuration algorithms needed to support rich future wireless networks.

In the rest of this chapter, I first explain the problem in further detail. I then present my hypothesis and explain my approach to solving this problem. I conclude this chapter by discussing the contributions of my work and the organization of the rest of this thesis.

## 1.1 The Problem

As stated above, a major challenge for Wi-Fi networks today is finding a good configuration in a changing world. To introduce the problem, I present the main configuration problems

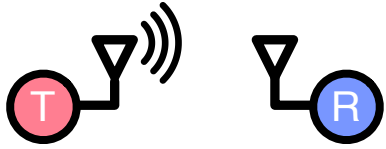

**Figure 1.1: A single Wi-Fi link, in which the transmitter** T **sends data to the receiver** R.
**No other wireless devices are present.**

in these systems, and briefly explain why today's Wi-Fi solutions are insufficient.

### 1.1.1   Configuring a Single Link

The most basic wireless network is a single link (Figure 1.1), in which a transmitter sends data to one receiver, with no other devices present. In this section, I will show that configuring a wireless link to work well involves choosing the right operating point in a large multi-dimensional space.

Perhaps the simplest configuration goal for a wireless link is *rate selection*: In most cases, the transmitter should send its data to the receiver using the fastest rate at which it will be successfully received. Sending data more slowly obviously means the transmission takes longer. At the same time, sending faster would be inefficient because the data would not be received, wasting all the airtime and energy consumed during the transmission.

In principle, selecting a rate for a wireless link should be trivial according to the well-known results of communications theory. Whether a transmission sent with a particular modulation and coding scheme is received is determined entirely by the amount of power delivered to the receiver and the noise level present. This factor is quantified in the *signal-to-noise ratio*, or *SNR*. The transmitter need only measure the channel SNR and apply textbook formulas that can compute the error rates of particular modulations. The fastest rate can then be easily selected. This approach is described in Figure 1.2(a).

In practice, this approach has never worked for Wi-Fi links. The 802.11 standard defines a channel metric related to the SNR, called the *receive signal strength indicator (RSSI)*, that captures the total amount of power in the channel. In most chipsets, RSSI is indeed a direct measure of the SNR. However, Wi-Fi systems have never used RSSI as more than a coarse indicator of expected performance. There have simply been too many ways in which the observed measurements and actual performance fail to match the predictions of theory. For example, hardware estimates of RSSI can be mis-calibrated, the wireless channel can vary over packet reception, and it can be corrupted by interference [17, 55, 95]. More fundamentally, the OFDM and MIMO (see Chapter 2) physical-layer techniques used in 802.11n send independent data on different subchannels with different subchannel SNRs, so that different bits of the packet can have different SNRs. This means that RSSI, which

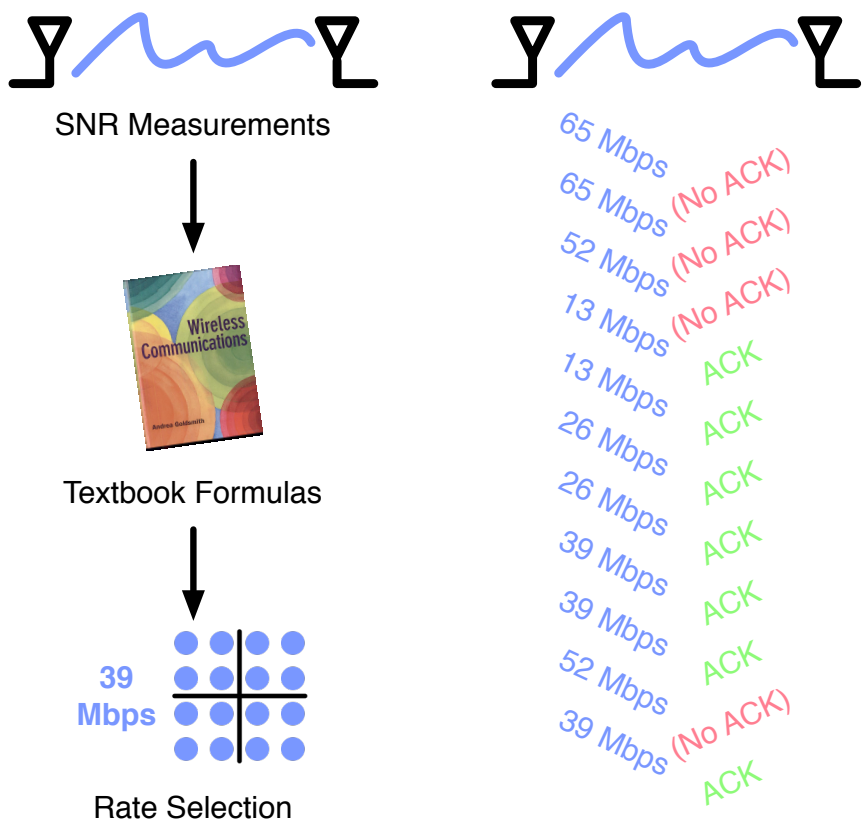

(a) **The theoretical approach to rate selection based on SNR measurement.**

(b) **The probe-based approach to rate selection.**

**Figure 1.2: Approaches to rate selection.**

captures only the average SNR of the signal, is fundamentally not a good indicator of performance for Wi-Fi.

Since rate selection based on RSSI has never worked for Wi-Fi, practical systems use *rate adaptation* algorithms instead [14, 109, 123]. These algorithms, exemplified by Figure 1.2(b), are guided search schemes that simply test individual rates to see how well they work. When the loss rate is too high, a lower rate is used; otherwise a higher rate is tested. This approach works well for slowly varying channels and simple links, since the best setting will soon be found.

However, remember the Wi-Fi trends discussed earlier: The transmit configuration of a single Wi-Fi link now includes not just rate, but additional dimensions that take into account the use of multiple antennas or channel widths, and these devices are increasingly being used while mobile. Thus algorithms to configure the rates of these links need to respond more rapidly to match changing conditions, while simultaneously choosing from among more possibilities. As a result, probe-based rate adaptation algorithms are becoming less efficient as these systems become more complex. (I provide results that illustrate this effect in Chapter 7.)

| Parameter | Number of values |
|---|---|
| Rate | 8 |
| Number of spatial streams | 4 |
| Operating channel | ≈30 |
| Channel width | 2 (some proposals use many more) |
| Transmit power level | ≈30 |
| Transmit antenna set | 6 (for up to 3 streams) |
| Receive antenna set | 6 (for up to 3 streams) |

**Table 1.1: A list of link configuration parameters.**

Thus far, I have described the challenges inherent to choosing an efficient rate to send data on a wireless link. On its own, this is a hard problem, for which there is a large body of prior work. In addition, I note that rate is only one of many parameters to optimize for a Wi-Fi link. For instance, a transmitter may want to trim excess transmit power to both save energy and reduce interference at nearby receivers. Or a sender might improve a link by selecting a different subset of its transmit antennas, or by applying beamforming techniques to better match the signal to the radio channel. Finally, note that these parameters are not generally independent—changing any one of them can affect the best operating point for another. For instance, switching the operating frequency (of which there are often 10 to 20 options) can dramatically change the RF channel, and this in turn can affect which transmit antennas provide the best link, and how the transmitted signal should be shaped for maximum performance. All of these factors contribute to determining the best way to configure a link. I provide a brief list of link configuration options in Table 1.1.

In practice, the solution taken by hardware/driver manufacturers (and by researchers) is to simply ignore most of these dimensions. For instance, only Intel's `iwlwifi` driver, out of all the 802.11n drivers in the Linux kernel driver, adapts the transmit antenna set in an online manner. Similarly, few access points and no clients adjust transmit power for ongoing links, instead opting to transmit at the maximum power and guarantee the best link. There are no known research solutions to these problems either. The solutions work well enough for wireless access point networks, mostly due to the simple way in which links are used. Still, these solutions are inefficient for a single link—and in the next section, we will see that the problem gets even more complicated when performing network-level configuration of multiple devices that operate in multiple links.

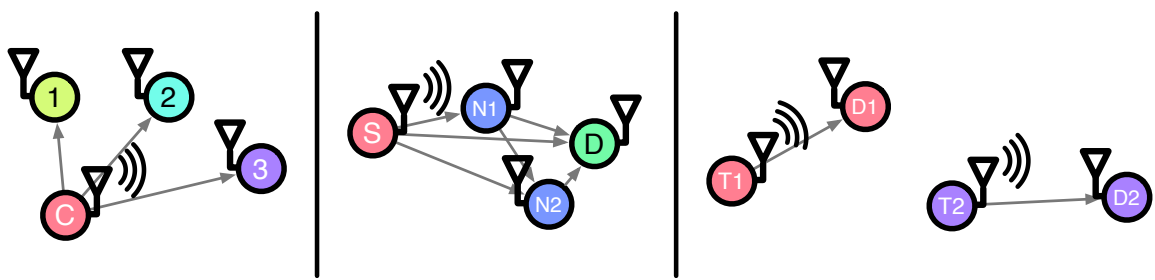

**Figure 1.3: The three key configuration problems in multi-device networks.** *Left:* **access point selection.** *Center:* **Multi-hop mesh routing.** *Right:* **Spatial reuse.**

### 1.1.2 Configuring a Network of Devices

In this section, I will illustrate how a network of devices has a significantly larger configuration space than a single link. I frame this discussion using the examples in Figure 1.3, which represent the three key network-level configuration problems that dense wireless networks like Wi-Fi Direct will have to solve to support device-to-device applications. Depending on the problem being solved, these configuration problems can have increased complexity that is linear in the number of devices (AP selection), quadratic (Multi-hop routing), or even exponential (Spatial reuse).

#### Access Point Selection

In Figure 1.3, on the left, the client C wishes to join the network offered by the access points $AP_1$, $AP_2$, and $AP_3$. The *access point selection* problem is simple: The client should connect to the access point that provides the link with the best rate. But in order to choose correctly, the client must accurately evaluate the rate offered by each access point. This in turn means that the client must have a way to assess its rate to each access point, i.e., a solution to the rate selection problem described above. Testing all access points using a rate adaptation-like approach would take too long and would take airtime away from ongoing connections. In practice clients simply connect the access point with the highest SNR. This heuristic approach provides only an approximation to the optimal solution, and clients would benefit from a better way to predict performance over measured wireless channels.

#### Multi-hop Mesh Routing

In Figure 1.3, in the middle, the source S wishes to send data to the destination D, and nodes $N_1$ and $N_2$ are also present in the network. The *multi-hop routing* problem is to choose the best path through the network by which to deliver data from S to D. In this case, many paths are available, such as the direct path S–D, the one-hop paths S–$N_1$–D and S–$N_2$–D, and finally S–$N_1$–$N_2$–D. To evaluate the different paths, we need to know the rate available on each hop, which in this case would require knowing the rates of six different links. Once again, measuring the ground truth rate of each link by testing each configuration would take too long, and it would add considerable overhead to the network.

Practical work in this area primarily takes one of two approaches. Most of the wireless mesh research in the past decade avoided this problem by simply ignoring many of the dimensions of the configuration space. These papers not only used single antenna systems at fixed transmit powers, but also typically fixed the entire network to a single rate (e.g., [15, 19, 59, 60, 61]). The alternative approach has been to collect statistics about packet delivery between all pairs of nodes for different rates, and estimate the rate from the measured SNR for links without sufficient statistics (e.g., [10]). These recent works have exclusively handled single-antenna 802.11a/b/g networks, and would likely be forced to rely on SNR-based rate predictions if the underlying links used multiple antennas (as in 802.11n).

*Spatial Reuse*

The third example, shown in Figure 1.3, is the *spatial reuse* problem. Here, two independent links both wish to communicate at the same time and in the same frequency, and so they need to share the wireless medium. If the links each use half the airtime, then each gets half of the rate available if they were operating alone. In certain situations, depending on the placement of the four devices and the amount of interference between the devices, throughput can be improved by sending concurrently, each using all of the airtime but maybe using a slightly lower rate.

Once again, deciding which of these two possibilities is better requires the system to predict the rate on multiple different links. In this case, the rate needs to be predicted not only for each link in isolation, but also for every possible pair of configurations of the links, since each transmitting device acts as an interfering signal for the other link. In this case, and unlike the prior two problems, the size of the resulting configuration space is the product of the sizes of the space of each individual link. As a result, practical works on spatial reuse for Wi-Fi [106, 121] have simply fixed the entire network to a single rate during experiments.

### 1.1.3 Summary

In this section, I first described how that the configuration problems for a single link have grown dramatically with the switch to 802.11n technology. I then presented the three key network-level configuration problems for future Wi-Fi networks and explained why configuration is even a bigger issue for networks than for a single link.

As wireless technology and architectures improve, network configuration algorithms will have to deal with increasingly large search spaces. To find good operating points if devices are mobile, we will need to search these large spaces quickly. The heuristic and adaptation-based approaches used today will not scale to these bigger problems. Instead, what we need is a way to accurately and rapidly assess the quality of links for all the factors

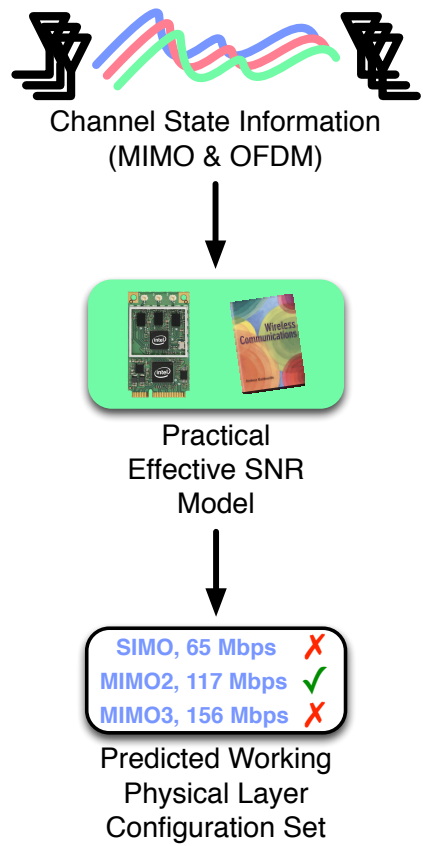

Channel State Information
(MIMO & OFDM)

Practical
Effective SNR
Model

SIMO, 65 Mbps  ✗
MIMO2, 117 Mbps  ✓
MIMO3, 156 Mbps  ✗

Predicted Working
Physical Layer
Configuration Set

**Figure 1.4: An Effective SNR-based approach to making application decisions in 802.11n networks.**

mentioned in Section 1.1.1. Then, we can use this process to easily solve joint optimization problems such as those described in Section 1.1.2.

## 1.2   Approach

In this thesis, I present a better way to inform Wi-Fi configuration algorithms that can solve problems like those described in the previous section. The key goal is to use a small number of measurements about the wireless channel and accurately extrapolate how well many different physical-layer configurations will work. This will provide a simple, unified, and fast way to evaluate potential operating points and lead to algorithms that are able to find good solutions, and do so quickly enough to adapt effectively to changing channels.

Figure 1.4 presents a pictorial summary of my approach. This approach is closely related to the "theoretical approach" presented in Figure 1.2(a), with a few key differences.

First, a client will measure the *channel state information (CSI)* for a wireless link, instead of the RSSI used to compute Packet SNR today. I described above the key problem with RSSI: It simply measures the total amount of power in a link, which does not capture the properties of the different frequency and spatial subchannels that Wi-Fi uses to send independent data. In contrast, the CSI is a fine-grained measurement that can capture details at the levels of frequency-selective fading (to understand performance under OFDM) and independent spatial paths (to understand performance when using MIMO). My thesis will show that the low-level measurements comprised by the CSI are fine-grained enough to be useful.

The second step uses the main contribution of my thesis: *A practical Effective SNR-based model for wireless packet delivery*. My model uses the measured CSI as input, and incorporates textbook algorithms, ideas from communications theory, as well as some implementation-specific details to handle a wide variety of channels, hardware devices, and applications. At the core of my model is the notion of *Effective* $E_b/N_0$, described in a seminal 1998 paper by Nanda and Rege [79]. The Effective $E_b/N_0$ is defined in the context of links with faded subchannels, either time-varying as in Nanda and Rege's initial work, frequency-selective as with 802.11n OFDM, or spatially-variant with 802.11n MIMO. The Effective $E_b/N_0$ for a faded link is a number, defined as the signal-to-noise ratio that describes the power of a flat link with the same error performance as the link studied. An Effective SNR[1] model aims to compute the Effective SNR using information about the variation of fading across the different subchannels.

The output of the model is a set of Effective SNR values, one for each studied physical-layer configuration. These can then be used to compute a predicted *set of working physical layer configurations*. For each physical layer configuration in the application space—which can span the choice of modulation, coding scheme, transmit or receive antenna set, and more—the model predicts how well that configuration is likely to deliver packets. The application can then choose among the configurations in a way that optimizes its objective function. My thesis includes a detailed description of how my model can be used to solve general classes of wireless network configuration problems, and comprehensive evaluations for the key applications that arise in device-to-device networks.

### 1.3 Hypothesis and Contributions

I use this approach to demonstrate my hypothesis that *it is possible to rapidly and accurately predict how well different configurations of MIMO and OFDM wireless links will perform in practice, using a small set of wireless channel measurements*.

My Effective SNR-based model takes as input only a single CSI measurement for a wireless link, and from this can compute a predicted Effective SNR value for every point in the configuration space of that link. Predictions can be used for rate selection, because they will indicate the fastest rate to use, but also support concurrent adjustment of factors such as antenna selection, spatial streams, and transmit power level. For multi-device problems such as access point or channel selection, my model requires only a single CSI measurement for each link involved, and thus it can cover a larger configuration space with only a small set of channel measurements.

---

[1] In electrical engineering literature, $E_b$ denotes the energy of a bit and $N_0$ denotes the noise floor, so $E_b/N_0$ is the signal-to-noise ratio of a single bit. In the context of 802.11, in which SNR is derived from RSSI, we use a slightly different definition of SNR that is not normalized by the number of bits.

The channel state information, like RSSI, is estimated by a receiver using only the preamble of a received packet. This means that unlike packet delivery probes, CSI measurements can be obtained sending only short packets with no payloads. My single-threaded implementation can compute all the Effective SNR values for a 3x3-antenna, 20 MHz 802.11n link in under 4 μs, less than a single Wi-Fi packet preamble. The combined process of CSI measurement and Effective SNR computation is so fast that predictions of how well links can work can be fed back into control algorithms at near-instantaneous timescales compared to sending a single data-laden packet. In addition to quick measurement and decision, there is little information sharing required in order to enable configuration decisions across links. As a result the network can rapidly respond to varying wireless channels.

I emphasize that my model is practical. To demonstrate my hypothesis, I prototype my Effective SNR-based model in the context of 802.11n using commodity Intel Wi-Fi devices. The model is designed to integrate into modern wireless systems, including the practical implementation aspects of real hardware. Using this practical model and prototype, I demonstrate that my model can make accurate predictions of packet delivery. Thus my model provides good performance in practice. My thesis includes an in-depth evaluation of my model in the context of many wireless link and network configuration problems.

### 1.3.1 *Contributions*

To summarize, the contributions of this thesis are as follows:

- I develop a model that accurately predicts the error performance of different MIMO and OFDM configurations on wireless channels. This model is flexible to support a wide variety of transmitter and receiver device capabilities, device implementations, and configuration problems. I present an implementation of my model using a commodity 802.11n wireless device that demonstrates its feasibility in practice and handles the practical considerations of operation over real links using real, non-ideal hardware. This includes a detailed experimental evaluation of my system that shows that this model accurately predicts packet delivery over real 802.11n wireless links in practice.

- I detail how to use this model in a system that can solve a large number and variety of configuration problems similar to those described in Section 1.1. I evaluate this system in the context of a wide variety of 802.11n configuration problems. These evaluations show that the predictions output by my model lead to good performance in practice, and exceed the performance of prior probe-based and RSSI-based approaches.

- As part of my thesis I have produced an 802.11n research platform based on open-source Linux kernel drivers, open-source application code, and commodity Intel 802.11n devices using closed-source firmware that I customized. I have released this tool publicly, and at the time of writing it is in use at 23 universities, research labs, and corporations.

### 1.4  *Organization of this Thesis*

The rest of this thesis is organized as follows. In Chapter 2, I provide background information on wireless signals and systems in general, and the IEEE 802.11 standards in particular. Chapter 3 introduces the problem with using channel measurements to predict wireless link performance in today's hardware and using today's techniques, and it introduces my Effective SNR-based approach to solving it. In Chapter 4, I develop my Effective SNR model for 802.11n link performance, and demonstrate its ability to handle a wide range of transmitter and receiver configurations as well as wireless applications. I describe my measurement tool and experimental apparatus in Chapter 5. I use this platform to evaluate the ability of my model to predict error performance over a single link in Chapter 6. Next, I conduct a detailed study of the model in the context of rate selection for 802.11n in Chapter 7, and then present brief results for a variety of other configuration problems in Chapter 8. I place this thesis in the context of related work in Chapter 9. Finally, I present concluding thoughts along with a brief discussion of the next steps for this work in Chapter 10.

Chapter 2

# BACKGROUND

In this chapter, I establish the fundamentals of wireless communication and the IEEE 802.11 standards to the extent needed to understand my thesis.

## 2.1 Digital Communication Principles

Electromagnetic (EM) communications, which send data using *electromagnetic signals*, form the basis of the technologies I will discuss in this thesis. One key aspect of each wireless technology is which part of the electromagnetic spectrum it uses, characterized by its *carrier frequency or center frequency*, denoted f. A fundamental property of radio waves is that the frequency of a wave determines its *wavelength* $\lambda$ according to the relationship $c = f\lambda$, where c is the speed of light. IEEE 802.11 networks typically use EM signals with a carrier frequency in the range of 2.4 GHz and 5 GHz and corresponding wavelengths of about 12 cm and 6 cm.

Data transmission using EM signals works by *modulating* a pure sine wave with frequency f, i.e. by transforming the sine wave to reflect the underlying data. The simplest modulation scheme might be to turn the sine wave on or off depending on whether the bit to be transmitted is a 1 or a 0. The rate at which the transmitter varies the signal—in this example, the rate the sine wave is turned on or off—is called the *symbol rate*, and determines the *bandwidth* of the channel B measured in Hertz (Hz).

The *amplitude* of the sine wave, e.g., how much the peak varies from the zero (usually measured in volts (V)), determines the *power* of the signal. These two quantities are related by a quadratic relationship: Doubling the amplitude of a signal results in a quadrupling of the signal power.

In a *link*, that is a sender communicating data to a receiver, the sender generates a signal with *transmit (signal) power* level T that propagates through the *channel* connecting the two. The channel could be a *wire* or it could be the free-space *radio frequency (RF)* environment in which signals propagate from the transmitter's antenna to the receiver's antenna over the air.

### 2.1.1 The Wired Channel

To simplify the discussion, I will start with the case of a wired channel. The transmitted signal propagates down the wire to the receiver and then is received with *receive (signal)*

| Variable | Meaning | Units |
|----------|---------|-------|
| $f$ | Frequency | Hz |
| $\lambda$ | Wavelength | m |
| B | Bandwidth | Hz |
| T | Transmit signal power | dBm (decibels relative to 1 milliwatt) |
| S | Receive signal power | dBm |
| $\alpha$ | Attenuation | dB (decibels, unitless) |
| $\theta$ | Phase | radians |
| N | Noise power | dBm |
| K | Temperature | kelvins |
| $\rho$ | Signal-to-noise ratio (SNR) | dB |
| R | Shannon Capacity | bits per second (bps) |
| d | Distance | m |
| $n$ | Path loss exponent | unitless |
| I | Interference power | dBm |
| $\rho_I$ | SINR | dB |
| M | Number of transmit antennas | (antennas) |
| N | Number of receive antennas | (antennas) |

**Table 2.1: Table of notation used in this chapter.**

*power* S. While propagating through the wire, the signal gets slightly weaker as a small amount of energy is absorbed. The net effect of this absorption is called *attenuation*, denoted $\alpha$, and is defined mathematically as the multiplicative decrease in power induced by the channel:

$$\alpha = \frac{T}{S}. \tag{2.1}$$

In addition to attenuation, the wired channel also induces a *phase shift* as the electromagnetic signal propagates. The value of this phase shift, denoted $\theta$, depends on factors including the length of the wire and the frequency of the signal, and is generally considered to be an unknown, uniformly random quantity between 0 and $2\pi$.

The signal measured by the receiver is also corrupted by broad-spectrum electromagnetic noise. This corruption is sometimes called *Johnson-Nyquist noise* after its identification in 1927 by Johnson [52] and explanation in 1928 by Nyquist [83], but it is more commonly known as *thermal noise*. Thermal noise can be modeled as a complex Gaussian with average *noise power* N (in Watts) equal to

$$N = kKB, \tag{2.2}$$

where $k \approx 1.38 \times 10^{-23}$ (in Joules/kelvin) is Boltzmann's constant, K is the temperature (in kelvins), and B is the bandwidth. This is called additive, white Gaussian noise (AWGN).

In the context of 802.11, we typically measure power-related quantities on a logarithmic scale to capture the wide range of possible values. Power levels such as the quantities $T$, $S$, and $N$ are usually measured in decibels relative to 1 milliwatt, or dBm, and typically take on values like $T = 20\,\mathrm{dBm}$ ($100\,\mathrm{mW}$) and $S = -80\,\mathrm{dBm}$ ($10^{-8}\,\mathrm{mW}$ or $10\,\mathrm{pW}$). To calculate $N$, we can use Equation 2.2: Wi-Fi links typically use bandwidths $B$ of $20\,\mathrm{MHz}$ or $40\,\mathrm{MHz}$, which correspond to thermal noise levels of $-101\,\mathrm{dBm}$ and $-98\,\mathrm{dBm}$ at room temperature. In practice, the total noise is assumed to be thermal noise plus a $5\,\mathrm{dB}$–$15\,\mathrm{dB}$ *noise figure*, which is a quantity that estimates additional error added by imperfect analog hardware used in receiver processing. The total noise for a $20\,\mathrm{MHz}$ Wi-Fi channel might then be in the range of $-91\,\mathrm{dBm}$.

Now that we have defined the signal and noise powers, we can discuss the limits of the communication channel. In their seminal works, Ralph Hartley [40] and Claude Shannon [104, 105] proved that the *capacity* of a channel—i.e., the maximum data rate $R$ at which the transmitter and receiver can communicate—is determined by the channel's bandwidth and its *signal-to-noise ratio (SNR)*. The SNR, denoted by $\rho$, is a unitless quantity typically measured in decibels and calculated as

$$\rho = \frac{S}{N}.$$

(2.3)

The example signal power of $-80\,\mathrm{dBm}$ and noise power of $-91\,\mathrm{dBm}$ correspond to an SNR of $11\,\mathrm{dB}$.

The Shannon-Hartley Theorem [105] establishes what is called the *Shannon capacity* to be

$$R = B \log_2(1 + \rho).$$

(2.4)

Figure 2.1 shows this relationship for the normalized quantity $R/B$.

The Shannon-Hartley Theorem determines a bound on the maximum rate achievable as a function of the bandwidth and signal strength. However, it does not give a practical scheme that realizes this bound. Instead, systems like 802.11 use many different modulations that achieve different points along the $y$-axis and choose among these in practice depending on the underlying channel conditions.

Note that for real links, the values of $S$ and $N$ are not known a priori. Instead, transmitters choose an encoding, and the receiver will be able to decode it successfully if the choice falls below the curve for the SNR experienced. The general problem of choosing the modulation to use, as well as the selection of other physical layer parameters, is the focus of my thesis. I describe this problem in more detail in the next chapter.

The binary modulation system I discussed above is a scheme called On-Off Key-ing (OOK). Each symbol conveys 1 bit, and since the symbol rate is directly tied to the bandwidth used by a scheme, OOK can deliver at most 1 bps/Hz. A generalized form of OOK is Amplitude Shift Keying (ASK), which can send more bits per symbol using multiple power levels. m-ASK, i.e., ASK with m power levels per symbol, can deliver up to $\log_2(m)$ bits per symbol and thus can achieve a higher capacity.

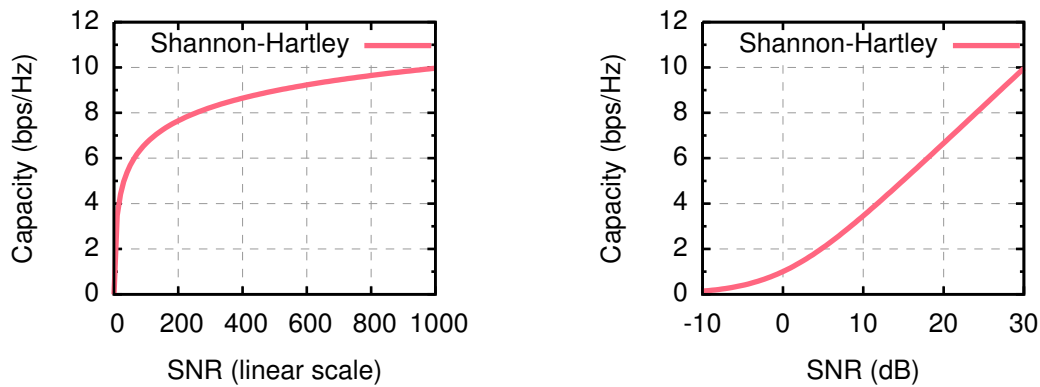

**Figure 2.1: The Shannon Capacity of a communications channel with Gaussian noise, presented in both linear and logarithmic (dB) scales.**

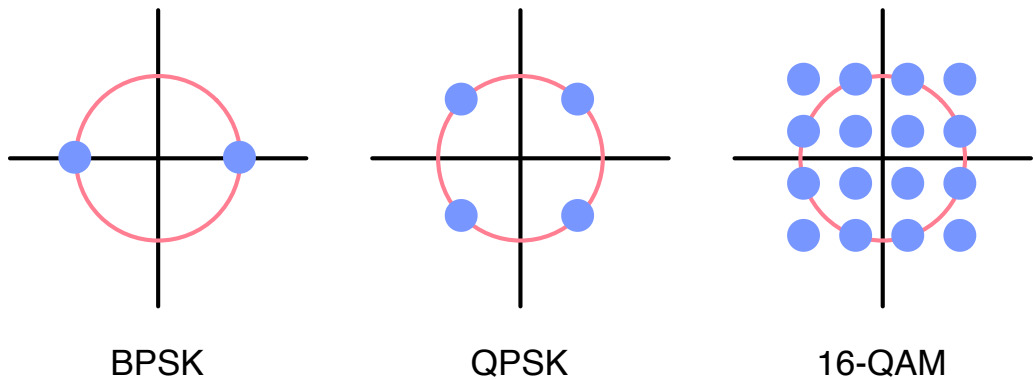

**Figure 2.2: Constellation diagrams for the BPSK, QPSK, and 16-QAM modulations. These constellations are normalized such that each modulation has equal average transmit power, indicated by the red circle.**

As mentioned above, electromagnetic signals actually have both an amplitude and a phase. Amplitude modulation varies one of these parameters, and a complementary scheme called Phase-Shift Keying (PSK) keeps the amplitude constant but varies the phase. A third scheme known as Quadrature Amplitude Modulation (QAM) varies both parameters simultaneously and results in a more efficient system when sending more than 2 bits per symbol. Noting that the polar coordinates given by amplitude and phase can equivalently be thought of as a complex number, m-QAM can be equivalently thought of as $\sqrt{m}$-ASK in both the real and complex dimensions simultaneously. Figure 2.2 shows the two-dimensional

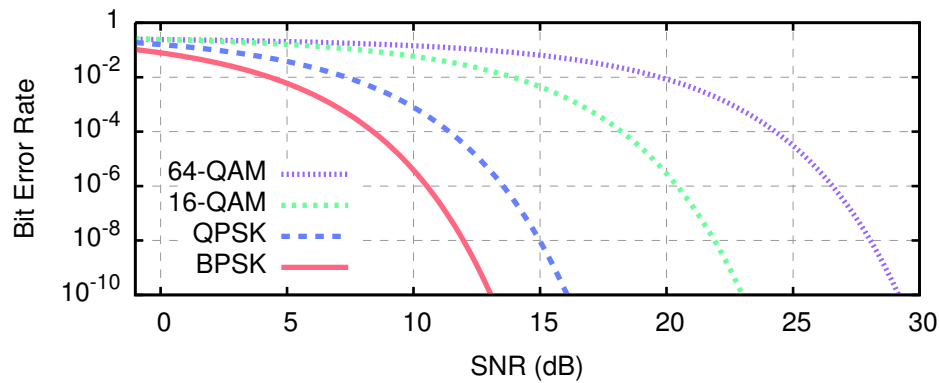

**Figure 2.3: The relationship between bit error rate and SNR for the four 802.11 modulation schemes.**

*constellations* that result from picturing the symbols sent in BPSK (i.e. 2-PSK), QPSK (i.e. 4-PSK), and 16-QAM modulation schemes.

There are many more modulation schemes than I have presented here, but PSK and QAM are the modulations applicable to 802.11. 64-QAM is the highest modulation currently used by Wi-Fi devices, though the future IEEE 802.11ac amendment [47] will add 256-QAM to this set.

Recall that the signal will be corrupted by noise when measured at the receiver. Under the standard AWGN model, we model this corruption as shifting the received symbol by a random complex vector whose length depends on the noise power. We see in Figure 2.2 that the different modulations have different constellation densities: The symbols of 16-QAM are clustered more closely than the symbols of QPSK or BPSK. This means that higher constellations which encode more bits per symbol are more vulnerable to noise. At low SNR, the receiver cannot easily distinguish between many symbols, so slower modulations with fewer constellation points should be used. At high SNR, the receiver can distinguish between more symbols and thus can use a denser constellation.

This property of the performance of different modulation schemes is closely related to the Shannon Capacity. Figure 2.3 illustrates the magnitude of this effect for the modulations used by 802.11 using textbook formulas [108] that relate the SNR to a bit error rate. We can also connect these different modulations directly to the Shannon-Hartley Capacity Theorem by examining the capacity achieved by each scheme as a function of SNR (Figure 2.4). In this graph, I assume an idealized coding scheme that delivers the maximum data rate for a given bit error rate; the practical schemes in widespread use today are somewhat less efficient in order to admit less expensive computation.[1]

---

[1] Though beyond the scope of this thesis, a number of recent proposals for practical *rateless codes* [33, 87]

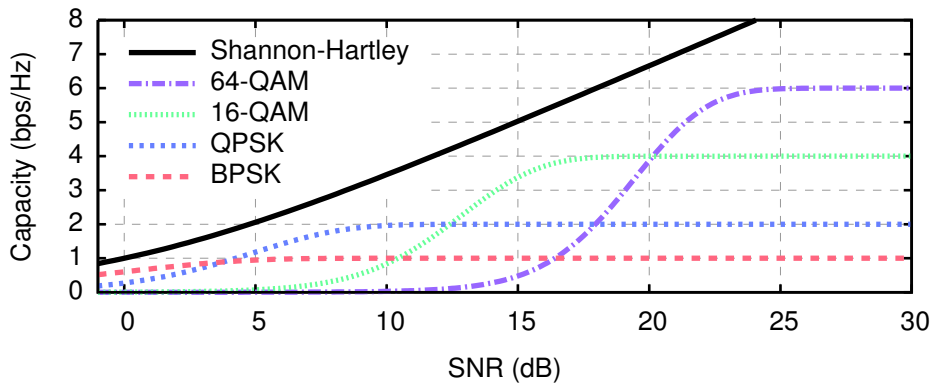

**Figure 2.4: The relationship between SNR and capacity for standard modulation schemes and idealized codes.**

### 2.1.2 The Wireless Channel

The previous section explained the basics of digital communications in the context of a wired link. Here, I expand to the significantly more complex case of a wireless channel.

In a wireless link, the electromagnetic signal is emitted from an antenna as a *radio wave* that then radiates through the *wireless medium*, i.e. the environment. The dominant source of attenuation in an wireless link is not absorption by the medium, but rather the diffusion of energy throughout the environment, of which a small fraction is captured by a receiver's antenna. This effect, called *path loss*, is captured by the Friis transmission equation, which yields the inverse relationship

$$S \propto \frac{T}{d^n},$$ (2.5)

where $d$ is the distance between transmitter and receiver, and $n$ is the *path loss exponent*. In free space, $n$ has a value of 2, reflecting the fact that the energy transmitted at a particular time is spread out over the two-dimensional surface of a sphere, an area that grows with $d^2$. (For a directional antenna, the energy is spread over a different geometric shape, e.g., a cone, but this shape will still have a two-dimensional surface area).

The path loss exponent varies in different indoor environments, but empirically tends to take on a value between 2 and 4 [108]. This empirical result is explained as the sum of many complex effects that result from the interaction of radio waves with objects in the environment. One such effect is *shadowing*, in which materials such as glass or metal prevent radio waves from passing through. I explain more additional, more complicated channel effects below.

In wireless systems, multiple devices might send at the same time and in the same

---

nearly achieve the Shannon Capacity bound by using much denser constellations and clever coding schemes across multiple transmissions.

frequency band; this *interference* causes a *collision* during which a receiver will measure the sum of both transmissions. There are a number of practical problems for operation during a collision, such as whether the receiver can properly lock onto the desired signal and estimate the effects of the channel on it. In general, however, we can model the reception probability by replacing the SNR with the *signal-to-interference-and-noise ratio (SINR)*, which treats the interfering signal of power I as another source of noise:

$$\rho_I = \frac{S}{I + N}. \tag{2.6}$$

While interference is an important problem, many systems, including 802.11, use *medium access control (MAC)* protocols to ensure that at most one device transmits at a time.

Beyond path loss, the most important channel effect is the inherent *multi-path* nature of indoor wireless environments. At 2.4 GHz and 5 GHz, RF signals bounce off metal and glass surfaces that are common indoors. This scattering leads to a situation in which many copies of the signal arrive at the receiver having traveled along many different paths. The net effect of this RF superposition depends on the phases of the individual signals. When these copies combine they may add constructively, giving a good overall signal, or destructively, mostly canceling the overall signal.

The phase-dependent nature of multi-path effects means that they vary over both frequency and space. For a given distance traveled d, the phase change is $2\pi d/\lambda$. Thus wideband channels may exhibit dramatically different received power levels for different frequencies; such channels are called *frequency selective*. Measurement studies of frequency-selective fading report signal variations as high as 15–20 dB [55]; in Chapter 3 I will present experimental evidence confirming these effects in the environments I studied.

With regard to spatial variation, the small 12 cm and 6 cm wavelengths of Wi-Fi signals means that small changes in path lengths can alter a situation from good to bad. Statistical models tell us that multi-path fading effects are independent for locations separated by as little as half a wavelength. This means that multi-path causes rapid signal changes or fast fading as the receiver moves, or in the case of a stationary node as the surrounding environment changes. Movement at fast speed also induces *Doppler effect*, which aggravates multi-path effects and makes the channel even more variable.

The net effect of multi-path fading is that the received wireless signal can vary significantly over time, frequency, and space. This is a problem for good performance because at any given time there is a significant probability of a deep fade that will reduce the SNR of the channel below the level needed for a given communication scheme.

However, an alternative way of looking at the effects of multi-path fading is that they provide *diversity*. In a sufficiently rich multi-path environment, there are so many combining copies of signals that the channel observed on different, nearby frequencies can be considered to be independently faded. For this reason among others, many systems including 802.11 use a scheme called *Orthogonal Frequency Division Multiplexing (OFDM)*. In OFDM, a wide frequency band is split into many *subcarriers* that each carry different modulated bits in parallel, with a higher level error-correcting code across them to take advantage of this *frequency diversity*.

To get fast rates while only using sending a single symbol at a time, a wideband system must have a fast symbol rate. Since OFDM sends many symbols in parallel on smaller subcarriers, an OFDM system sends each symbol for a longer period of time. Thus by turning a single fast channel into many parallel slower channels, OFDM allows more time for the channel to average out temporal fades and provides *time diversity*.

One more type of diversity is *spatial diversity*: Antennas separated by at least half a wavelength see independently-faded channels. Devices with multiple antennas can use schemes that take advantage of the spatial diversity these antennas provide. For example, a multi-antenna receiver measures multiple independent copies of each transmitted signal. Thus with clever signal processing, such a receiver can align the phases of these copies and add them together, which averages out the noise and improves overall channel performance. In a complementary manner, a multi-antenna transmitter with knowledge of the fading properties of the individual paths between pairs of antennas can steer its signal such that the multiple copies arriving at the receiver's antenna combine optimally. This process, in which the gain and phase of the signal emitted by each antenna are adjusted (with OFDM, this adjustment may be different for each subcarrier) is called *beamforming*.

Finally, suppose that both the transmitter and receiver have multiple antennas. The foundational work by Foschini and Gans [29] and Telatar [111] in the mid 1990s introduced *spatial multiplexing*, which uses this new spatial degree of freedom to improve capacity. Instead of sending the same data out each antenna as above, a transmitter with $M$ antennas can use its different antennas to send up to $M$ independent *spatial streams* of data. An $N$-antenna receiver will then measure $N$ different signals, each signal an independent linear combination of the $M$ transmitted streams. If $M \leqslant N$, the receiver has enough information to solve the linear system and separate the streams, thus providing an $M$-fold gain in performance. Thus spatial multiplexing, with $N$ antennas at each side, results in a modified capacity theorem:

$$R = BN \log_2(1 + \rho). \tag{2.7}$$

Together, spatial diversity and spatial multiplexing techniques form a set of what are called MIMO (multiple-input, multiple-output) techniques.

I conclude this discussion by mentioning one last channel effect relevant to 802.11:

*Inter-symbol interference.* In multi-path environments, some spatial paths can be so long that the delayed copies of the signal substantially overlap with the next symbol and make it harder to receive. The delay between the earliest and latest copies is called the *delay spread* of the channel, and it can be substantial. OFDM systems that use longer symbol times are more resilient to this effect, but still repeat each symbol for a period of time called a *guard interval*. If the guard interval is at least as long as the delay spread, the receiver can ignore the inter-symbol interference and still receive a complete symbol.

### 2.1.3   Summary

In this section, I have presented the fundamental principles of digital communication of wired and wireless channels, including the limits of noisy RF channels and how data is encoded. I have also described the most relevant channel effects that communicating devices must overcome, and the primary techniques used to do so. In the next section, I make this discussion concrete in the context of Wi-Fi by describing the specifics of the IEEE 802.11n standard.

## 2.2   The IEEE 802.11n Standard

The IEEE 802.11 (Wi-Fi) standard [44] is targeted towards defining a mode of operation for a *wireless local area network (WLAN)*, intended to provide medium-range connectivity ($\approx$100 m) using low transmit power (at most 1 W). It was first introduced in 1997, and has been amended many times since. In this thesis, I limit my discussion to the features of 802.11n, the newest physical layer amendment, and 802.11a, its predecessor.

Wi-Fi devices use unlicensed spectrum in the 2.4 GHz and 5 GHz bands, and must coexist with consumer electronics such as microwaves, cordless phones, and baby monitors. In addition to this cross-device interference, nearby Wi-Fi networks in separate administrative domains—such as neighboring apartments—may need to share the same channel. As a result, Wi-Fi networks are not planned in a centralized fashion, but rather use decentralized protocols that work towards a good solution in a distributed fashion. For instance, 802.11 includes a *carrier-sense multiple access (CSMA)* protocol to manage which devices send: In essence, a transmitter listens to ensure no other devices are transmitting before sending a packet, and reduces its sending probability exponentially (via *exponential backoff*) if its transmission is not acknowledged.

At the physical layer, 802.11 uses the modulation schemes and OFDM I described above, operating over 20 MHz channels. In conjunction with different modulations, 802.11 also uses error-correcting codes with different *coding rates* to achieve different operating points in the rate-robustness tradeoff space. I summarize the specific single-stream configurations in 802.11n as well as the resulting link data rates in Table 2.2.

| MCS | Modulation | Coding Rate | Data Rate (Mbps) |
|-----|-----------|-------------|------------------|
| 0 | BPSK | 1/2 | 6.5 |
| 1 | QPSK | 1/2 | 13.0 |
| 2 | QPSK | 3/4 | 19.5 |
| 3 | 16-QAM | 1/2 | 26.0 |
| 4 | 16-QAM | 3/4 | 39.0 |
| 5 | 64-QAM | 2/3 | 52.0 |
| 6 | 64-QAM | 3/4 | 58.5 |
| 7 | 64-QAM | 5/6 | 65.0 |

**Table 2.2: The single-stream 802.11n modulation and coding schemes (MCS). These are only slightly different than the 802.11a MCS combinations that achieved up to 54 Mbps. The increase in maximum rate comes from slightly more efficient use of OFDM subcarriers and a new, less redundant 5/6-rate code.**

The standard link metric is the *receive signal strength indicator (RSSI)*. The RSSI was included in the 802.11 standard from the beginning as "a measure by the [physical layer hardware] of the energy observed at the antenna used to receive the current [packet]" [44: §17.2.3.2]. There are no specified requirements on its accuracy, instead, it is only required to be "a monotonically increasing function of the received power" [44: §17.2.3.2], and is generally used by the hardware to tell whether another device is transmitting. In practice, however, the RSSI reported by commercial Wi-Fi chipsets is an estimate of the received signal power and can be meaningfully translated into units of dBm. In this case, RSSI can be used in combination with noise measurements to compute the SNR of the link.

The 2009 standard amendment to IEEE 802.11n [45] added functionality and protocols for multi-antenna techniques such as spatial diversity, spatial multiplexing, and beamforming. The 802.11n enhancements are shown in Table 2.3. Most of improvement in the maximum data rate—from 54 Mbps in 802.11a to 600 Mbps in 802.11n—comes from the ability to use wider channels and multiple spatial streams. Together, these enhancements add $2 \cdot 2 \cdot 4 = 16$ times as many configurations to the space of a single link. Beamforming is effectively an analog parameter and adds nearly unbounded options.[2] The gains of beamforming vary depending on the channel—for strong links, they tend to be small, but for weak links beamforming can provide dramatic performance improvements [7].

The hardware/software interface in 802.11n operates at the level of individual packets or continuously-transmitted batches of packets. Packets are sent to the hardware and transmitted over the air. The receiver detects a new transmission from the increase in energy, estimates the parameters of the wireless channel from the packet's standard, known preamble, and then decodes the packet. The standard behavior for 802.11 links is that all bits—after error correction—must be correct in order for a packet to be received, otherwise the packet is dropped by the hardware. Correctly received packets are delivered to the software layer in conjunction with physical layer configuration information about the transmission (e.g., what MCS in Table 2.2 was used) and reception (e.g., which receive antennas were used) of the packet, plus physical layer metrics of link quality.

---

[2] For transmitter and receiver each using 4 antennas on a 40 MHz channel, representing the beamforming matrices at maximal resolution takes 29,184 bits.

| Enhancement | Capacity Gain | Description |
|---|---|---|
| Short OFDM guard interval | 1.11× | Data can be more efficiently encoded when the multi-path delay spread is low. |
| Spatial multiplexing | 2× to 4× | Up to 4 concurrent spatial streams. |
| 40 MHz channels | 2.08× | More bandwidth, higher capacity (Eq. 2.4). |
| Beamforming | ?? | A sender with multiple transmit antennas can shape its signal to match the RF channel, improving both performance and reliability. |

**Table 2.3:** IEEE 802.11n adds a number of enhancements to the base single-stream configurations depicted in Table 2.2. The performance improvement from beamforming varies depending on the properties of the wireless channel.

## 2.3  Summary

In this chapter, I have presented the background information to provide a basic understanding of wireless channels and the specific IEEE 802.11n technology used to operate in them. As I described in Section 2.2, there are many different techniques that a transmitter and/or receiver can use to achieve robust operation in indoor wireless channels. However, the challenge—and the focus of most Wi-Fi research—is to decide which techniques to use, when to use them, and how to configure them to obtain the best operating point given the actual properties of the wireless channel. This is the primary problem I tackle in this thesis; in the next chapter I describe this problem in detail and given an overview of my approach.

Chapter 3

# PROBLEM AND APPROACH

The problem I study in this thesis is how to inform configuration decisions for wireless networks. I begin this chapter by presenting some of the primary problems in this space.

Next, I discuss how we handle these challenges today. There are two primary classes of techniques: (1) *statistics-based* schemes which only use packet reception or loss as an indicator of the suitability of a particular configuration for a specific channel, and (2) *channel-based* schemes which use measurements of the RF channel to predict packet delivery. Generally, packet delivery or loss is too specific, because a packet probe in one configuration says little about whether another configuration—say, at a different rate or antenna mode—-will work similarly. This means that many configurations must be tested. On the other hand, RF channel measurements offer the potential for predicting the performance of a configuration without testing it. However, the aggregate data such as RSSI recorded today and the way they are applied only coarsely reflect the true behavior of wireless operation over the underlying RF channel. Thus RF channel measurements are not presently accurate enough to provide good predictions.

My hypothesis is that it is possible to rapidly and accurately predict how well different configurations of MIMO and OFDM wireless links will perform in practice, using a small set of wireless channel measurements. To do so, I develop a comprehensive system that uses low-level RF channel measurements in conjunction with a simple but powerful model to predict the performance of every operating point in the 802.11n configuration space.

## 3.1 Problem: Rate Control for a Single Link

The problem of rate control is to find a rate configuration that can successfully deliver packets while maximizing performance. With 802.11n and its modern multi-antenna and physical layer techniques, this problem has become significantly more complex. To illustrate this, Figure 3.1 shows the available rate configurations in 802.11n for a device with three antennas. These configurations use the eight modulation and coding scheme (MCS) combinations described in Table 2.2 and the 802.11n enhancements shown in Table 2.3.

At the bottom of the figure, the SIMO line shows the eight single-stream configurations, which provide rates ranging from 6.5 Mbps to 65 Mbps. These are precisely the eight choices for rate that algorithms controlling a legacy 802.11a/g system must choose from.

This space expands by a factor of 12 when using 802.11n with three antennas. Adding a second (MIMO2) and third (MIMO3) spatial stream increases the maximum rate to 195 Mbps, for a total of 24 different configurations. For each of these configurations, 802.11n adds the optional use of double-width (40 MHz, HT40) channels, which raises the maximum rate to 405 Mbps with 48 choices. Finally, a physical layer tweak to a shorter OFDM guard interval (SGI) adds another ≈ 11% and pushes the fastest configuration to 450 Mbps among 96 possibilities.

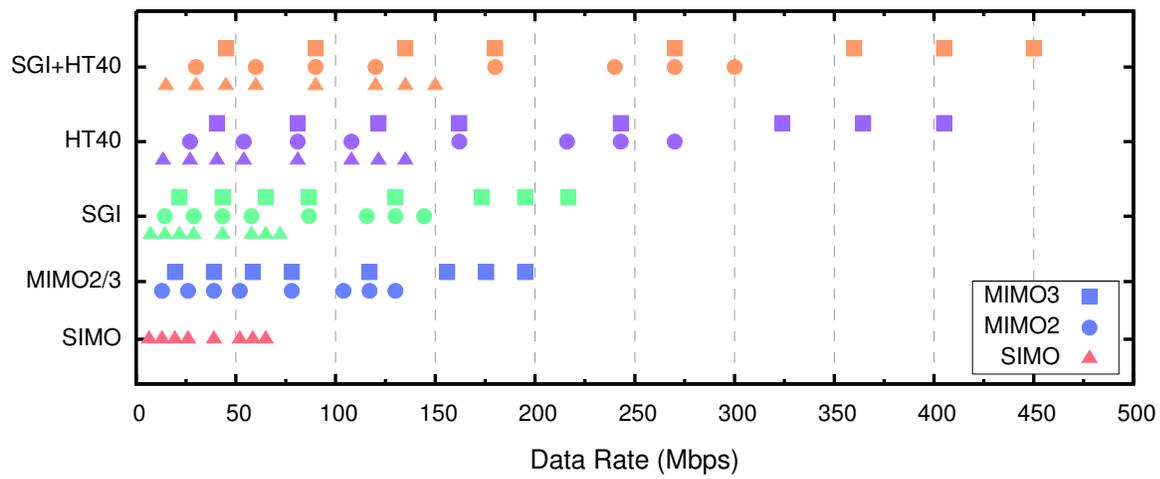

**Figure 3.1: The different rate-related 802.11n configurations that use three antennas and 802.11n physical layer enhancements.**

Looking forward, though three antennas are common today 802.11n can support up to four. The next amendment (802.11ac) will add two new single-stream rates using the 256-QAM modulation, plus channels of up to 160 MHz bandwidth. Combined, 802.11ac will comprise 320 configurations for rate alone—a factor of 40 more than 802.11a. Note that this analysis ignores related configurations such as antenna selection and beamforming, which exacerbate the problem. For the foreseeable future, the dramatic expansion in the rate space will continue as wireless technology evolves.

Having defined the basic problem of rate control, in the next two sections I describe the statistics-based and the channel-based approaches to solving it.

### 3.2   Existing Statistics-based Approaches

The majority of rate control algorithms today rely on packet loss statistics to adapt the operating rate. These algorithms use the number of losses as a signal of link quality. With too many losses, the channel is too poor to support the current rate, and the system should fall back to a lower rate. Conversely, when a link experiences a very small number of losses using its current rate, it sends some *packet probes* at a new faster rate, and switches to the faster rate if those probes succeed. Figure 3.2 shows a typical 802.11a adaptation search pattern, where each box corresponds to an 802.11a rate and the arrows show the faster (solid) and slower (dashed) rates that might be probed.

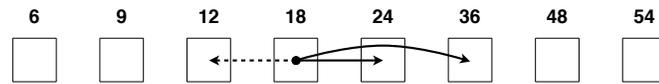

**Figure 3.2: A typical rate adaptation search pattern for 802.11a. When the current 18 Mbps rate works well, a rate control algorithm might probe one or two faster rates (solid lines). If the current rate results in many losses, the algorithm may fall back to the slower 12 Mbps rate (dashed line).**

The first algorithms (ARF [56] and AARF [66]) would only switch between a rate and the next fastest or slowest; later implementations look up to two rates ahead [14]. The dominant 802.11a rate adaptation algorithm used today is minstrel [109], which performs intelligent (biased) sampling of *all* rates to keep up-to-date estimates of the global rate space and can thus take discontinuous jumps. Recent revisions to these algorithms have focused on better handling of corner cases, such as improving performance when there is interference via adaptive control over RTS/CTS [109, 123].

Some algorithms have been proposed to use bit error rate statistics instead of loss rate statistics to adapt rate. SoftRate [120] estimates the bit error rate using a soft-output Viterbi decoder for error correction, and Chen et al. [22] designed a coding scheme called Error Estimating Coding (EEC) to enable accurate BER estimation at a higher layer.

### 3.2.1   Complication: Multi-dimensional Search Space

All of the statistics-based approaches, which walk up or down the list of rates based on whether the current rate works well, implicitly rely on the following basic assumption (outlined by Vutukuru et al. [120]):

> **Assumption:** *BER is a monotonically increasing function of the bit rate.*

But 802.11n rate configurations are *non-monotonic*. That is, it is not necessarily true that faster configurations are generally less likely to work than slower ones. This violates the axiom of these statistics-based approaches, and hence the *multi-dimensional* search space must be treated as such. I explain why in the following example.

Figure 3.3 shows three plausible *rate maps* for 3-antenna 802.11n links. In these rate maps, each row represents a different number of spatial streams, and each column represents a different MCS. A cell is shaded if a link can reliably deliver packets using that rate at that number of streams. The black box corresponds to MCS 12—2 streams at 39 Mbps each—which is the highest working 2-stream rate for these three hypothetical links.

In 802.11n, each of the scenarios (A), (B), and (C) illustrated in Figure 3.3 is possible. In particular, if the link can reliably deliver packets using MCS 12 then it is likely that MCS 4— the same encoding, but fewer streams—works as well. The same holds for MCS 11, since it uses the same number of streams but a less dense encoding. Similar logic implies all the shaded cells in link (A), which represents the most conservative situation in which MCS 12 works well. Conversely, link (B) exhibits the best corresponding situation; higher-encoding single-stream configurations may also deliver most packets, and there may be little penalty from using 3 streams, thus resulting in the same set of 3-stream links working. Finally, link (C) exhibits an average case, in which lower rates must be used as the number of spatial streams increases.

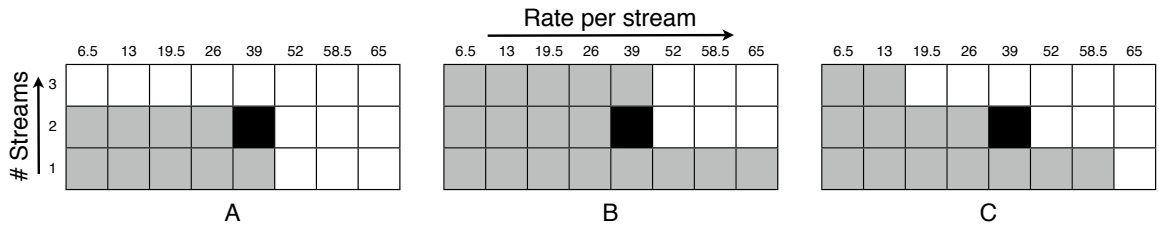

Figure 3.3: Three different rate maps for 802.11n links. On these links, MCS 12 (the black box) is the highest reliable 2-stream rate, and gray boxes indicate other reliable transmit configurations. A (*left*): The worst possible situation in which no 3-stream rates and no higher single-stream configurations work. B (*middle*): The best case in which all single stream rates work and all 3-stream rates work up to 39 Mbps each. C (*right*): An average case in which the set of reliable rates decreases as more spatial streams are used.

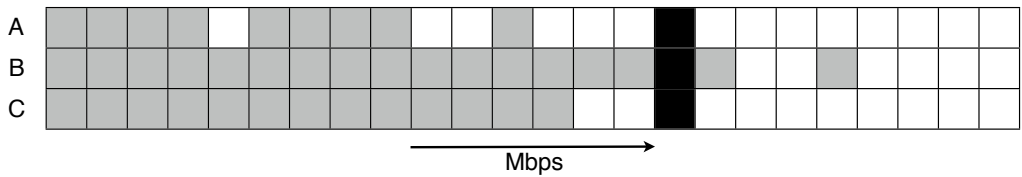

Figure 3.4: Rate maps for the links in Figure 3.3 now mapped into one dimension and sorted by total link speed (Mbps). In this view, the strict monotonicity assumed by rate adaptation algorithms is violated, and the violations occur in a channel-dependent way. Thus, 802.11n rate adaptation requires optimization along multiple dimensions.

The key is that each of these three links is plausible, which means that the search space for rate is non-monotonic in 802.11n. In Figure 3.4, I have redrawn the rate maps for these three links along a single-dimension, sorted by data rate. Ties in pure data rate—e.g., MCS 1 vs MCS 8, both at 13 Mbps—are broken such that fewer streams is lower in the search. Here we can see that for all three links there exist higher rates that work well where lower rates do not. I also plot the measured rate maps for 166 wireless testbeds links in Figure 3.5. These results show that with MIMO, 802.11n rates are not monotonic in practice. Thus a rate configuration algorithm for a multi-antenna link needs to consider a multi-dimensional search space.

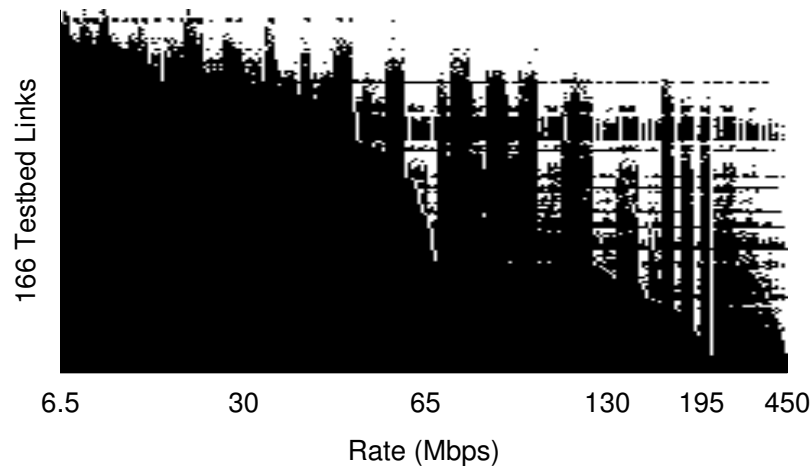

**Figure 3.5: Rate maps for 166 wireless links in my 802.11n testbeds show that with MIMO, the rate space is not monotonic in practice.**

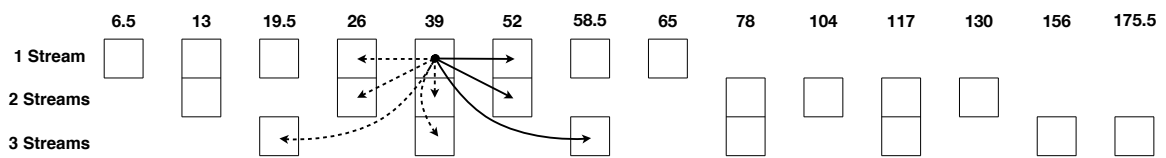

**Figure 3.6: A typical rate adaptation search pattern for 802.11n. The current rate configuration is MCS 4, which uses one 39 Mbps stream. There are three different immediately faster configurations, indicated by the solid lines. There are also 5 rate configurations as potential fallback rates, indicated by the dashed lines, that do not offer better bitrates but may be faster if MCS 4 experiences loss. These configurations offer different trade-offs between the number of spatial streams and the density of the modulations, and all are valid options for the next choice of rate.**

### 3.2.2 Current 802.11n Statistics-based Approaches

Figure 3.6 illustrates the multi-dimensional search challenge with a concrete example. Each row corresponds to a transmit configuration with one, two, or three spatial streams. The eight boxes correspond to the eight 802.11n MCS combinations, placed in columns that reflect the aggregate link speed.

As we saw in Figure 3.2, the single-dimensional search algorithm might try one or two rates higher, and during periods of loss it might fall back to the next lowest rate. In contrast, when increasing 802.11n rate from a single stream at MCS 4 (39 Mbps), the configurations MCS 5 (52 Mbps), MCS 11 (MIMO2-52 Mbps), and MCS 18 (MIMO3-58.5 Mbps) are all transmit configurations with better link speed, and each might work or not work depending on the channel. When MCS 4 experiences loss, there are five choices of fallback configuration. This includes the higher-stream MCS 10 (MIMO2-39 Mbps) and MCS 17 (MIMO3-39 Mbps) which both have the same link speed and might work, as they use more robust modulation and coding combinations (but require good separation between streams).

MiRA [84] is a recent research algorithm that implements a version of this multi-probing scheme, as does the minstrel_ht [28] algorithm used by the Linux kernel to do 802.11n rate selection in practice. A third approach is used by Intel's `iwlwifi` driver [49], which uses an 802.11a-like algorithm to select between MCS using a fixed number of streams, and adjusts the number of streams at coarse intervals.

### 3.2.3 State of the Art in Statistics-based Approaches

The dominant algorithms used in the Linux kernel today are minstrel [109] (for 802.11a/g) and minstrel_ht [28] (for 802.11n). In general, statistics-based schemes provide good performance for static links in which the devices do not move and the surrounding environment does not change. For such cases, the algorithms may be inefficient at first but will converge to a good operating point. The challenge, as pointed out by several works [43, 55, 120], is that—depending on the speed at which devices move or the environment changes—these algorithms may be slow to react to varying conditions for mobile links, resulting in significant performance degradation. In Chapter 7, I evaluate loss-based rate selection algorithms for 802.11n and confirm that the increased size of the search space and the larger number of rates that must be probed do indeed result in poor performance in fast changing channels.

SoftRate [120] and EEC [22] are the newest BER-based adaptation algorithms. Both algorithms provide faster adaptation than their loss-based counterparts because by using the BER they can distinguish between a rate that is barely working with marginal BER (in which case the next fastest rate will not work) and a rate that has a lot of headroom (in which case it is worth probing the next fastest rate). These algorithms perform well at shifting up and down within a monotonic rate space of 802.11a/g; however, their BER estimations do not apply across the orthogonal dimensions such as multiple spatial streams of 802.11n. To handle 802.11n, these algorithms would need to be amended to do multi-dimensional search as well.

## 3.3 Channel-based Approaches

The second class of approaches to configuring rate use channel information to guide rate selection or adaptation. As I described in Chapter 2 (Figure 2.3), textbook analyses of modulation schemes give delivery probability for a single signal in terms of the signal-to-noise (SNR) ratio [30]. These theoretical models hold for narrowband channels with additive white Gaussian noise. They predict a sharp transition region of 1–2 dB over which a link changes from extremely lossy to highly reliable. This feature in theory makes the SNR a valuable indicator of performance.

This gives rise to a simple SNR-based configuration scheme, at least for selecting rate:

Upon receiving a packet, a device can use the measured RSSI to compute the *Packet SNR* and predict the fastest rate supported. It can then feed this information back to the transmitter, which will use the newly selected rate for subsequent transmissions. This approach was explored in simulation by Holland et al. [43] with an algorithm called RBAR and found to work well.

### 3.3.1   Complication: Packet Delivery versus SNR in Practice

Though SNR-based rate control algorithms may work well in simulation, subsequent practical work found that the Packet SNR computed from RSSI was unreliable [3, 95, 127]. In very early devices, the RSSI was found to vary wildly over time or device temperature, providing unreliable thresholds; this was corrected by calibration in later devices (e.g., confirmed by Zhang et al. [126] and by my measurements). Reis et al. [95] found that RSSI estimates were corrupted by interfering transmissions. Finally, even in the absence of these latter effects, several studies found that the same RSSI value gives dramatically different performance for different links.

To understand which effects still hold for 802.11n hardware, I generated performance curves using an Intel Wireless Wi-Fi Link 5300 a/g/n wireless network card (I describe my experimental platform in Chapter 5). I connected two network cards together via a wire, and configured them to operate in a mode that uses a single antenna to transmit or receive. Using an inline variable attenuator I varied the amount of power received, and for each power level I sent around 1,000 packets using each of the eight 802.11n single-stream rates (Table 2.2) and measured the fraction of delivered packets, the *packet reception rate (PRR)*. With these measurements, I plotted the PRR against the link's SNR (computed from RSSI measurements at the receiver), and present the result in Figure 3.7.

This figure shows a characteristic sharp transition region between SNR values at which the link goes from lossy to working, 2 dB at low modulations up to 4 dB for the fastest 65 Mbps rate. There is also a clear separation between rates: At a given SNR value, it is clear which rate should be used. This wired link provides a good approximation of a theoretical narrowband channel despite the relatively wide 20 MHz channel, the use of 56 OFDM subcarriers, coding and other bit-level operations. This is the behavior we would want from a link metric in order to predict packet delivery.

In contrast, packet delivery over real wireless channels does not exhibit the same picture. Figure 3.8 shows the measured PRR versus SNR for three sample rates (6.5 Mbps, 26 Mbps, and 65 Mbps) over all wireless links in two wireless testbeds, using the same 802.11n NICs. The SNR of the transition regions can exceed 10 dB, so that some links easily work for a given SNR and others do not. There is no longer clear separation between rates. This is consistent with the measurements from prior work mentioned above [3, 55, 95, 126, 127].

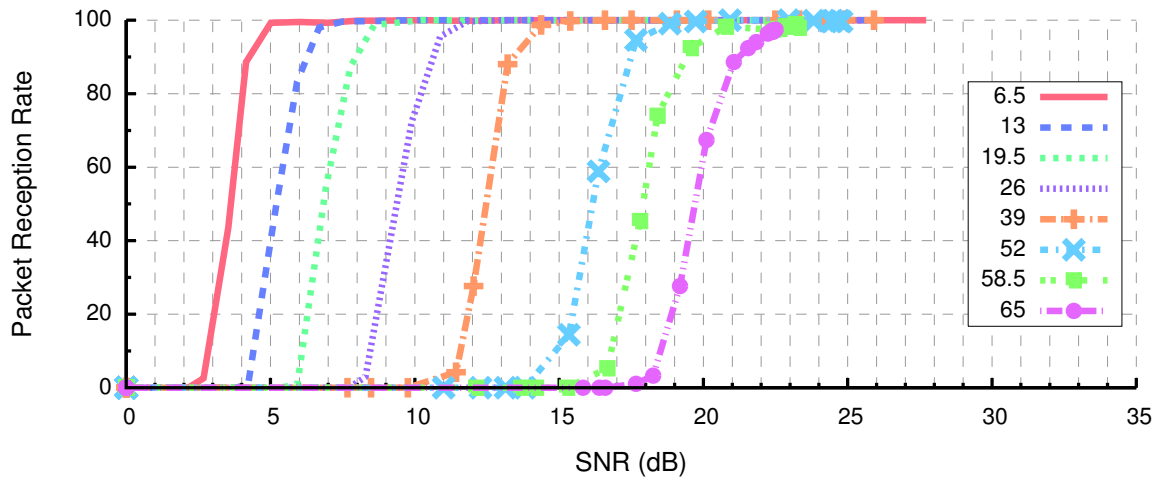

**Figure 3.7: A wired 802.11n link with variable attenuation has a predictable relationship between SNR and packet reception rate (PRR) and clear separation between rates.**

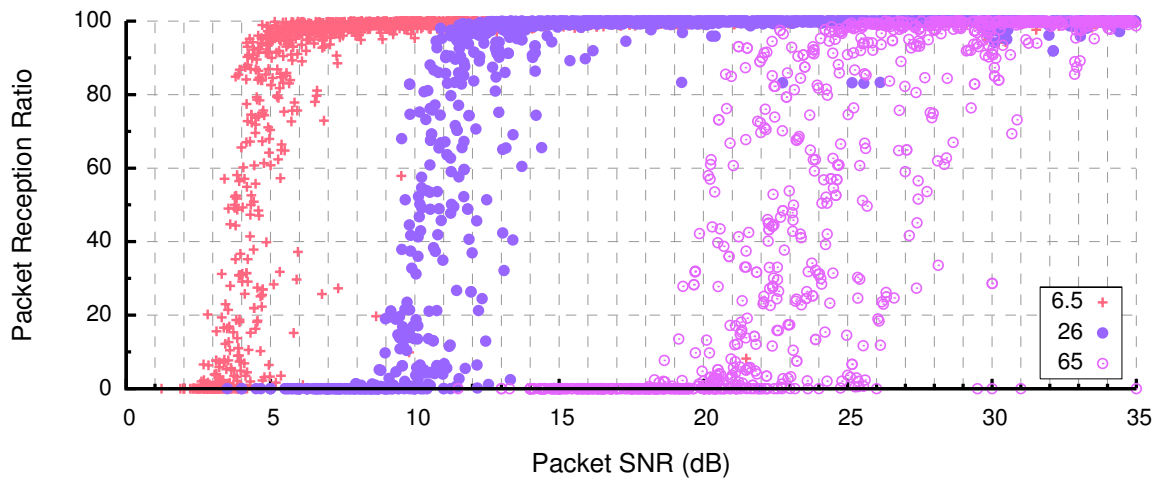

**Figure 3.8: Over real wireless channels in my testbeds, the transition region varies by 10 dB or more. The wireless channel loses the clear separation between rates. Only three rates are shown for legibility.**

### 3.3.2 State of the Art in SNR-based Approaches

Although prior studies and my measurements showed that Packet SNR does not accurately predict delivery across links, they also found that for a particular link a higher SNR generally has higher packet delivery for a given rate [3,55,126]. Consequently, there are two algorithms, SGRA [126] and CHARM [55], that use SNR feedback from the receiver in conjunction with packet loss statistics in order to learn the relationship between SNR and packet delivery online. Like statistics-based approaches, these algorithms work well for static links. Additionally, they provide good performance for fixed devices in mobile environments [55], because the learned relationship between SNR and PER is only slightly affected by moving objects and the learned calibration is generally valid. However, successive measurements by Vutukuru et al. [120] showed that CHARM tends to under-select in mobile links because it is unable to adapt its thresholds quickly enough to respond to the changing channel.

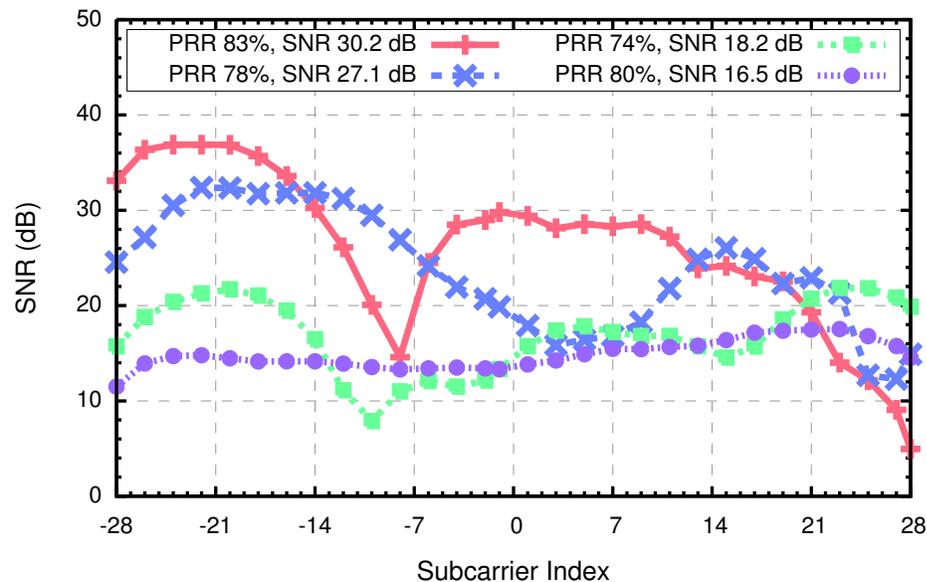

**Figure 3.9: Channel gains on four links that perform about equally well at 52 Mbps. The more faded links require larger RSSIs (i.e., more transmit power) to achieve similar PRRs.**

### 3.3.3 Complication: High-level Measurement of Low-level Subchannel Effects

I listed above several reasons that Packet SNR calculated from RSSI has historically been a poor predictor of performance. In the modern era of calibrated hardware, however, measurements no longer vary significantly with changing temperature or power level, or across devices. Instead, the dominant factor is likely to be the use of OFDM, and the presence of frequency-selective fading in the RF channel.

To illustrate this fact, I chose four representative links in my 802.11n testbed. These four links have SNRs ranging from 16 dB to 30 dB and yet they each perform similarly, delivering around 80% of packets sent using single-stream MCS 6 (52 Mbps). Figure 3.9 shows the packet reception rates and SNRs for these four links; it also includes the SNRs of the individual OFDM subcarriers for each links.

With this detailed picture, we can see that multipath causes some subcarriers to work markedly better than others although all use the same modulation and coding. These channel details, and not simply the overall signal strength as given by SNR, affect packet delivery. The fading profiles vary significantly across the four links. One distribution is quite flat across the subcarriers, while the other three exhibit frequency-selective fading of varying degrees. Two of the links have two deeply-faded subcarriers that are more than 20 dB down from the peak.

Because of these different fading profiles, these links harness the received power with different efficiencies. The more faded links are more likely to have errors that must be repaired with coding, and they require extra transmit power to compensate. Thus, while the performance is roughly the same, the most frequency-selective link needs a much higher overall Packet SNR (30.2 dB) than the frequency-flat link (16.5 dB). This difference of almost 14 dB (more than 20×) highlights why Packet SNR based on RSSI does not reliably predict performance.

To exacerbate this issue, 802.11n adds MIMO techniques to the network. During a multi-stream transmission, the receiver still records only one RSSI value per antenna. This RSSI (and the resulting SNR) reflects the total received power combined across all subcarriers and all spatial streams. The total power received will vary with the number of streams, and the actual performance of the link will depend on how well this total power is balanced across spatial streams and how well the receiver can separate the two spatial streams. Thus in 802.11n Packet SNR is likely to be even less accurate when predicting link performance.

### 3.3.4 Approach using Low-level RF Measurements

AccuRate [103] takes an alternative approach to using physical layer information to predict performance. Instead of measuring information about the *signal power*, AccuRate measures the *error vectors* (described in Chapter 2) of the received symbols when demodulating a packet. To make predictions about bit error rate, AccuRate then replays those same error vectors to a physical layer simulator, which models the reception of a packet using each of the different rates and selects the fastest successfully received packet. Though it would be impractical to implement a full physical layer simulator for each received packet, AccuRate was shown to be significantly more accurate than SNR-based algorithms with performance comparable to SoftRate. At the same time, AccuRate suffers from the same 802.11n-related flaws as the remaining algorithms: The magnitude of the error vectors will change depending on different numbers of spatial streams or channel widths or the use of a short guard interval, and AccuRate can handle none of these cases without implementing a multi-dimensional search.

## 3.4 Further Wireless Configuration Problems

In the previous section, I outlined the rate configuration problem, the current approaches, and the multi-dimensional aspects of OFDM and MIMO technologies that make them less effective. In this section, I briefly mention several other configuration problems and how they are solved in Wi-Fi networks today. (For a deeper discussion of each, see Chapter 9.) Together, these problems show the richness of the wireless network configuration space. Today, they are handled separately, if at all; in future wireless networks we will want the ability to configure them all concurrently.

### 3.4.1   Antenna Selection

A transmitter sending fewer spatial streams than it has transmit antennas may use only a subset of the available antennas to save power or pick the best subset to improve performance. Among production 802.11n drivers in Linux, only Intel [49] has an antenna selection algorithm; within a fixed number of streams it occasionally probes the different transmit antenna configurations. In this algorithm, switching between transmit antenna sets occurs on timescales of seconds or longer.

In the analogous scenario on the receive side, some chips automatically use only a subset of receive antennas when receiving a packet instead of fully maximizing available spatial diversity. These algorithms typically use Packet SNR measurements of the antennas to determine when additional antennas will add little gain. (Of course, this determination can be erroneous in the face of subchannel fading effects.)

### 3.4.2   Channel Width Selection

When both devices in a link support multiple channel widths, they may obtain better performance or power savings depending on the bandwidth they choose. Wider channels generally offer higher ideal Shannon Capacities, but since the total power is constrained this is not necessarily true at low SNR. Links may also wish to use smaller channels to avoid badly faded OFDM subcarriers.

SampleWidth [21] uses a probing algorithm to determine which width gives the best performance, aiming to achieve spectral isolation from interferers. For choosing between 20 MHz and 40 MHz bands, the algorithms in Linux tend to integrate channel width selection into the rate selection algorithm. When both widths are available, Intel's driver, which coarsely switches between different numbers of spatial streams, doubles the set of modes it probes by instantiating one copy for each potential bandwidth.

### 3.4.3   Transmit Power Control

Some devices have the ability to adapt their transmit power levels to save energy or to reduce interference with other nodes. In practice, however, all drivers seem to aim to optimize performance of their link, and simply use the maximum output power. The effects of transmit power control on a link are unpredictable (because RSSI does not capture OFDM fading), so research proposals to control transmit power typically rely on sampling performance and/or interference at various power levels.

### 3.4.4 Access Point Selection

When clients select which access point to connect to today, a number of factors are important, including both the quality of the link to the access point and also the load on that access point. Clients today typically choose access points solely based on the measured Packet SNR, on the assumption that it is a good proxy for downlink performance. Some proposed enterprise AP systems (e.g., DenseAP [78]) can take load into account as well.

One interesting note is that today's algorithms are not heterogeneity-aware. Consider a choice of two APs: (1) with 30 dB SNR and 3 antennas, and (2) a closer AP with 35 dB SNR but only 2 antennas. Today's clients will choose AP (2) with the larger SNR, even though the first will likely provide better bandwidth.

### 3.4.5 Channel Selection

Channel selection is the problem of choosing the best operating frequency for a pair (or set) of wireless devices. This is not a runtime choice in today's access point networks, because the frequency is chosen by the AP. However, channel selection is likely to be an important problem in device-to-device networks, e.g., Wi-Fi Direct. (I note that algorithms have been proposed for the related problem of distributed access point channel assignment to manage load and interference (e.g., [4]).

### 3.4.6 Multi-Hop Routing

Multi-hop paths will be needed in device-to-device networks, but are not needed in today's AP networks. Work in this space uses a combination of probing, Packet SNR-based heuristics, and state space reduction (e.g., assuming single-antenna devices and a single fixed rate network-wide). One more practical recent proposal by Bahl et al. [10] uses relays to address the rate anomaly problem in Wi-Fi access point networks, and uses Packet SNR to predict bitrate and calculate how well paths work. The performance of these solutions depends on the accuracy of these heuristics, which have not yet been adapted for heterogeneous devices or been made MIMO-aware.

### 3.4.7 Spatial Reuse

Spatial reuse is the problem of managing concurrent transmissions that occupy the same frequency. In today's access point networks, algorithms typically aim to have only one transmission at a time and adaptively turn on RTS/CTS when necessary to eliminate hidden terminals that hurt performance. This effectively avoids spatial reuse entirely.

CMAP [121] is a state-of-the-art algorithm to promote spatial reuse that determines whether two links can operate concurrently on the same frequency. However, CMAP only works by fixing the entire network to homogeneous single-antenna devices using the same rate, and even so it requires a complex distributed probing algorithm in a static environment to achieve good performance.

### 3.4.8 Beamforming

Finally, there is the problem of beamforming, that is, for a transmitter to shape the combined signal sent out its antennas so that the spatial paths best combine at the receiver. This is not yet used in 802.11n, so I discuss the theoretical work and challenges to practical implementations.

Theoretical gains from beamforming are evaluated by measuring the increase in Shannon Capacity using ideal hardware and optimal receiver algorithms, rather than by the practical constraints of real hardware. I do not know of any practical systems for Wi-Fi that use this type of beamforming, but such a system would need a way to ground the output of the theoretical model to evaluate how the link would work in practice. Rather than an abstract complaint, there are real issues such as the tension between the optimal "water-filling" algorithms for beamforming [115] (which allocate power unequally across antennas or subcarriers) and practical hardware constraints such as amplifier peak-to-average-power limits.

In practice, I imagine that this grounding would be done by probing the link performance, as has been the case with analog beamforming strategies [73] that operate in a fundamentally different way than 802.11n-like beamforming.

### 3.4.9 Summary

I have described several configuration problems for wireless networks. In Wi-Fi today, these tasks are complicated to implement and have not been adapted to new technologies like the use of multiple antennas in 802.11n. In practice, these challenges mean that these tasks are often simplified drastically or avoided entirely. In my thesis, I aim to make these configuration problems practical solve.

## 3.5 My Approach: An Effective SNR-based Model for Wi-Fi

My hypothesis is that *it is possible to rapidly and accurately predict how well different configurations of MIMO and OFDM wireless links will perform in practice, using a small set of wireless channel measurements*. I develop a framework to evaluate how well a particular physical layer configuration works that is flexible enough to handle all the problems discussed in this section. My system uses low-level RF channel measurements in conjunction with a simple but powerful model to predict the performance of each operating point in a large physical-layer configuration space.

In particular, I develop a practical methodology that uses low-level measurements of the RF channel and the concept of an Effective SNR [79] to predict performance for wireless channels that use modern physical layer technologies such as OFDM, multiple antennas, variable channel widths, and spatial diversity and multiplexing. I also explain how to apply this prediction engine to a wide variety of link and network configuration problems such as those described in this chapter. To demonstrate that this methodology is practical, I prototype a working system in the context of IEEE 802.11n, which is the dominant consumer wireless networking technology today and includes these state-of-the-art RF techniques. Finally, to show that my model works, I use my prototype to evaluate the accuracy of the choices made using my techniques. I find that my model accurately predicts packet delivery over hundreds of indoor wireless links in two environments, and that this level of accuracy is sufficient to lead to good configurations for many link and network problems. Because channel measurements can be obtained rapidly and Effective SNR uses lightweight computation,

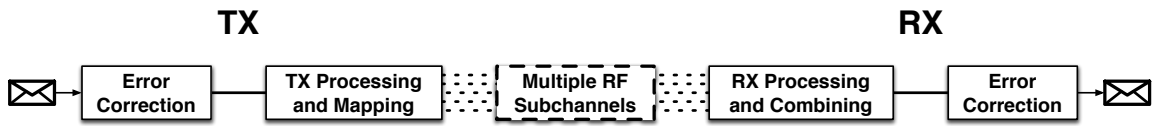

**Figure 3.10: Simplified overview of an RF link operating over multiple subchannels.**

### 3.5.1 *Effective SNR-based Model*

The central component in my thesis is a model for packet delivery that uses low-level RF measurements called CSI (see Section 3.5.2 below) to predict packet loss over real wireless channels. To be useful, this model must accurately predict the packet delivery probability for a given physical layer configuration operating over a given channel. It must also be simple and practical, so that it can be readily deployed, and it must cover a wide range of physical layer configurations, so that it can be applied in many settings and for many tasks.

In this thesis, I scope my model to devices that use MIMO and OFDM, which captures the fundamental technological primitives for many current and future networks. In particular, the scope of my model is 802.11n including all the enhancements described in Section 2.2. My model is based on the concept of Effective $E_b/N_0$ developed by Nanda and Rege [79], and described as follows:

Figure 3.10 shows a simplified overview of a link operating over an RF channel that has multiple subchannels, such as MIMO spatial paths or OFDM subcarriers. The transmitter applies error correction to the original data packet, and then processes the coded bitstream and maps the resulting symbols onto the multiple subchannels. The receiver processes the noisy signal to recover the (potentially errored) coded bitstream, and then uses error correction to attempt to recover the original data bits.

The key hypothesis introduced by Nanda and Rege is that error correction—in conjunc-tion with mechanisms like frequency- and spatial-aware interleaving in 802.11n—works to spread the errors caused by faded subchannels across the entire channel. If this assumption holds, the link can be modeled as performing with an aggregate error rate equal to the average error rate across subchannels. This average bit error rate is called the *Effective BER* of the channel, and from it we can compute the *Effective SNR* of the channel. Since the four links displayed in Figure 3.9 have similar error performance, they should have similar Effective SNRs. Then the Effective SNR can be used as a metric of link quality and hence can provide accurate estimates of packet delivery.

My model is based on Nanda and Rege's work, which was framed in the context of time-varying channels for CDMA. I extend their work to handle OFDM and MIMO subchannels. I include practical considerations such as implementation constraints in real hardware and the effects of operating over real channels. I also present the first experimental evaluation of Effective SNR as applied to Wi-Fi.

### 3.5.2   Model Input: Fine-grained RF Measurements

As described above, the input to my Effective SNR-based system is a set of low-level RF channel measurements. The particular measurements I use in this thesis are called *Channel State Information (CSI)*. For an OFDM link, the CSI comprises the channel gain coefficient (amplitude and phase,[1] see Chapter 2) for each OFDM subcarrier. For an NxM MIMO link, the CSI is an NxM matrix where each entry reflects the channel gain coefficient from one transmit antenna to one receive antenna. For a MIMO-OFDM link such as in 802.11n, the CSI comprises a three-dimensional NxMxS matrix that reflects the NxM MIMO link for each of S subcarriers.

A single comprehensive CSI measurement captures the low-level channel details that enable my model to calculate the Effective SNR for a wide configuration space. I next summarize how these measurements are used.

### 3.5.3   Model Output, and how to Apply it

I describe the model and how it is used in complete detail in the next chapter, but the basic structure of the model is simple: Given (1) a current CSI measurement of the RF channel between transmitter and receiver, and (2) a target physical layer configuration of the transmit and receive NICs, it predicts how well that link will deliver packets in that configuration.

This simple decision primitive integrates easily into higher-layer optimization protocols. These include solutions to all of the problems mentioned in this chapter, such as selecting the best rate, number of spatial streams, or transmit antenna set; whether to use 20 MHz or the entire 40 MHz channel; or choosing the lowest transmit power at which the link supports a particular rate. The key is that my model makes the most complicated step of those protocols—evaluating how well a link will work in a particular configuration—trivial.

---

[1] Note that in Figure 3.9, I plot only the amplitude as the phase offset does not affect packet delivery for a SISO link (assuming it is properly equalized by the receiver).

### 3.5.4   Problems Addressed

| Application of Effective SNR | Described in |
|---|---|
| Bitrate/MCS selection | Chapter 6, Chapter 7 |
| Channel width selection | Chapter 6, Chapter 7 |
| Antenna selection | Chapter 6, Chapter 7 |
| Transmit power control | Chapter 6 |
| Channel selection | Chapter 8 |
| AP selection | Chapter 8 |
| Multi-hop path selection | Chapter 8 |
| Interference planning/Spatial reuse | *Future work* |
| Partial packet recovery/FEC | Bhartia et al. [13] |
| Beamforming | *Future work* |
| Multicast rate selection | *Future work* |

**Table 3.1: A variety of applications of Effective SNR. I describe and evaluate how to solve many of these problems with Effective SNR in this thesis, some have been addressed by other researchers using my research platform, and I leave some problems for future work.**

Table 3.1 shows a list of several potential applications of Effective SNR. These cover all the problems described above and range from optimizing various parameters of a single Wi-Fi link, such as the MCS or antenna set used, to coordinating many nodes in a dense wireless network. Additionally, I identify applications that can be implemented by looking at other aspects of the Channel State Information in Table 3.2. These provide useful primitives that can enable systems to adapt behavior based on the location and movement of the user.

Combined, I believe these applications form the critical building blocks for configuring dense future wireless networks like Wi-Fi Direct. In particular, my Effective SNR model provides the information needed to select rates or configure the network topology, among other things. The CSI can be used to supplement these schemes, particularly by using mobility classification to determine when a device starts to move and trigger reconfiguration of the wireless network in response. I implement and evaluate many of these applications in the rest of this thesis, several have been investigated by other researchers in follow-on work, and some are left for future research.

| Application of CSI | Described in |
|---|---|
| Mobility classification | Chapter 8 |
| Guard interval selection | *Future work* |
| Indoor localization | FILA [124], PinLoc [102], SpinLoc [100] |

**Table 3.2: A variety of applications of Channel State Information. I describe and evaluate how to classify mobility in this thesis. Some problems have been addressed by other researchers using my research platform. I leave guard interval selection for future work.**

### 3.6 Summary

In this chapter I have presented a detailed overview of wireless link and network configuration problems, and the statistics-based and channel-based approaches used today. Generally, packet loss statistics are too specific, applying only to a single or a few configurations. These approaches therefore require packet probes of many different operating points, and are slow to converge in changing channels. Conversely, channel measurements used previously have been too general, not capturing the low-level details of the channel. Thus they do not provide accurate predictions and cannot be used to select operating configurations in practice.

I then described my channel-based approach, which uses low-level channel measurements of the MIMO and OFDM subchannels in conjunction with an Effective SNR-based model to provide a way to predict performance over the broad configuration space. In the next chapter, I flesh out this model and how to use it, and argue that it is indeed flexible and has low overhead. In the remainder of the thesis, I will show that my model makes accurate predictions that lead to good choices of operating points in practice for many of the problems I described in this chapter.

Chapter 4

**EFFECTIVE SNR MODEL**

In this chapter, I detail my Effective SNR-based model for Wi-Fi. I set the stage for the discussion with a detailed description of the operation of a MIMO-OFDM link.

I then present the core Effective SNR model, showing how the input channel state is processed to determine how well a particular transmitter/receiver configuration set will deliver packets. This process is conceptually simple, but each step is complex, because the model must be able to handle a wide range of transmitter and receiver techniques and their implementations. The challenge is to capture these complexities in a relatively simple model.

Having described my model and how it meets this goal, I then explain how an algorithm can use it make configuration decisions. I also describe what the wireless on-air protocols for using my model actually look like. I conclude this chapter by comparing my Effective SNR approach to other state-of-the-art techniques for understanding and predicting the performance of wireless channels.

### 4.1 Overview of a MIMO-OFDM Link

I begin with a detailed description of a MIMO-OFDM link in the context of 802.11n in order to explain the configuration space and implementation-specific choices that my model my model must support (Figure 4.1). The first block shows standard transmitter processing that generates $S$ spatial streams of modulated symbols from the bits of a packet. The internals of this block scramble the original packet to randomize the bits, add error correction, then split the coded bits across the $S$ spatial streams and interleave them between the OFDM subcarriers. These steps spread bits that are coded together across frequency- and spatially-diverse subchannels, after which the transmitter modulates the spread bits into the S streams.

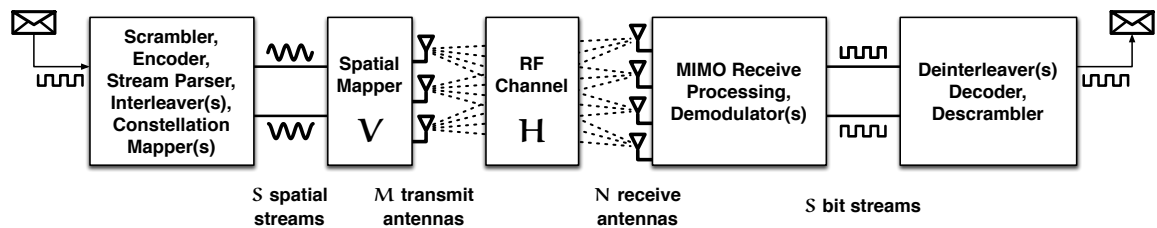

**Figure 4.1: A detailed view of a MIMO-OFDM link in the context of 802.11n.**

The second block is the *spatial mapper* **V**, which maps the S spatial streams to the M transmit antennas. Different spatial mapping algorithms might map each stream to a single antenna, send a linear combination of each stream to each antenna, or (when $S < M$) use *Space-Time Block Codes (STBC)* to take advantage of the extra spatial diversity provided by excess transmit antennas.

These signals then propagate across the RF channel **H** to the N receive antennas. Note that although the figure shows a single RF channel **H**, the channel can actually be different for every OFDM subcarrier. Consequently, transmitters with the ability to use beamforming can choose for each subcarrier $i$ a different spatial mapping $\mathbf{V}_i$ designed to make the best use of the channel $\mathbf{H}_i$.

After reception, the receiver employs one of many MIMO processing algorithms (i.e., MIMO equalizers) to disentangle the S streams from the N received signals, and demodulates the symbols to recover S (potentially errored) streams of bits. This can be *hard demodulation* that simply outputs bits, or *soft demodulation* ( [50, 108: §5.3.1.3]) that includes a confidence value for each decoded bit based on the amount of noise in the channel.

In the last block, the receiver deinterleaves and decodes the coded bits, and then descrambles them to undo the transmit processing and recover the original bit stream. At this point, IEEE 802.11n devices typically compute the checksum of the received data and if correct, deliver the packet to the network stack on the host. This completes the description of the most important operations in the sending of a packet from transmitter to receiver in an 802.11n link.

I separated the link into the blocks shown to reflect the considerations that a practical model must handle. The transmitter operation in the first block is completely specified by the IEEE 802.11n standard and the selected rate and channel width. In contrast, the transmitter has a wide choice of spatial mapping matrices to implement unspecified antenna selection, transmit power control, and beamforming algorithms, among others. The receiver can adapt its configuration in response to the channel, for instance by disabling certain antennas and receive chains to save power. To process the received signals there are several MIMO equalizers, demodulation techniques, and error correction decoders that trade off complexity, cost, and performance, and the standard leaves these choices to the implementor. A practical model must be general and flexible to support these many algorithmic and implementation concerns.

### 4.2   Effective SNR Model Overview

Figure 4.2 gives an overview of my Effective SNR-based model for wireless links, designed to handle the cases described in the previous section. At the left, the primary input to the model is an RF measurement of the ground truth *Channel State Information (CSI)* for the wireless link. In the context of MIMO-OFDM technology such as IEEE 802.11n, the CSI is a set of NxM matrices $\mathbf{H_i}$, where each matrix describes the MIMO channel between the M transmit and N receive antennas for one OFDM subcarrier.

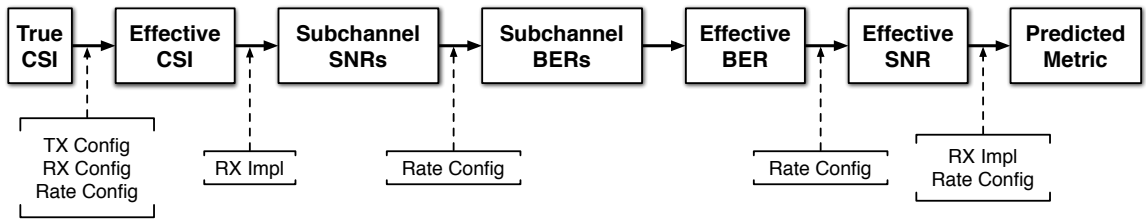

**Figure 4.2: Overview of my Effective SNR-based model for wireless links. The model takes as input a CSI measurement (the "true CSI") along with a transmitter configuration, receiver configuration, rate configuration, and some information on the receiver implementation. The output is a prediction of how well the link will work using the specified rate and device configurations, using a task-specific metric.**

The other inputs to the model are the configuration of the transmitter and receiver devices, the rate configuration, and some information about the receiver implementation. The final output of the model is a prediction of how well the link will work using the specified rate and device configurations, using a task-specific metric. This model can flexibly handle a wide variety of configurations: By varying the input transmitter, receiver, and rate configurations, we can solve all the problems described in Chapter 3.

In the rest of this chapter, I explain each step of my model. To ease exposition, I start with the core Effective $E_b/N_0$ algorithm from Nanda and Rege [79], by which I convert Subchannel SNRs to an Effective SNR. I then explain how a configuration algorithm can use this Effective SNR to predict how well this configuration will work. I make the model concrete in the context of IEEE 802.11n by explaining how to calculate the Subchannel SNRs from the *Effective CSI*, which is the CSI that would be measured if the link operated in the specified configuration. Next, I explain how to compute the Effective CSI as a function of the ground truth CSI and the specified configuration, thus showing how this model can support a variety of configuration problems. Finally, I conclude by discussing how this model can be used in practical scenarios, including which side of the link performs the computation and what information is communicated.

### 4.3  Computing Effective SNR from Subchannel SNRs

At the core of my model is the Effective $E_b/N_0$ algorithm from Nanda and Rege [79], which uses information about the fading properties of the individual subchannels in a link to compute an SNR value for that link that is directly tied to its overall error rate. This process

| Variable | Meaning |
|---|---|
| $M$ | Number of transmit antennas |
| $N$ | Number of receive antennas |
| $S$ | Number of spatial streams |
| $T$ | Number of OFDM subcarriers (tones) |
| $\mathbf{H}$ | Channel state information (CSI) matrix |
| $\mathbf{H}'$ | Effective CSI matrix |
| $\mathbf{V}$ | Spatial mapping matrix |
| $C = ST$ | Number of subchannels |
| $i, j$ | Subchannel indices |
| $\rho$ | Signal-to-noise ratio (SNR) |
| $\beta$ | Bit error rate (BER) |
| $k$ | Number of bits per symbol |
| $\rho_{\text{eff}}, \beta_{\text{eff}}$ | Effective SNR or BER |
| $\mathbf{V}_i, \mathbf{H}_i$ | Per-subcarrier spatial mapping or channel state matrix |
| $m$ | Modulation and coding scheme (MCS) index |
| $\mathcal{M}(c, \rho_{\text{eff},c})$ | Function that computes the metric for configuration $c$ |
| $\tau_m$ | Effective SNR threshold for MCS $m$ |

**Table 4.1: Table of notation used in this chapter.**

| Modulation | Bits/Symbol (k) | $\text{BER}_k(\rho)$ |
|---|---|---|
| BPSK | 1 | $Q\left(\sqrt{2\rho}\right)$ |
| QPSK | 2 | $Q\left(\sqrt{\rho}\right)$ |
| 16-QAM | 4 | $\frac{3}{4}Q\left(\sqrt{\rho/5}\right)$ |
| 64-QAM | 6 | $\frac{7}{12}Q\left(\sqrt{\rho/21}\right)$ |
| 256-QAM* | 8 | $\frac{15}{32}Q\left(\sqrt{\rho/85}\right)$ |

**Table 4.2: Bit error rate as a function of the symbol SNR $\rho$ for narrowband signals and OFDM modulations. Here $Q$ is the standard *Q-function*, the tail probability of the standard normal distribution. *IEEE 802.11ac will add 256-QAM.**

works as follows:

Suppose that we are given a set of Subchannel SNRs, indexed such that $\rho_i$ corresponds to the SNR for the $i$th subchannel, $i \in 1 \ldots C$. The first step is to convert the Subchannel SNRs to Subchannel BERs. In Table 4.2, I give the formulas that relate SNR to BER for the modulations used in 802.11. These are adapted from textbook formulas [108: §3.7.1 and §7.9.3.1] to use the SNR that is measured by wireless NICs instead of the $E_b/N_0$ that is traditionally used in textbooks. Because different modulations have distinct constellations, each modulation has a slightly different error rate function identified as $BER_k$, where $k$ is the number of bits encoded by one symbol. I use $BER_k^{-1}$ to denote the inverse mapping from BER to SNR.

Because modern technologies use narrowband subchannels (such as OFDM subcarriers), we can assume that these formulas are accurate with respect to subchannel SNR and BER, unlike the packet-level SNR and BER for the entire link. Then we can compute the Effective BER (denoted $\beta_{eff,k}$)

$$\beta_{eff,k} = \frac{1}{C} \sum_i^C BER_k(\rho_i). \tag{4.1}$$

That is, the Effective BER is the average bit error rate across subchannels. The Effective SNR ($\rho_{eff,k}$) is then defined as the SNR for a narrowband channel with that error rate:

$$\rho_{eff,k} = BER_k^{-1}(\beta_{eff,k}). \tag{4.2}$$

Note that the BER mapping and hence Effective SNR depend on the modulation ($k$), because different constellations have different densities (recall Figure 2.2). That is, unlike the RSSI, fixing all parameters but rate for a particular 802.11n wireless channel will still result in four different Effective SNR values, each one describing performance for each of the modulations. (IEEE 802.11ac will have five.) In practice, the interesting regions for the four Effective SNRs do not overlap because at a particular Effective SNR value only one modulation will be near the transition from useless (BER $\approx$0.5) to lossless (BER $\approx$0). When graphs in this paper are presented with an Effective SNR axis, I use all four values, each in the appropriate SNR range.

Here, I use the Effective SNR $\rho_{eff,k}$ to represent the Effective SNR for a particular modulation scheme. To generalize to configuration problems beyond rate, I use $\rho_{eff,c}$ to denote the Effective SNR of a link in a particular configuration $c$. Note that $c$ includes both the choice of modulation as well as the other Wi-Fi physical layer parameters.

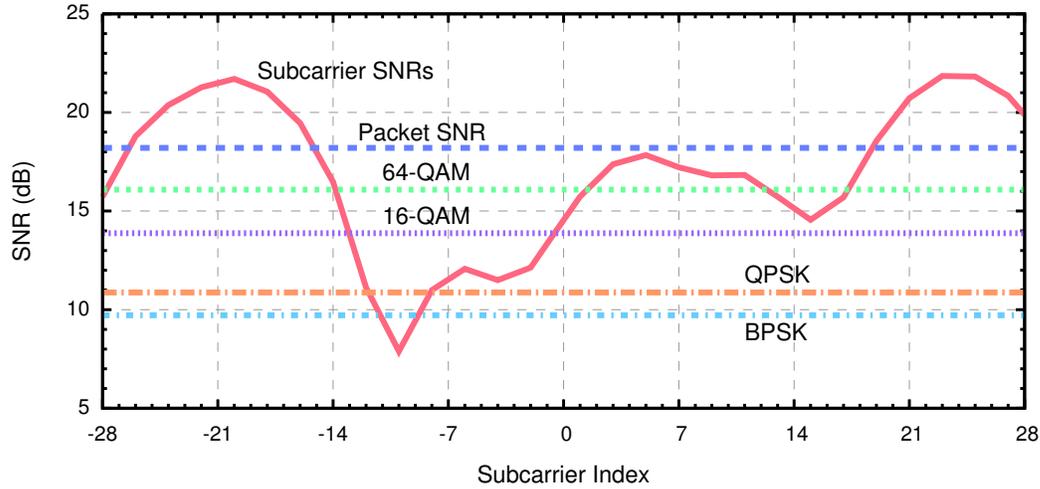

**Figure 4.3: Sample faded link showing the Packet SNR and Effective SNRs for different modulations. BPSK has the lowest Effective SNR, but it needs less energy to decode.**

| Modulation | $\rho_{eff}$ (dB) | $BER_{eff}$ |
|------------|-------------------|-------------|
| BPSK       | 9.7               | $8 \times 10^{-6}$ |
| QPSK       | 10.9              | $2 \times 10^{-4}$ |
| 16-QAM     | 13.9              | 1%          |
| 64-QAM     | 16.1              | 5%          |

**Table 4.3: Effective SNRs and Effective BERs for the example link in Figure 4.3.**

### 4.3.1 *Effective SNR Example*

To make this description concrete, I present an Effective SNR calculation using this core model for an example SISO 802.11n link in Figure 4.3. The subchannels of this single-antenna link are simply the OFDM subcarriers, illustrated by the solid line. There is also one line for the Packet SNR, i.e., the average SNR across all subchannels. The figure includes four lines that each represent the Effective SNR for a different modulation. Table 4.3 shows the Effective SNRs and Effective BERs for this link.

Note that the Effective SNRs are well below the Packet SNR, which is biased towards the stronger subcarriers (note the logarithmic y-axis scale). This link does a poor job of harnessing the received power because it is badly faded, so its Packet SNR is a poor predictor of its rate. The simpler modulations have lower Effective SNRs than the denser modulations, but also perform much better at low SNR (recall Figure 2.3).

### 4.3.2 *Predicting Performance with Effective SNR*

Recall that the model output is a prediction of how well the link will work in a particular configuration. Generically, this prediction can be generated using a metric function

---

**Algorithm 4.1** Threshold-based Effective SNR Metric: $\mathcal{M}_\tau(c, \rho_{\text{eff},c})$

---

1: Extract the modulation and coding scheme $m$ used in configuration $c$
2: **if** $\rho_{\text{eff},c} \geqslant \tau_m$ **then**
3:     **return** the bitrate of the link in configuration $c$       *// predict 100% packet reception*
4: **else**
5:     **return** 0                            *// for low Effective SNR, predict 0% packet reception*
6: **end if**

---

$\mathcal{M}(c, \rho_{\text{eff},c})$. This function takes as input a configuration $c$ and the Effective SNR of the link in that configuration, and it outputs a task-specific prediction of how well the link can work. I define this metric function in a generic way because the form of the output depends on the objective function of the configuration algorithm.

*Threshold-based Effective SNR Metric*

In this thesis, I use the simple *threshold-based* metric function described in Algorithm 4.1. In this approach, I assume that each configuration has an Effective SNR threshold that separates lossy links (with low SNR values below the threshold) from links that deliver packets reliably (with SNR values above the threshold):

$$\text{works?}_c = (\rho_{\text{eff},c} \geqslant \tau_m). \tag{4.3}$$

Note that the configuration $c$ specifies transmitter, receiver, and rate configurations. These in turn determine the Effective CSI, the modulation and coding scheme $m$, and all the other information used to calculate $\rho_{\text{eff}}$. The configuration-specific thresholds $\tau_m$ are exclusively a function of MCS. The thresholds are determined by the *receiver implementation*, but are not link- or device-dependent. We can choose $\tau$ in a variety of ways; I explain my method in the next chapter. As in CHARM [55], this model can support different packet lengths with different SNR thresholds.

For simplicity, this function assumes a sharp transition from 0% to 100% packet reception of the link. In practice, this metric will over- or under-estimate performance for links within the transition region that deliver packets at intermediate rates; in the next chapter, I evaluate the width of this transition region to quantify this inaccuracy.

*Alternative Effective SNR Metrics*

There are many options for more advanced metrics. One extension is to map Effective SNR to PRR in a continuous way instead of the binary classification of links I use. Another might use these predicted PRRs in conjunction with knowledge of the 802.11 MAC protocols to predict the link's *goodput* taking into account protocol overheads, the expected overhead of backoff at a particular PRR, and the effects of 802.11n packet aggregation. A metric could also account for the effects of proposed higher-layer coding schemes [33, 71, 87].

A metric can also have a different output than throughput. For instance, suppose the metric can compute PRR and then MAC airtime for a particular configuration. Then, using measurements of the power consumed by the chipset in different configurations [34], the metric could compute the expected power consumption of each side of the link. These estimates can be combined in many ways, such as to optimize power consumption across devices or to minimize consumption for the battery-operated device if the other device is plugged in.

I leave exploration of these more advanced Effective SNR metrics for future work. In this thesis, the simple threshold-based metric (Algorithm 4.1) is sufficient to implement the configuration algorithms studied and provides good performance, as my results will show.

### 4.3.3    A Note About Coding

The astute reader will notice that I did not mention error-correction in the calculations I laid out above. One key decision that makes my model practical is that it is coding-agnostic.

I made this choice because coding interacts with the notion of Effective SNR in a way that is difficult to analyze. One challenge is that the ability to correct bit errors depends on the position of the errors in the data stream. To sidestep this problem, I rely on the interleaving that randomizes the coded bits across subcarriers and spatial streams. Assuming perfect interleaving and robust coding, bit errors in the stream should look no different from bit errors for flat channels (but at a lower SNR). Thus the estimate of the Effective BER in Equation 4.1 will accurately reflect the uncoded error performance of the link. In other words, my model assumes that coding works—but does not care about the details.

Note that this procedure differs from the typical approach of simulation-based analyses [57, 72, 82], which instead map the *uncoded* BER estimate such as I compute to a *coded* BER estimate by means of a log-linear approximation parameterized by the receiver implementation. They then use this coded BER estimate and packet length to directly compute the packet delivery rate of the link. But changing the coding parameters involves changing the internals of the calculation, and with some implementations these parameters can be different for different configurations. (I describe why in the next subsection.) In contrast, all of these effects can be easily expressed, albeit approximately, as (perhaps modulation-dependent) shifts in the Effective SNR thresholds. Thus I believe my method of thresholding the Effective SNR is a better design point because it efficiently accommodates these variations.

### 4.4   Modeling the Receiver: Computing Subchannel SNRs from Effective CSI

In the last section, I showed how the model of Nanda and Rege [79] can determine the Effective SNR from a list of Subchannel SNRs. In this section, I explain how to compute the Subchannel SNRs given the Effective Channel State Information of the link.

The *Effective CSI* is a NxS channel state matrix $\mathbf{H}'$, where each entry describes the gain and phase of each spatial stream as measured at each receive antenna. The columns of the Effective CSI represent *spatial streams*, not *transmit antennas* as in the CSI matrix $\mathbf{H}$. From the point of view of receiver processing the number of transmit antennas used is not relevant, just how the transmitted streams are received after sender-side processing and propagation through the RF channel. I discuss how to compute the Effective CSI in the next section.

Note that the IEEE 802.11 standards do not specify which algorithms the receiver uses to process received signals. However, a small set of techniques are likely chosen in practice: optimal algorithms or near-optimal techniques that have known, good complexity/optimality tradeoffs. My contact with Intel engineers and marketing documents from other chipset manufacturers confirm that the techniques I present here are the ones used in practice.

Of course, which techniques the receiver might use to process the Effective CSI depends on the dimensionality of the effective channel. In a SISO system, there is little processing to be done, while MIMO systems that use spatial multiplexing and spatial diversity offer lots of options. Hence, I frame my presentation of these techniques around the number of streams and antennas available.

#### 4.4.1   SISO Links: $S = 1, N = 1$

As in the example of Figure 4.3, the SNR of a 1x1 channel matrix is simply the power of the single entry in the CSI matrix. There is one Subchannel SNR for each OFDM subcarrier:

$$\rho_i = \left| \mathbf{H}'_i \right|^2.  \tag{4.4}$$

#### 4.4.2   SIMO Links: $S = 1, N > 1$

When there is a single stream and more than one receive antenna, there are two dominant algorithms used. The simpler technique is *antenna selection*, in which the antenna with the largest SNR is chosen, and then the SISO algorithm is used on that antenna's signal. The same antenna selection applies to all OFDM subcarriers.

$$\rho_i = \left| \mathbf{H}'_{i,\hat{a}} \right|^2, \text{where antenna } \hat{a} \text{ has maximal Packet SNR.}  \tag{4.5}$$

Some older chipsets, such as the Intel PRO/Wireless 3945 [48: §3.5] and Atheros AR5007 [74] use antenna selection by default.

The optimal SIMO technique is *maximal-ratio combining (MRC)* [115], which combines the multiple copies of the signal from the different receive antennas. The resulting subcarrier SNR is simply the sum of the SNRs of the individual entries:

$$\rho_i = \sum_{a=1}^{N} \left| \mathbf{H}'_{i,a} \right|^2. \tag{4.6}$$

MRC requires more hardware than antenna selection—one receive chain per antenna, instead of one receive chain total. But since spatial multiplexing techniques require multiple receive chains, 802.11n chipsets already have the hardware necessary to perform MRC. The additional computation is minimal, and the resulting gains are large. Production documentation for Atheros (e.g., the AR6004 [8]) and Realtek (e.g., the RTL8192SU [93]) chipsets and the Intel 802.11n driver source code [49] confirms that they use MRC.

Devices with more antennas than receive chains can use *hybrid selection-combining* algorithms. Such a device might first select the $n < N$ antennas that have the strongest Packet SNRs and then apply MRC to those.

### 4.4.3  MIMO Links: $S > 1$

When the transmitter uses spatial multiplexing, i.e. $S > 1$, the receiver uses a MIMO equalizer to disentangle the streams from its $N \geqslant S$ receive antennas. For a MIMO link, there are a total of $C = TS$ values for Subchannel SNRs, one per subcarrier (tone) per stream. I denote these $\rho_{i,j}$ where $i$ is an index over tones and $j$ an index over streams. There are several possible techniques in use; here I discuss the two algorithms most likely to be used in practice.

A *Minimum Mean Square Error (MMSE)* MIMO receiver is used by the Intel Wireless Wi-Fi Link 5300 and perhaps other 802.11n devices. For a single stream, MMSE is optimal and equivalent to MRC, and MMSE is near optimal for spatial multiplexing.

The SNR of the $j$th stream after MMSE processing for subcarrier $i$ is given by

$$\rho_{i,j} = \frac{1}{\mathbf{Y}_{jj}} - 1, \text{ where } \mathbf{Y} = \left( (\mathbf{H}'_i)^\dagger \mathbf{H}'_i + \mathbf{I}_S \right)^{-1} \tag{4.7}$$

for $j \in [1, S]$ and SxS identity matrix $\mathbf{I}_S$ [115]. The operator $(\cdot)^\dagger$ indicates the *Hermitian*, or conjugate transpose of the input matrix.

The advantage of MMSE is that it performs well in most cases and has low computational cost, roughly the cost of matrix multiplication and inversion. An alternative is the *Maximum Likelihood (ML)* decoder. This is an optimal decoder that uses channel estimates to predict every possible received symbol, and then finds the most likely match. However, the ML decoder has added complexity that scales with the size of the constellation, i.e., is exponential in the number of bits per symbol k. In the initial 802.11n chipsets, ML decoding was considered too complex to be practical, but this outlook has recently changed. Today, Intel's IWL6300 (successor to the IWL5300 I use) uses ML decoding for the smaller constellations for which it is practical. Atheros uses a "simplified maximum likelihood detector" [7,8] for all modulations that approximates ML performance but has lower complexity.

The task of estimating the per-stream SNR of the output from a maximum likelihood decoder is non-trivial. Existing estimation techniques, such as the approximation by Redlich et al. [94], have the same complexity as the full ML receiver, and are hence impractical for software implementation (though they may be suitable for use if implemented in hardware).

Instead, I observe that maximum likelihood decoding can be approximated as a $\approx 3$ dB gain over MMSE for typical MIMO systems [65]. Thus, the practical solution I propose for my model is to always compute Effective SNR estimates using MMSE, and instead adapt the decision thresholds $\tau_m$ to reflect the use of ML decoding. This approximation is imprecise—the gains of ML will vary depending on the particular channel—but allows the model to flexibly handle many device implementations (even hybrid MMSE/ML such as in the IWL6300). The error is likely to be small for most channels.

### 4.4.4 Summary

In this section, I explained how to determine the Subchannel SNRs given Effective CSI matrices for Wi-Fi links. The SISO matrices give the Subchannel SNRs directly, and I assume the use of MRC when modeling SIMO processing. For MIMO links, the model will always use the MMSE formula given by Equation 4.7 to compute the Subchannel SNRs, though some receivers may use ML instead. But since the difference between ML and MMSE is often just a few dB, the model will assume MMSE and instead represent the gains from using ML by having lower Effective SNR thresholds. I describe how the sender and receiver might communicate these thresholds later in this chapter. Before doing so, I will next describe the last component of my model needed to make application decisions over a space of configurations.

### 4.5 Applications: Adapting CSI to Compute Effective CSI

Recall that for a given CSI measurement and a target configuration, the Effective CSI is the channel state information that would be measured if the link operated in that configuration. The last few sections have described how to compute the model output given the Effective CSI; the final missing piece is to actually compute the Effective CSI ($\mathbf{H}'$). To compute $\mathbf{H}'$, I adapt the CSI ($\mathbf{H}$) to reflect the difference between the configuration in which $\mathbf{H}$ was measured and the target configuration. Here, I explain how to perform this step for an illustrative set of configuration problems.

**Antenna Selection.** There are M columns and N rows in the CSI matrix **H**; each corresponds to one transmit or receive antenna. To compute the Effective CSI that would be used under a particular antenna selection, we simply pick the subset of rows and columns that correspond to the desired antennas. Note that this includes both transmit antenna selection (e.g., when S < M, pick the best S of the M transmit antennas to send with) and receive antenna selection (e.g., when N > S, turn off the least useful of the excess antennas in order to reduce power consumption).

**Channel Width Selection.** In 802.11n, a full 40 MHz CSI measurement comprises 114 matrices $\mathbf{H}_i$. To implement channel width selection, we use only the $\mathbf{H}_i$ covered by the potential channel. For the lower 20 MHz channel, we would use the matrices corresponding to $i \in [1, 56]$; for the upper 20 MHz channel we would use the matrices for $i \in [59, 114]$. (Subcarriers 57 and 58 are not part of either 20 MHz channel because the 40 MHz channel does not need the same guard band.) This method can also predict performance using 5 MHz or 10 MHz channels as in 802.11j or SampleWidth [21], and could support a potential future version of 802.11 that allows the devices to use non-contiguous tones as in SWIFT [90].

**Transmit Power Control.** If the transmitter changes the amount of power it radiates, this change will be reflected as simple scaling of the received CSI. Recall that each entry in **H** represents an amplitude (also a phase), which has a quadratic relationship with transmit power. Thus when the transmitter reduces the transmit power uniformly across subcarriers, say by 3 dB (a factor of 2), we multiply each matrix by the square root of the power change, in this example by $\sqrt{2}$. To model the effects of asymmetric power control across subchannels, e.g., water filling across streams or subcarriers, we can apply scaling differently to different parts of the channel matrix.

**MCS (Rate) Selection.** For any of the above transmitter and receiver configurations, my model can predict whether a particular MCS will reliably deliver packets. The CSI matrix may need to be slightly modified depending on the MCS, however. For instance, when using multiple streams the total amount of power is generally kept constant, so the Effective CSI must include a division of the transmit power across antennas. Similarly, some chipsets (e.g., IWL5300 and the Atheros-based Ubiquiti SR71-A [116]) are unable to send at full power when using the highest modulations, because the peak-to-average power ratio (PAPR) is higher than the transmit amplifier(s) can support. For these chipsets, the Effective CSI must be scaled to compensate for the reduced transmit power when predicting the performance of the relevant modulations.

**Spatial Mapping.** Spatial mapping algorithms determine how the different transmit streams are mapped to transmit antennas. These are implemented by the MxS matrices $\mathbf{V_i}$, such that the Effective SNR is

$$\mathbf{H_i'} = \mathbf{H_i} \mathbf{V_i}. \tag{4.8}$$

I discuss the four types of spatial mapping techniques from the simplest to the most general.

- In *direct mapping*, the transmitter maps each spatial stream directly to one of S antennas; excess antennas go unused. After choosing the appropriate rows of $\mathbf{H}$ as described above under transmit antenna selection, the matrix $\mathbf{V}$ is the identity matrix $\mathbf{I_S}$. We can combine transmit antenna selection and direct mapping using a MxS matrix $\mathbf{V}$ consisting of zeros except for the S entries in different rows and columns that indicate which streams are mapped to which antennas. For $S = 2$ and $M = 3$, a direct mapping matrix using the first and third antennas would be

$$\mathbf{V} = \begin{pmatrix} 1 & 0 \\ 0 & 0 \\ 0 & 1 \end{pmatrix}. \qquad \text{(Combined 3x2 TX Antenna Selection and Direct Mapping)}$$

- In *indirect mapping*, the S streams are mapped to S antennas but in such a way as to spread over multiple antennas. For instance, the IWL5300 uses the following indirect mapping for 3 streams and 20 MHz channels:

$$\mathbf{V} = \begin{pmatrix} e^{-i2\pi/16} & e^{-2i\pi/(80/33)} & e^{-2i\pi/(80/3)} \\ e^{-i2\pi/(80/23)} & e^{-2i\pi/(48/13)} & e^{-2i\pi/(240/13)} \\ e^{-i2\pi/(80/13)} & e^{-2i\pi/(240/37)} & e^{-2i\pi/(48/13)} \end{pmatrix}. \qquad \text{(IWL5300 3x3)}$$

Atheros uses Walsh (also called Hadamard) spreading. The 2x2 mapping is:

$$\mathbf{V} = \begin{pmatrix} 1 & 1 \\ 1 & -1 \end{pmatrix}. \qquad \text{(Atheros 2x2)}$$

- When using *spatial expansion*, S spatial streams are mapped to $M > S$ transmit antennas. Here, $\mathbf{V}$ is the product of an expansion matrix and one of the direct or indirect mapping matrices above. I show the case of $S = 2$ and $M = 3$ using a sample expansion matrix from the 802.11n standard:

$$\mathbf{V} = \mathbf{V}' \sqrt{\frac{2}{3}} \begin{pmatrix} 1 & 0 \\ 0 & 1 \\ 1 & 0 \end{pmatrix}, \qquad \text{(3x2 Spatial Expansion)}$$

where $\mathbf{V}'$ is a direct or indirect mapping matrix. The $\sqrt{2/3}$ factor above represents the even spreading of the power from 2 streams across 3 antennas.

- Finally, *beamforming* is a general technique for mapping streams to antennas. Beamforming can be thought of as a synthesis of indirect mapping, spatial expansion (if applicable) and transmit power control. The distinctive feature of beamforming compared with the above is that it is *channel-aware*; in other words, the beamforming matrix is chosen to improve performance based on the known current channel $\mathbf{H}$. This means that beamforming can use different mapping matrices $\mathbf{V}_i$ for each subcarrier, unlike the prior techniques which are applied uniformly across tones. There are a large variety of techniques that use beamforming to improve the channel, but all can be represented in matrix form.

**Higher-level Tasks.** To solve configuration problems that require the composition of multiple measurements, such as channel selection, path selection, or access point selection, we can repeat the steps above. For instance, two devices that wish to find the operating channel on which their link is fastest need only frequency-hop in unison and take CSI measurements for each frequency. They can then use the model to predict the achievable bitrate of each measured operating channel and select the fastest.

### 4.6   Protocol Details

In this section I make concrete the actual protocols and algorithms by which my Effective SNR model can be used. Decisions about which configuration to choose can be made by either end of the link, and each has its advantages. I begin by presenting an algorithmic overview of my model and the set of information needed to compute it. I then discuss transmit-side and receiver-side procedures for calculating the Effective SNR, including what information needs to be communicated between endpoints (and when that exchange happens) for these approaches to work.

#### 4.6.1   Protocol Overview

I present the skeleton of a procedure to use my Effective SNR model as Algorithm 4.2. We can define a configuration problem as a set of configurations under consideration ($\mathcal{C}$) and some metric ($\mathcal{M}$) that enables us to choose between configurations.

Consider the example problem of rate selection over a three-antenna 802.11n link. For this application, we define $\mathcal{C}$ to be the 24 modulation and coding scheme (MCS) combinations that span all SIMO, MIMO2, and MIMO3 rates. (This assumes only 20 MHz channels and that the receiver is fixed to use all three antennas.) For metric $\mathcal{M}$, we use the threshold-based metric given in Algorithm 4.1. Then Algorithm 4.2 will use my Effective SNR procedure on a measured 3x3 channel matrix $\mathbf{H}$ and return the configuration predicted to be the fastest.

---

**Algorithm 4.2** EFFECTIVE SNR ALGORITHM SKELETON

---

1: Define an application by a set of configurations $\mathcal{C}$ and a metric $\mathcal{M}$
2: Measure the CSI **H**, covering the subchannels used in all configurations
3: **for** configuration $c_i \in \mathcal{C}$ **do**
4:     Use the Effective SNR model and **H** to determine $\rho_{\text{eff},c}$ // *Note that* $c_i$ *specifies a choice of rate, sender configuration (including channel), and receiver configuration.*
5:     Set metric $m_i$ to $\mathcal{M}(c_i, \rho_{\text{eff},c_i})$
6: **end for**
7: **return** the configuration $c_i$ with the best metric $m_i$

---

Note that Algorithm 4.2 uses the brute-force approach of checking all configurations to find the best one. For extremely large search spaces, it may be necessary to prune the search space if computation time dominates the rate at which the channel changes. However note that this inflection point comes for much larger spaces than the probe-based approach, since Effective SNR can predict the performance of many configurations using a single packet preamble, while probe-based algorithms must send one or more full-length packets to test a configuration.

In order to actually run this procedure, three pieces of information are necessary in practice:

1. Up-to-date CSI measurements including all relevant subchannels.

2. The receiver's MCS-specific Effective SNR thresholds.

3. Knowledge of how to compute the Effective CSI **H**′ from the CSI **H**.

A challenge is that this information is spread across the transmitter and receiver. In particular, CSI measurements are naturally measured at the receiver in the course of receiving 802.11n packets, and I envision that the receiver's Effective SNR thresholds will be programmed in at the factory since they are fixed and independent of channel conditions. At the same time, computing the Effective CSI requires knowledge of transmitter behavior, including spatial mapping matrices as well as any implementation-specific oddities. The latter can be quite broadly defined, such as Intel's reduction of transmit power by up to 5 dB for the highest 64-QAM single-stream encodings to compensate for amplifier non-linearities. Finally, which endpoint applies the result of the computation depends on the application in mind. Often the adjustment will be at the transmitter side, such as choice of rate, but some applications, such as receive antenna selection, are applied at the receiver.

In the next few subsections, I first discuss how CSI is obtained, as this is the main run-time measurement needed to run this algorithm. I then describe the transmitter- and receiver-side ways to implement Algorithm 4.2 and the tradeoffs of the various approaches. The primary tradeoff is between the amount of data shared (because the party performing the calculation needs up-to-date CSI) and protocol complexity.

### 4.6.2   CSI Collection

As with RSSI, receivers gather channel state information automatically while receiving a packet; the OFDM and MIMO equalizers described in Section 4.4 fundamentally rely on CSI to decode the received signal. This *passive collection* of CSI via overheard transmissions suffices for most applications, such as rate selection. When a more comprehensive CSI measurement is needed, the transmitter can send a sounding packet [45: §20.3.13.1] at a wider bandwidth or using more streams to probe the extra dimensions. For the special case of measuring additional spatial streams, 802.11n includes a mechanism to send a modified packet with *extension spatial streams* [45: §20.3.9.4.6] such that the receiver can measure the full channel during the preamble but that does not change the rate or number of streams used for the data portion.

### 4.6.3   Transmitter-side Computation

The transmitter seems to be a natural place to perform these computations, as most configuration points are determined by transmitter configurations. To do so, it needs an up-to-date CSI measurement as well as the receiver's MCS thresholds. The latter is easy: Since the thresholds are fixed for a particular model of NIC, they can be shared by the receiver once, e.g., during association.

The transmitter can obtain CSI via feedback or estimate it implicitly from the reverse path. When the receiver feeds CSI back to the transmitter, this process adds latency and reduces airtime efficiency. 802.11n includes mechanisms to limit these effects via rapid feedback protocols [45: §9.19.2] and compressed CSI feedback formats [45: §20.3.12.2.5]. Additionally, my measurements taken in conjunction with experimenters at Intel show that in slow-fading channels, such as static nodes with indoor mobility, CSI estimates are valid for a few hundred milliseconds [85] and this may require less frequent feedback.

Alternatively, the transmitter can attempt to estimate the CSI implicitly when receiving packets, such as ACKs, sent by the receiver. This is the standard approach taken in Packet SNR-based schemes such as CHARM [55] proposed for single-antenna 802.11a systems. For systems with multiple antennas, like 802.11n, this approach mandates that the receiver inform the transmitter of its spatial mapping matrices and use all its antennas to send these packets. Another approach is for the receiver to send ACKs using fewer streams, but alternate which antenna set it uses so that the transmitter can gradually build up the full CSI; this process is called CSI Sampling and Fusion [25]. The use of implicit feedback also requires devices to use the 802.11n calibration process to compensate for hardware differences in the measured transmit and receive chains and those that will be used when the two devices swap roles. Some receivers, such as the Intel IWL5100, have more receive antennas than transmit chains and cannot send on all antennas. For these devices, implicit feedback would simply not work.

Finally, the implicit feedback approach induces all the problems of receiver-side computation (see below) with few of the benefits, and I do not find it to be practical. Instead choosing to make decisions at the transmit side of the link, I recommend that the transmitter obtain CSI feedback explicitly when making an Effective SNR decision.

### 4.6.4 Receiver-side Computation

The other approach is for the receiver to perform the computations and make the application decisions, feeding them back to the transmitter when necessary. This obviates sending CSI and speeds up the protocols, but instead requires that the receiver understand the transmitter's behavior and capabilities.

Today, the transmitter shares some of these details, such as number of transmit antennas and whether it supports optional 802.11n features, with the receiver during association. But the standard does not include a method by which a transmitter can share its spatial mapping matrices, and it is not immediately clear that this would sufficiently capture all implementation artifacts such as those described above.

The other drawback to this approach is that it complicates receiver algorithms. For instance, if the transmitter is an access point and the client a cell phone, the former device is likely to have much more sophisticated silicon. We have seen this today as many 802.11 manufacturers have targeted systems with asymmetric capabilities such that the access point shoulders all the computation load, reducing the complexity and hence cost of the client which need only support basic functionality. The receiver-side approach requires that the receiver shoulder the computational load of running the model and making decisions; in contrast for the transmitter-side approach it must only support CSI feedback.

Still, I believe that the receiver-side approach is fundamentally the best. This approach has the primary advantage of putting the computation near the data, so that rather than feeding back the full CSI measurement to the transmitter, the receiver need only send the application decision over the air. This is similar to the approach taken by RBAR [43], and support for this has been added to 802.11 with the new 802.11n Link Adaptation Control field [45: §7.1.3.5a] that can embed feedback in packet headers, including ACKs, to directly request a particular MCS, select transmit antennas, or request a particular beamforming (or other spatial mapping) matrix.

To make the receiver-side algorithm work best, I suggest a simple new contract be added to 802.11n as a standard requirement. Since today transmitters may use fundamentally different spatial mapping matrices for different numbers of streams, it is hard to compute the Effective CSI for a different number of streams. Indeed, the Intel IWL5300 uses spatial mapping matrices such that the 2-stream CSI bears no obvious relation to the 3-stream CSI without knowledge of the possible mappings, and uses different matrices for 20 MHz and 40 MHz channels. In my experiments using the IWL5300, I assumed that the receiver knows the spatial mapping matrices a priori, but sharing the full extent of these mappings is hard to fit into a clean protocol. As a cleaner solution, I propose that transmitter spatial mapping matrices be restricted such that when a receiver subsets a large (e.g., MIMO3) CSI measurement to simulate a smaller (e.g., MIMO2) configuration, the result indeed corresponds to the channel that would be measured if the transmitter used that mode. I think this new contract would best resolve the situation without greatly compromising transmitter performance or flexibility.

### 4.6.5 Summary

In this section, I discussed some practical details of implementing my Effective SNR-based decision procedure. When comparing the transmitter-side and receiver-side options for implementing this functionality, I concluded that receiver-side computation likely represents the best tradeoff, because it has the benefit of low overhead without compromising on flexibility, accuracy, or agility.

In certain situations, such as when selecting beamforming matrices, the decision may be best made at the transmitter, which better understands the limitations of its own hardware. And for multiple links to coordinate, CSI measurements may need to be shared in a local neighborhood. One example of a system that does this lightweight local sharing in order to achieve better spatial reuse is IAC [31].

## 4.7   Comparison to Other Techniques

I conclude this chapter by comparing my Effective SNR to other techniques that use physical layer or other low-level information to predict the performance of a wireless link. I compare algorithms along two axes: (1) their applicability to many configuration problems, and (2) their efficiency in terms of overheads and responsiveness.

### 4.7.1 Accuracy

Table 4.4 summarizes the comparison between Effective SNR and other recent algorithms that aim to predict the performance of a wireless link. The results show that Effective SNR can handle a much broader problem space because it looks at the raw channel details, while other algorithms primarily apply only to single-stream links with fixed antennas and transmit power.

| Algorithm | 802.11a/b/g | 802.11n (MIMO) | Antenna Selection | TX Power | Channel Width | Real Wi-Fi Devices |
|---|---|---|---|---|---|---|
| Packet SNR | ? | | | | | ✓ |
| SoftRate [120] | ✓ | | | | | |
| AccuRate [103] | ✓ | | | ✓ | ✓ | |
| EEC [22] | ✓ | | | | | ✓ |
| Effective SNR | ✓ | ✓ | ✓ | ✓ | ✓ | ✓ |

Table 4.4: A comparison of Effective SNR to other recent algorithms that purport to predict the performance of a wireless link. Effective SNR can predict packet delivery in the largest space because it looks at the underlying subchannel response, whereas the other techniques mostly apply to single-stream links with fixed antennas and transmit power.

| Algorithm | Measurement | Communication Overhead | Compute | Response Time |
|---|---|---|---|---|
| Probe-based | Multiple packet loss rate | None | Low | High |
| Packet SNR | Preamble | Low | Low | High |
| SoftRate [120] | A few full symbols | Low | Low | Low |
| AccuRate [103] | Preamble | Low | Very High | High |
| EEC [22] | Full packet | Low | High | Medium |
| Effective SNR | Preamble | Low (RX) Medium (TX) | Medium | Low |

Table 4.5: Comparing overheads and response times of Effective SNR and other algorithms. In addition to being more flexible and accurate than the other algorithms (Table 4.4), Effective SNR has low measurement and communication overheads, little computational cost, and low response time. This matches the best aspects of the other algorithms.

As I presented in the previous section, Packet SNR based on RSSI is available in today's devices, but is does not accurately predict performance for 802.11a/g due to its inability to capture frequency-selective fading; the use of MIMO in 802.11n only exacerbates this deficiency.

SoftRate [120] and EEC [22] both use information from the error correction hardware to accurately measure the Effective BER of the *current* configuration. They have been shown to work well for rate selection using adjacent 802.11a/b/g rates, but they cannot predict the performance of other configurations such as using different numbers of antennas.

AccuRate [103] uses the measured receive error vectors as input to a full channel simulator; like my model this procedure can estimate the Effective BER of all modulations in the current configuration. However, these computations are computationally expensive and do not hold across antenna modes. One case where it may work is changing channel width, by only using the error vectors for the desired subcarriers. AccuRate may also be able to support transmit power control: I believe this can be approximated by scaling the error vectors, if the hardware distortion does not change across transmit powers. However, neither of these tasks were explored by the authors.

In contrast, Effective SNR applies in more settings than the other techniques because its computations are aware of the low-level physical-layer effects of the RF channel, including frequency- and spatially-selective fading. My model is broadly compatible with 802.11n and requires no hardware changes.

### 4.7.2   Overheads and Response Time

Table 4.5 summarizes my qualitative comparison of these algorithms in terms of their practical overheads and the achievable response times.

Packet SNR-based algorithms, AccuRate, and Effective SNR all require only a packet preamble to record their measurements. SoftRate, which uses the output of the error correction blocks to estimate Effective BER, requires a packet that contains at least a few MIMO-OFDM symbols that fully utilize the available subchannels. EEC requires slightly more bits to accurately estimate the BER (around a full packet), and the standard probe-based approaches today require multiple full-length packet probes to estimate the loss rate of a particular configuration.

None of the schemes have much additional communication overhead beyond the existing 802.11 protocol, feeding back only configuration decisions or concise metrics such as SNR or BER estimates. When using Effective SNR to make decisions that require transmit-side computation (which other algorithms cannot handle at all) Effective SNR must feed CSI back to the transmitter, though the feedback can be compressed and may only need to be sent periodically.

In terms of computation, Packet SNR and SoftRate require very little, either directly feeding back the hardware's estimate to the other side of the link or performing a simple local threshold test. Today's probe-based algorithms require a small amount of computation to update loss rate estimates. Among the rest, AccuRate has such a large overhead as to make it impractical, because it requires a full wireless channel simulator of a full-length packet in order to compute its output. EEC's custom codes require a medium amount (about $250\,\mu s$) of computation [22], which is longer than a packet sent at 58.5 Mbps but still quicker than a full 4 ms packet batch. In contrast, I will show in the next chapter that the Effective SNR computations for rate selection among 24 MCS combinations can be computed in under $4\,\mu s$, around the time for a single MIMO-OFDM symbol. Both EEC and Effective SNR computations would be faster if implemented in hardware instead of in software on a desktop computer, but Effective SNR can already be calculated fast enough for real-time adaptation in fast-changing channels.

Finally, I compare the algorithms on the basis of their responsiveness. The algorithms that are more accurate—SoftRate, AccuRate, EEC, and Effective SNR—can respond quickly to changing conditions. However, this applies only to those configurations they can solve (see Table 4.4), and the response time of AccuRate and EEC are limited by their computation time. In contrast, because probe-based algorithms must search a large space and Packet SNR-based algorithms require online training of SNR thresholds, they cannot respond in an agile way in mobile environments.

### 4.8  Summary

This section has presented my Effective SNR model and how to use it. The key idea is to use CSI measurements made by the receiver from a single packet preamble to efficiently infer how well other configurations would work. When used with my Effective SNR model, the CSI can easily and efficiently handle a wide range of transmitter, receiver, and rate configurations. My model provides a flexible API to support many different configuration tasks and includes considerations for real receivers with practical implementations operating over real wireless channels.

I concluded with a comparison of my Effective SNR-based model to other techniques with similar goals. Effective SNR can handle a much larger problem space than comparable algorithms, because my model is able to capture the underlying RF details of the MIMO-OFDM channel, rather than high-level observations of the channel, packet delivery, or error rates. In terms of overheads and response time, Effective SNR comes close to matching the best of the other algorithms, with the upside of having wider applicability to new Wi-Fi technologies.

Of course, I have not yet shown that my Effective SNR model is accurate. In order to understand whether this is the case, I built a prototype implementation of my model using a commercial Intel wireless chipset. In the next chapter I describe this implementation. I use this implementation in the rest of this thesis to demonstrate that my Effective SNR model works well and supports a wide space of applications.

Chapter 5

# EXPERIMENTAL PLATFORM

In this chapter, I describe my experimental 802.11n platform, which comprises a prototype implementation of a CSI measurement tool based on commodity Intel Wi-Fi chipsets and experimental 802.11n wireless testbeds in two indoor office environments. This experimental setup is the foundation for the experimental results in the remainder of my thesis.

## 5.1  Experimental 802.11n Wireless Testbeds

I conducted experiments in two 802.11n stationary wireless testbeds. (In addition to these two stationary testbeds, I use three laptops for mobile experiments; I describe their configuration in the next section.)

The first testbed, pictured in Figure 5.1(a), contained 10 nodes spread over one floor of Intel Labs Seattle covering 8,100 square feet. In the second testbed, I deployed 24 nodes across 3 floors in UW CSE (Figure 5.1(b)), where each floor measures approximately 20,000 square feet in size. Both locations are indoor office buildings, the former mostly a wide open area with cubicles and a few conference rooms, the latter consisting primarily of 5-person offices.

Each node runs the experimental platform described below in Section 5.3. In both testbeds, I placed the nodes to ensure a large number of links between them, a variety of distances between nodes, and diverse scattering characteristics. Devices are located on desktops, under tables, on a cart in the server room, and even mounted on the ceiling. The UW CSE testbed also includes a more dense concentration of nodes in one area of the building (pictured in the upper right corner of Figure 5.1(b)), making it a highly diverse testbed and increasing the challenge of configuration.

## 5.2  Node Configuration

Each node is a stationary desktop (Figure 5.2) or portable laptop (Figure 5.3) equipped with an Intel 802.11n wireless *network interface card (NIC)* that supports three antennas. As the antenna geometry of a multi-antenna device is important for spatial diversity, I mount the three antennas per node on custom stands. Each antenna achieves 5 dBi gain for the 2.4 GHz band, and 3 dBi for the 5 GHz band.

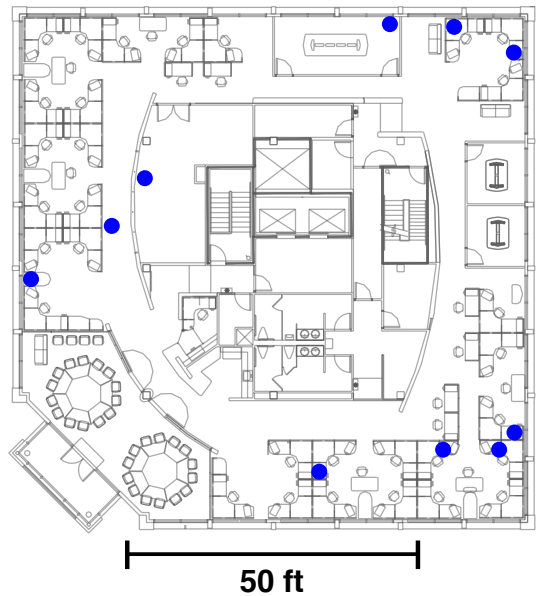

**(a)** The testbed at Intel Labs Seattle contained 10 nodes spread over 8,100 square feet.

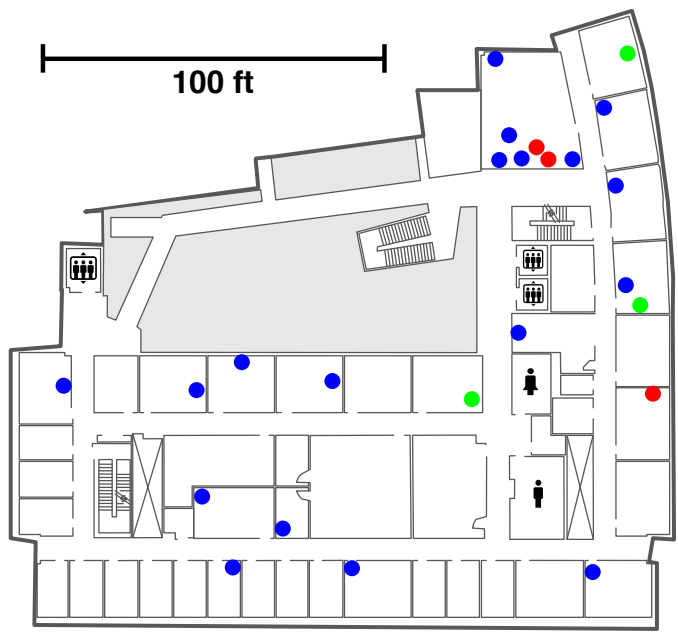

**(b)** The testbed at the UW CSE comprises 24 nodes spread over 3 floors (indicated by color) of 20,000 square feet.

**Figure 5.1:** My two indoor 802.11n testbeds. In both testbeds, the nodes are placed to ensure a large number of links between them, a variety of distances between nodes, and diverse scattering characteristics.

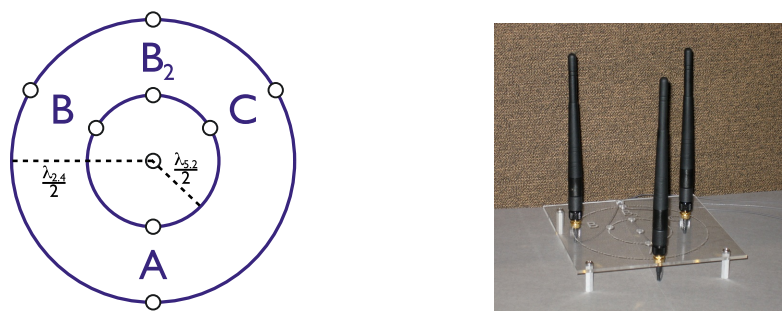

**Figure 5.2: The antenna stand I use to achieve consistent spatial geometry for desktop machines. It supports circular and linear arrays of two or three antennas with the correct $\lambda/2$ separation at either 2.4 GHz or 5 GHz.**

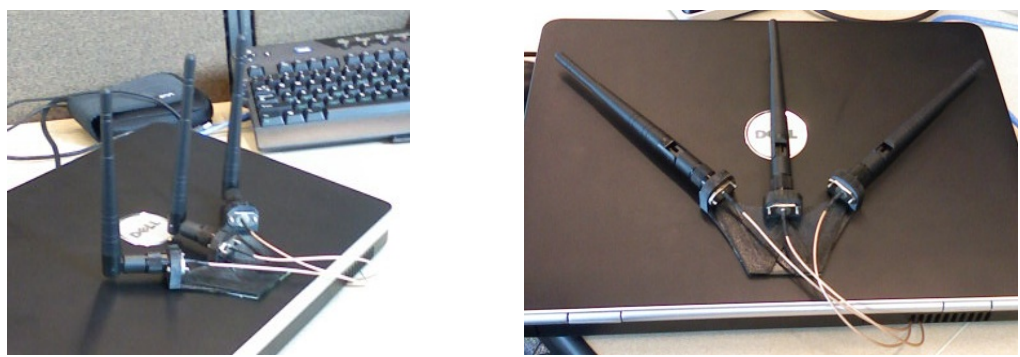

**Figure 5.3: One of the three laptops, pictured here, includes a custom antenna mount for a 2.4 GHz linear array.**

The desktop stands (Figure 5.2) allow for a circular array (using ports ABC) or a linear array of two or three (AB$_2$ and optionally center) antennas, with antenna separations of half the wavelength for either 2.4 GHz Channel 6 ($\lambda/2 = 6.15$ cm) or 5 GHz Channel 48 ($\lambda/2 = 2.86$ cm). In desktop experiments in this paper, I use the circular three-antenna configuration for the 2.4 GHz band. It is robust and suited to dual-band chips that need the wider 2.4 GHz antenna separation.

The three laptops use three different antenna configurations. One laptop has all three antennas embedded internally like in a commercial laptop deployment; one laptop uses the same antenna stand as the desktop machines to mimic their behavior in portable experiments. The third laptop, pictured in Figure 5.3, includes a custom antenna mount for a 2.4 GHz linear array. For mobile experiments, I use the laptop with internal antennas; the other two are used as portable (but not mobile) testbed nodes to supplement the stationary testbed.

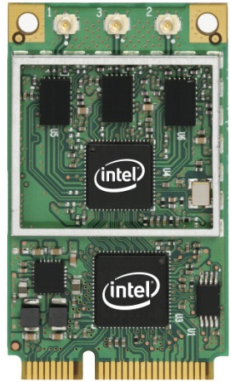

**Figure 5.4: The Intel Wireless Wi-Fi Link 5300. This 802.11n device has three trans-mit/receive antennas, operates on both 2.4 GHz and 5 GHz frequency bands, and supports up to three spatial streams for a maximum bitrate of 450 Mbps.**

### 5.3    Node Software: 802.11n CSI Tool and Research Platform

In conjunction with Intel Labs Seattle, I built an experimental 802.11n platform that uses the Intel Wi-Fi Wireless Link 5300 (IWL5300) 802.11a/b/g/n network cards (Figure 5.4). These 802.11n MIMO chipsets have three antennas and support many new features of 802.11n. I modified the closed-source firmware and open-source `iwlwifi` driver for Linux to add a number of experimental features and, crucially, to measure the 802.11n CSI.

**802.11n CSI Measurement.** The channel sounding mechanism added in 802.11n defines a management frame used to report the CSI from the receiver of a frame back to the transmitter. This mechanism is intended for calibration or to inform transmit beamforming, and I co-opt it for my experiments. In standard operation, the CSI is reported only when the sounding procedure is initiated by the transmitter, though the receiver measures CSI for every frame in order to receive the packet. In my tool, I configure the NIC with a debug mode to compute this feedback packet for every received frame, rather than just during sounding. For correctly received packets, the firmware will send the measured CSI up to the driver on the receiving node, which forwards it to a userspace application that can log or process the CSI.

In my tool, CSI feedback is not returned to the transmitter by default. Instead, the userspace application that processes CSI can optionally generate feedback (potentially including Effective SNR or rate selection information rather than simply the measured CSI) and return it to the transmitter.

The IWL5300 provides CSI in a format that reports the channel matrices for 30 subcarrier groups, which is about one group for every 2 subcarriers at 20 MHz or every 4 subcarriers at 40 MHz. Each channel matrix entry is a complex number, with signed 8-bit resolution each for the real and imaginary parts. It specifies the gain and phase of the spatial path between a single transmit-receive antenna pair. Intel's implementation of the 802.11n CSI does not include per-subcarrier noise measurements, so I assume the noise floor is uniform across all subcarriers to compute SNRs. This is consistent with white noise observed on other OFDM platforms [89].

**RSSI Measurement.** For each received packet the NIC reports the traditional metrics of RSSI per receive antenna, noise floor and the setting on the automatic gain controlled (AGC) amplifier. These combine to define the per-receive-chain Packet SNR ($\rho_{\text{packet}}$):

$$\rho_{\text{packet}} = \text{RSSI (dBm)} - \text{Noise (dBm)} - \text{AGC (dB)} \tag{5.1}$$

The IWL5300 calculates the quantities RSSI and Noise as the respective sums of average signal strength and average error vector magnitude in each OFDM subcarrier [49]. This is exactly the traditional definition of SNR applied to OFDM.

Note that with multiple receive antennas, there is a different RSSI and hence a different Packet SNR for each antenna. As described below in Section 5.4.1, I compute the Packet SNR for the link by summing the Packet SNR of the antennas.

**Transmit Power Control.** I modified the driver and firmware to enable transmit power variation. With these changes, I can vary the transmit power level from $-10$ dBm (100 µW) to $+16$ dBm (40 mW) in steps of 0.5 dB. For all modulations, the IWL5300 divides power equally across transmit antennas. Additionally, the IWL5300 reduces the transmit power slightly when using the highest single-stream rates to avoid distortions caused by passing 64-QAM symbols with high peak-to-average power ratio through the transmit amplifier.

**Rapid Rate Variation.** In normal operation, the IWL5300 decouples queuing packets for transmission from selecting rates for these packets, since queues must be kept large to take advantage of 802.11n block transmissions. This makes it difficult to control the rate at which individual packets are transmitted. I modified the firmware and driver to support the transmission of individual packets at predetermined rates, and added driver-level code to rapidly iterate through a user-configurable set of available rates.

**Userspace Connector.** I used the Linux kernel `connector` framework to implement a low-latency socket-based communication channel between the kernel driver and userspace utilities. This enables userspace utilities to log CSI and other output from the driver, and to send messages that adapt behavior of either end of the link online, e.g., by changing the currently selected rate or antennas or adjusting the transmit power level.

**Publicly Released Tool.** I have publicly released the experimental platform and CSI collection tool in the form of open source drivers, userspace utilities, MATLAB data processing code, and binary firmware image [37]. At the time of writing, I am aware many users of the tool: 20 universities in 7 countries, multiple research and product groups within Intel, one industrial research lab, and one startup. The users of my tool have published at least 7 papers [13, 25, 32, 85, 100, 102, 124].

### 5.4    *Computing 802.11n SNR and Effective SNR using IWL5300 Measurements*

In the rest of this thesis, I use measurements from my experimental platform to evaluate how well Effective SNR (and other algorithms) work for real 802.11n wireless channels. In most experiments, I computed the packet reception rate (PRR), Packet SNR, and Effective SNRs for all the measured configurations. The packet reception rate is easy to compute by simply counting the number of correctly received frames compared to the number sent. The rest of this section describes practical considerations when computing Packet SNR and Effective SNR values.

#### 5.4.1    *Processing Multiple RSSIs to Compute Packet SNR*

When receiving a transmission with multiple antennas, there is one RSSI value per antenna; how should these measurements be combined into a single SNR value for the link? As described above, I first convert the per-antenna RSSI and noise measurements to SNRs (Equation 5.1) and then sum the SNRs. This is a straightforward choice for a single spatial stream as it corresponds to receiver processing using maximal-ratio combining (Equation 4.6). It is also reasonable for 2- and 3-stream MIMO because the symbols carried on different spatial streams are interleaved coded bits [45].

#### 5.4.2    *Processing CSIs to Compute Effective SNRs*

The 802.11n standard mandates that CSI measurements, which are stored as matrices of 8-bit complex numbers indicating gain and phase of each subchannel, include a per-subcarrier SNR reference that enables the grounding of each per-subcarrier channel matrix, such that each entry in the CSI can be treated as a magnitude relative to the noise floor. It is under this model that the equations I presented in Chapter 4 hold.

In attempting to use the measured CSI from the Intel NICs, I discovered a practical issue that needed to be taken into account: Quantization error. This issue is a fundamental problem that applies to all uses of CSI in the 802.11n standard. Consider a very strong link, say with a signal-to-noise ratio of 50 dB. Most actual receiver processing occurs after this signal has been digitized by the *analog-to-digital converter (ADC)* at each receive chain; this conversion from analog to digital cannot be perfect, and thus induces *quantization error* in the digital signal. For instance, a 12-bit ADC can represent signal levels from $-2048$ to $+2047$, with an error up to $\pm 0.5$. Note that the *analog gain control (AGC)* hardware on the chip aims to amplify the received analog signal to fully utilize the ADC range.

An error of $\pm 0.5$ in a measured magnitude of up to 2048 corresponds to a quantization error of about $-72\,$dB. Roughly, each bit that the ADC outputs gives a 6 dB reduction in quantization error, because it halves the relative error in magnitude, dividing the power of that error by a factor of 4, i.e., 6 dB.

How does quantization error compare to the noise floor? The answer is that it depends on the strength of the link and the ADC output. For the link above with an SNR of 50 dB, a 12-bit ADC corresponds to a quantization error 72 dB below the signal strength, hence 22 dB below the noise floor. This is essentially no additional error at all. In contrast, a 6-bit ADC would cause quantization error 36 dB below the signal strength, which is 25× larger than noise power for this very strong link! So, in some cases it can be important to take quantization error into account when using CSI measurements.

My results later in the thesis show that accounting for quantization error indeed improves the accuracy of Effective SNR predictions considerably. I achieved the best results when I assumed that the IWL5300 NIC in my testbed used a 6-bit ADC.

### 5.4.3 Computing Effective SNR

Taking into account quantization error, I used the processed and corrected CSI[1] to compute Effective SNR values for links in my testbed according to the model described in Chapter 4. I parameterized the model with known properties of the Intel IWL5300 devices: They use minimum mean square error (MMSE) MIMO equalizers, and have known, fixed spatial mapping matrices that I detail in the tool source code [36: `matlab/sm_matrices.m`]. These are the CSI measurements I use in the remainder of this thesis.

### 5.5 Summary

This chapter described my 802.11n experimental platform and CSI measurement tool, and detailed a number of practical considerations that are important when computing Packet SNR and Effective SNR values on real hardware. In the rest of this thesis, I use these measurements to evaluate whether my Effective SNR model is accurate and how well it works relative to other algorithms for a variety of configuration problems.

---

[1] A second issue is an implementation artifact of the way that the Intel chip reports channel state information. The SNR reference is missing from each subcarrier's CSI matrix, and instead all matrices are normalized to an unknown reference. I determined via trial-and-error that this reference value is the total RSSI and developed a procedure to normalize it [36: `matlab/get_scaled_csi.m`].

Chapter 6

# EVALUATING EFFECTIVE SNR FOR MIMO-OFDM CHANNELS

In this chapter, I experimentally evaluate how well my Effective SNR model predicts packet delivery for 802.11n wireless links.

To do so, I use my CSI measurement tool to gather a wide range of channel and performance information across 200 wireless links in both testbeds. This captures a wide variety of fading environments, from line-of-sight links in the same room to links between nodes in different rooms with RF barriers and reflectors spread around and between them. I use this data to evaluate the accuracy of predictions made using Packet SNR and Effective SNR.

The primary study in this chapter determines how accurately Effective SNR predicts whether packets will be delivered using different modulation and coding schemes. The goal is that for every modulation and coding scheme (MCS), a clear Effective SNR threshold separates those links that do not deliver packets at that rate from those that do, and thus that my model is accurate for practical links. Using these thresholds, I determine how well Effective SNR can identify the MCS with the highest throughput. I also compare how well Packet SNR works when used in the same way.

I next consider how well my model enables predictions about the effects of transmit power control on rate. This joint optimization problem highlights my model's flexibility. I conclude this chapter by evaluating the resilience of my Effective SNR system to interference, so that it can still be used to make predictions in contested wireless environments.

Combined, these three studies lay the foundation for showing that Effective SNR is accurate, flexible, and practical.

## 6.1 Experimental Data

I measured packet delivery over a 20 MHz channel on my two 802.11n testbeds, using links with four different antenna configurations:

1. The **SISO** configuration uses a single antenna at each node. This configuration corresponds to 802.11a.

2. The **SIMO** configuration uses a single transmit antenna but three receive antennas. This is an 802.11a/g/n configuration that uses spatial diversity techniques.

3. The **MIMO2** configuration uses two spatial streams and three receive antennas. This

employs both 802.11n techniques of spatial multiplexing and spatial diversity.

4. The **MIMO3** configuration uses three antennas at each node to send and receive three spatial streams. This configuration uses spatial multiplexing but does not benefit from spatial diversity.

For each of these configurations, I measured the packet delivery for each link using each MCS, at each transmit power level between $-10$ dBm and $+16$ dBm in steps of 2 dB. I sent 1,500-byte packets as constant bit-rate UDP traffic generated by `iperf` at 2 Mbps for 5 seconds, about 860 packets total. The receiver also recorded the CSI and per-antenna RSSIs and noise floors to measure the RF channel for each correctly received packet. In these experiments, I turned off 802.11's link layer retransmissions in order to observe the underlying packet delivery rate. The experiments were conducted at night on unused 802.11 channels in order to minimize the effects of environmental movement and RF interference on these results.

The above tested across 200 links, 26 dB of transmit power, four antenna configurations ranging from SISO to MIMO3, and 8 MCS values per configuration. This covers all of the key variables needed to implement and evaluate my Effective SNR model.

### 6.2   *Packet Delivery with Effective SNR*

The first study in this chapter aims to understand whether Effective SNR is a good metric, i.e., whether it is an accurate predictor of packet delivery. In this section, I evaluate the model in three ways. The first is via the *transition window*, i.e., the SNR regime in which packet delivery for all links goes from near-zero to near-perfect. We saw in Chapter 3 that this transition occurs rapidly for a wired link (Figure 3.7), but occurs over a wide range for wireless links (Figure 3.8) when using the Packet SNR. A narrow transition window that matches measurements of Packet SNR over a wire would be one indicator that Effective SNR works well.

The second evaluation metric is *rate confusion*, i.e., how many rates might be best at a particular SNR value. The example wired link showed clear separation between rates, such that at every SNR value there is a clear best rate. Conversely, because the transition regions of different wireless links overlap, links with the same SNR might support very different rates.

Finally, I determine the SNR thresholds for each MCS and use them to predict the rate configuration with the *highest throughput* for each link. The accuracy of this prediction is a core measurement of how well an SNR metric can inform decisions.

For all of these analyses, I evaluate the predictions made by my Effective SNR model independently and as compared to Packet SNR.

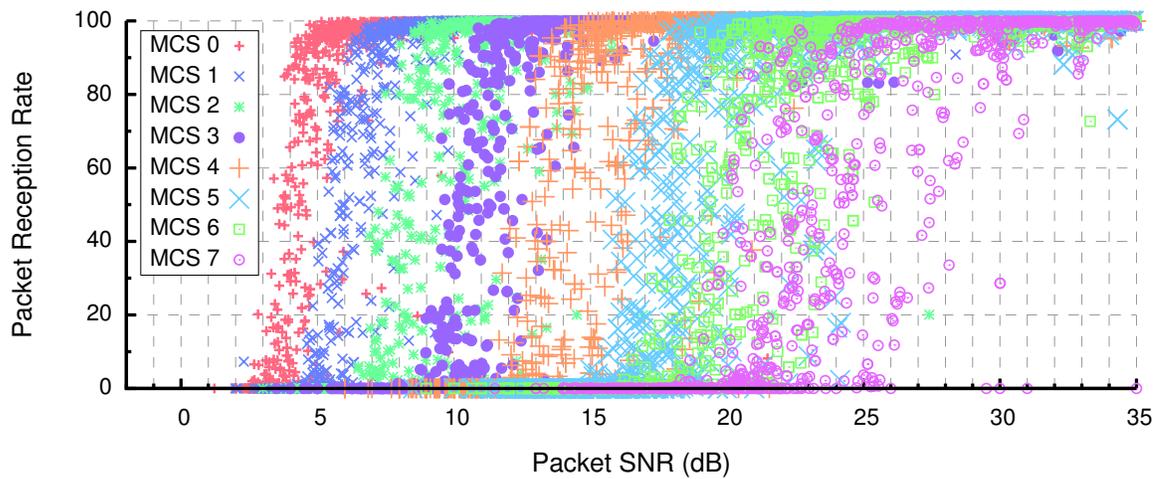

**Figure 6.1: A scatterplot of packet reception rate versus Packet SNR for wireless links in the testbed. There are wide transition regions and significant overlap between rates.**

### 6.2.1 Transition Windows

I begin with a visual comparison of the transition regions for real wireless links, and then present a quantitative evaluation of the difference between Packet SNR and Effective SNR in transition window width. I focus on SISO links, for which the most testbed links transition from being lossy to reliable. I examine the remaining configurations in the next section.

#### Visual Comparison

Recall Figure 3.8 from Chapter 3, which shows a scatterplot of packet reception rate (PRR) as a function of Packet SNR for three sample single-antenna modulations in the testbeds described in Chapter 5. This graph demonstrated that for real wireless links, the width of the transition region is 10 dB or more, so that there is a large range of power levels for which Packet SNR does not predict performance.

In Figure 6.1, I present a version of that plot that now includes data points for all eight modulation and coding scheme (MCS) combinations. In this graph, we can see that there is a correlation between Packet SNR and rate, but there is also a significant overlap between rates. For most MCS values, the transition region is at least five and often ten dB wide. For a large SNR range, many links will be lossy using one rate, while other links will work well at the next higher, or even the second higher, rate. This illustrates why Packet SNR computed from RSSI does not provide a good indicator of performance across testbed links in practice.

Contrast this with Figure 6.2, which shows the exact same set of data points but using Effective SNR instead of Packet SNR along the x-axis. This picture now shows a much clearer separation between rates. Especially for lower MCS values, only a few outlier links overlap with the next higher rate. Note also that, as shown in Chapter 4, the Effective SNR is several decibels lower than the Packet SNR because it measures the amount of power that is actually harnessed by the link, rather than the total power.

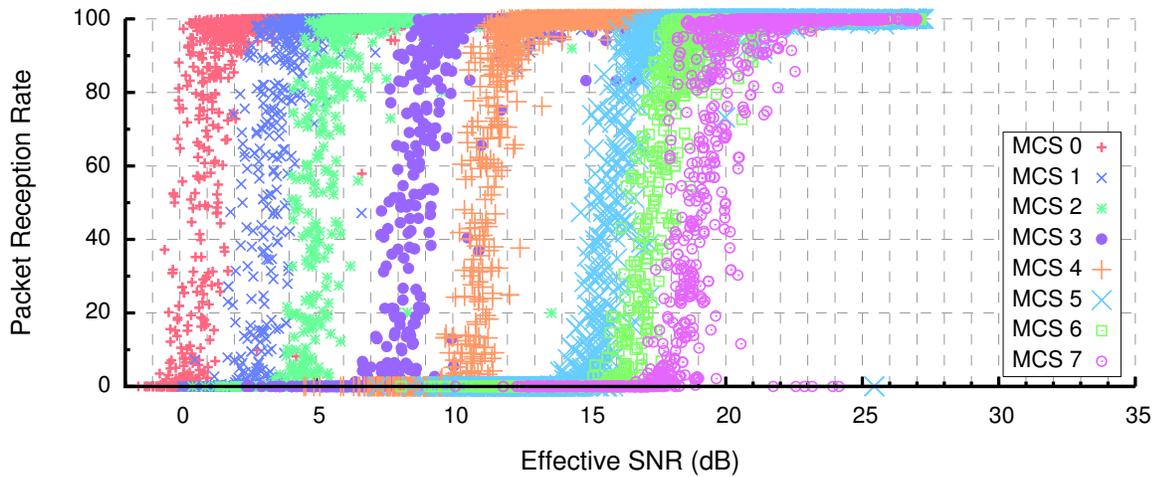

**Figure 6.2: A scatterplot of PRR versus Effective SNR for the same wireless links. Relative to Packet SNR, there are narrower transition regions and cleaner separation between rates. The overlap that is present is generally between links that use the same modulation but different coding schemes (see Table 2.2).**

In both graphs the separation is generally larger between rates that use different modulations, e.g., between MCS 2, which uses QPSK, and MCS 3, which uses 16-QAM. In contrast, rates that only differ in coding rate overlap to a more substantial degree. This effect is worst for the highest rates (MCS 5–MCS 7), which use 64-QAM modulation. I believe this artifact is fundamental, as it matches the results from the wired link (Figure 3.7); I attribute it to the fact that these three MCS use coding rates 2/3, 3/4, and 5/6 that are much closer together than the 1/2-coded and 3/4-coded combinations used for the QPSK and 16-QAM rates.

*Quantitative Evaluation*

The visual comparison presented above shows qualitatively that Effective SNR provides a more compact transition region and clearer separation between rates, but note that scatterplots can be misleading because they obscure density and distributions—each plot above has 16,188 points. To understand the data quantitatively, I analyzed the SISO measurements to find the transition window for each of the measured links.

Formally, I define the transition window of a particular rate to be the set of SNR values between which packet delivery rises from 10% (lossy) to 90% (reliable) for any link. Table 6.1 gives the width of the transition window (denoted $\Delta\rho$) for SISO rates using the Packet SNR and Effective SNR metrics. I show the 25th–75th percentile range of points in the transition window as a measure of the typical link, and the 5th–95th range as a measure of most links. A good result here is a narrow 1 dB–2 dB window like that measured over a wire (Figure 3.7).

| MCS | Rate (Mbps) | $\Delta\rho_{\text{packet}}$ (dB) | | $\Delta\rho_{\text{eff}}$ (dB) | |
|---|---|---|---|---|---|
| | | 5%–95% | 25%–75% | 5%–95% | 25%–75% |
| 0 | 6.5 | 3.08 | 1.29 | 2.05 | 0.81 |
| 1 | 13.0 | 3.45 | 1.44 | 2.38 | 0.89 |
| 2 | 19.5 | 6.27 | 3.12 | 2.30 | 0.85 |
| 3 | 26.0 | 3.93 | 1.98 | 3.02 | 0.94 |
| 4 | 39.0 | 7.05 | 3.49 | 2.19 | 0.93 |
| 5 | 52.0 | 7.16 | 3.20 | 2.29 | 1.06 |
| 6 | 58.5 | 7.25 | 3.37 | 2.92 | 1.41 |
| 7 | 65.0 | 7.24 | 2.81 | 2.92 | 1.35 |
| | Average | 5.68 | 2.59 | 2.51 | 1.03 |

**Table 6.1:  A table of the widths of SISO transition windows.**

The table shows that the transition widths are consistently tighter with my model than with Packet SNR. Most links transition within a window of around 2 dB for most rates. The width of the SNR-based transition windows is typically two to three times looser, especially for the denser modulation schemes like 64-QAM and higher code rates. At higher rates, it is easy for a sub-ideal channel to degrade packet reception. However, while the transitions for the last four rates are high with Packet SNR, they remain tight with Effective SNR.

*Limits on Accuracy*

In Table 6.1, the transition regions for Effective SNR range from 1 dB to about 3 dB, depending on the MCS value. In fact, these results for Effective SNR are about the best that can be obtained because they are close to textbook transitions for flat-fading channels and those measured over a wire (Figure 3.7). A small improvement is surely possible, but this is probably limited by the precision of my measurement data. The IWL5300 gives RSSI, AGC and noise values in dB to the nearest integer, and outputs at most 8-bit CSI over a 48 dB range for only 30 out of 56 subcarriers. With this combination of factors, a CSI quantization error of at least 1 dB is likely.

*Summary*

This section showed that my Effective SNR model provides a channel metric that can narrow transition windows and shows a separation between rates. The larger significance of narrow transition windows is that, by reducing them enough that they do not overlap, I can unambiguously predict the highest rate that will work for nearly all links nearly all of the time. In contrast, Packet SNR transition windows overlap such that for a given SNR there may be many different best rates for different links in the testbed. I explore this next.

### 6.2.2 Rate Confusion

To understand whether my Effective SNR model accurately predicts packet delivery, I analyze the fastest working rate (PRR ⩾ 90%) for each link and all NIC settings. If we consider the set of all links that have the same SNR value (binned in groups of 1 dB), the *best link* is the link with the fastest working rate, and the *worst link* is the slowest. Ideally, SNR would perfectly indicate rate and all links with the same SNR would have the same best rate, but in practice there is a gap; this gap is the *rate confusion*.

In Figure 6.3, I show the rate confusion by plotting the rate versus SNR for the best and worst links, broken down by configuration. Recall that to measure PRR I sent over 800 packets for each link in transmitter configuration. To assign a single Packet SNR or Effective SNR value to the trace, I choose the median measurement over all successfully received packets. The SISO experiment (Figure 6.3(a)) shows links for both testbeds combined. The remaining graphs (Figure 6.3(b)–6.3(d)) show rates for SIMO, MIMO2 and MIMO3 configurations for the Intel testbed only; it is denser than UW and supports MIMO experiments over the IWL5300's transmit power range. Note that the SIMO figure does not include data for the lowest 6.5 Mbps rate because, with the high degree of spatial diversity, very few links experience loss at that rate within the transmit power range of the IWL5300.

For the SISO (Figure 6.3(a)) and MIMO3 (Figure 6.3(d)) cases, the figures show that using Packet SNR results in a large spread between the best and worst lines. Except for extremely low and high SNRs, nearly all SNRs have at least two—and up to five different—rates as suitable choices for the best rate. That is, Packet SNR often poorly indicates rate.

In sharp contrast, the two Effective SNR lines overlap almost all the time, and mostly appear to be a single line. This is almost an ideal result. Effective SNR is a clear indicator of best rate. When there is slight separation, the spread is only between rates that use the same modulation but different amounts of coding, just as I described in the last section. These combinations are also close together in our wired experiments.

Interestingly, these results show that Packet SNR predictions are much better for the SIMO and MIMO2 cases, though still not as accurate as Effective SNR, particularly for the highest rates. The reason is *spatial diversity*: Spare receive antennas gather the received signal and combine it to make the channel more frequency-flat [35], thus bringing the Packet SNR closer to the Effective SNR. This effect is well-known, though typically not observable using real 802.11 NICs which, except my prototype implementation, do not export CSI. This result suggests that Packet SNR *is* a reasonable predictor for an 802.11 configuration with significant diversity, using RSSI measured in that configuration. Still,

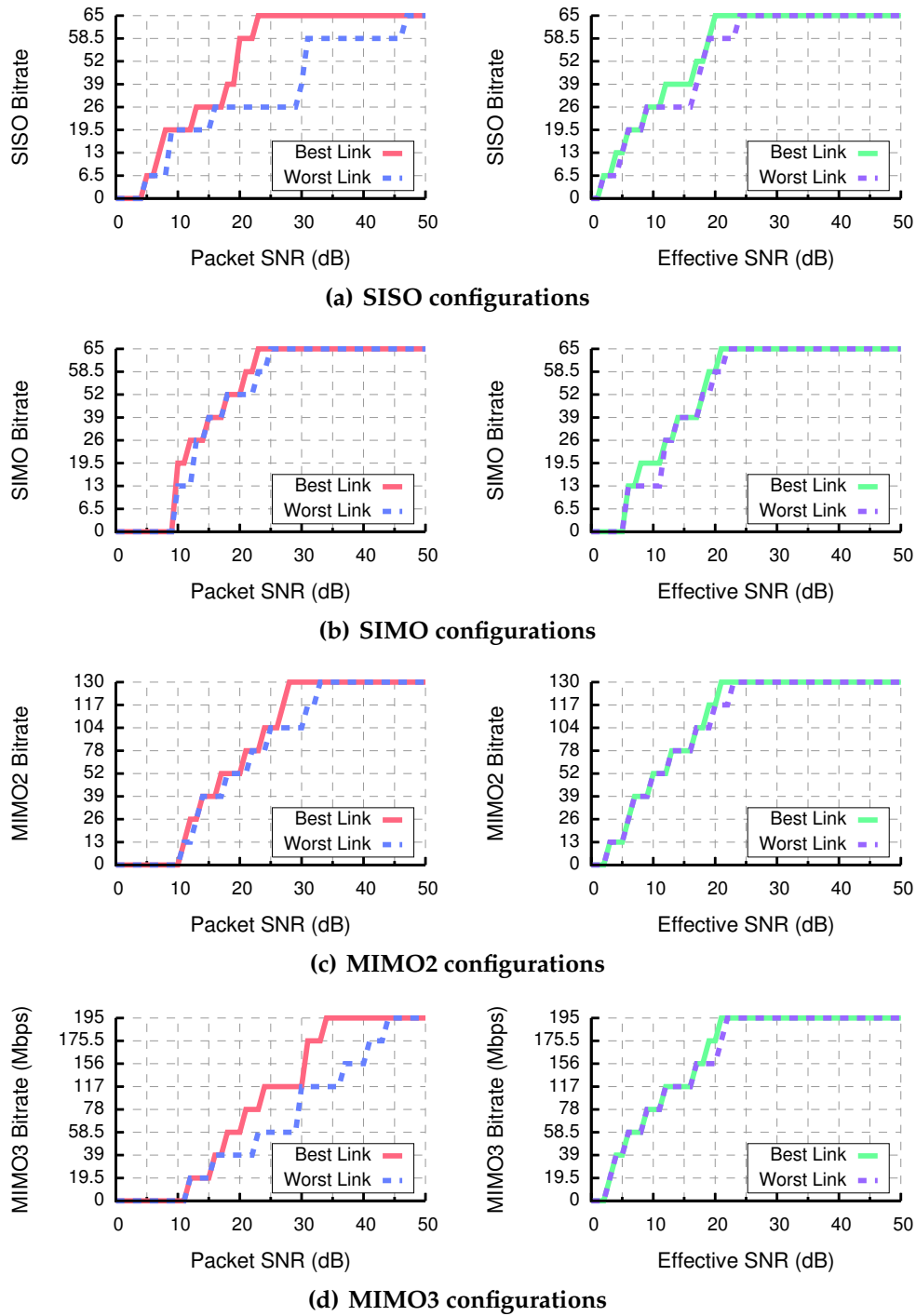

**(a) SISO configurations**

**(b) SIMO configurations**

**(c) MIMO2 configurations**

**(d) MIMO3 configurations**

Figure 6.3: Rate confusion with Packet SNR and Effective SNR. Excepting very low and high SNRs, one Packet SNR value maps to multiple best rates for different links, though it works better in configurations that use spatial diversity (SIMO and MIMO2). For the same data, Effective SNR provides a clear indicator of the best rate for nearly all links.

Packet SNR does not transfer well across the antenna modes, because diversity gains and inter-stream interference change unpredictably. This makes Packet SNR less useful as a method of selecting rates, as we will see in the next section.

### 6.2.3 Selecting Rates

In Section 6.2.1, I showed that transition regions for Effective SNR are generally tighter than for Packet SNR. In Section 6.2.2 I showed that Effective SNR as a channel measure has less rate confusion relative to Packet SNR. These results indicate that a threshold-based Effective SNR algorithm should be better able to distinguish whether a particular MCS combination works for most links. In this section, I show how to choose SNR thresholds and analyze whether these thresholds lead to good choices for real wireless links in my testbeds.

#### Choosing SNR Thresholds

The first question is how we should choose the SNR thresholds needed in Equation 4.3 to decide whether a link will reliably deliver packets in a particular configuration. A low SNR threshold will lead to aggressive rate selection, with many *false positives* as transmitters send at a rate faster than their links support, and widespread packet loss. A high SNR threshold will cause many *false negatives*, in which transmitters select rates conservatively and send more slowly than necessary. In this thesis, I use a simple heuristic: Choose thresholds to balance the prevalence of false positives and false negatives.

To find the threshold for a particular MCS value, I use the following procedure. Define *good links* to be those links with PRR ⩾90%. The CDF of the SNR values for these links then indicates the false negative rate; choosing the 25th percentile SNR as the threshold means that the 25% of links that have lower SNR values will be falsely classified as not working. Similarly, we can define *bad links* to be those with PRR <80%. Then the CDF of the SNR values for these links shows the inverse of the false negative rate; the 25th percentile SNR would falsely classify the 75% of links with larger SNR values as good links. To balance false negatives and false positives, we can plot the CDF of SNR values for good links and the complementary CDF of SNR values for bad links on the same graph. The x-coordinate of the intersection point of the two lines is the SNR threshold that balances false positive and false negative error rates, and the y-coordinate of that point gives the *balanced error rate*. An ideal result is a balanced error rate of 0%, so that the SNR threshold perfectly classifies all links as working or not.

Figure 6.4 shows the error rate versus Packet SNR for the SIMO, MIMO2, and MIMO3 configurations, and Figure 6.5 shows the same data using Effective SNR. Unlike in Section 6.2.2, the Packet SNR or Effective SNR value used to make these graphs is taken from a single MIMO3 packet. To compute Effective SNR for SIMO and MIMO2 configurations, I compute the SIMO and MIMO2 Effective CSI as described in Chapter 4. For Packet SNR, I use the measured Packet SNR for all antenna configurations, as the transmitter keeps the total radiated power constant across antenna modes.

The results show that both SNR metrics have a balanced error rate under 14% for all configurations. Effective SNR has a lower error rate for all but the lower MIMO2 rates, often much lower. We also see that both schemes perform worst for the highest MCS combinations in each antenna mode that use 64-QAM with close coding rates. This matches the expectations from earlier sections.

I plot the CDF of the balanced error rate for each scheme in Figure 6.6. The balanced error rate curve is generally 1%–2% less for Effective SNR than for Packet SNR. I also plot the relative balanced error rate of Effective SNR compared to Packet SNR in Figure 6.7. This graph shows that the Effective SNR balanced error rate is as low as 40% of that achieved with Packet SNR, and less than 10% worse for the single rate where that is the case. Thus, it seems likely that my threshold-based Effective SNR model will be a better predictor of link performance than using thresholds with Packet SNR. I investigate this next.

*Using Thresholds to Select Rates*

Having computed the Packet SNR and Effective SNR thresholds for each MCS value, I now evaluate how well the chosen rates work in my testbed.

I used the thresholds computed above to choose rates for 2163 links in my wireless testbed. These are the links that support at least 6.5 Mbps, so they can deliver packets, but are not so strong that the fastest rate works perfectly, so that rate selection is meaningful. I follow the procedure in Algorithm 4.2: Given a single Packet SNR or Effective SNR measurement from each trace, predict how well each configuration works and then choose the fastest one. I compare to the ground truth optimal rate for each link, computed by using the packet reception rates described earlier to compute the actual performance of each rate.

Figure 6.8 shows the results when selecting rates using Packet SNR or Effective SNR. The graph contains one line for each algorithm. A point $(x, y)$ on a line means that the corresponding algorithm achieves at least an $x$ fraction of optimal performance for a $y$ fraction of links. An ideal result would be a vertical line $x = 1$, showing that all links achieve optimal performance.

In this graph, we see that Effective SNR dramatically outperforms Packet SNR. The median link in our testbed achieves 83% of optimal performance using rates selected with Effective SNR, while using Packet SNR for rate selection results in a median performance of 13%. With Packet SNR, nearly half the links choose rates that deliver no packets at all, while Effective SNR is so excessively aggressive for less than 10%. Though there is still a sizable gap from optimal, this experiment highlights that Effective SNR performs much better than the Packet SNR, even when using only a single channel measurement and no adaptation of

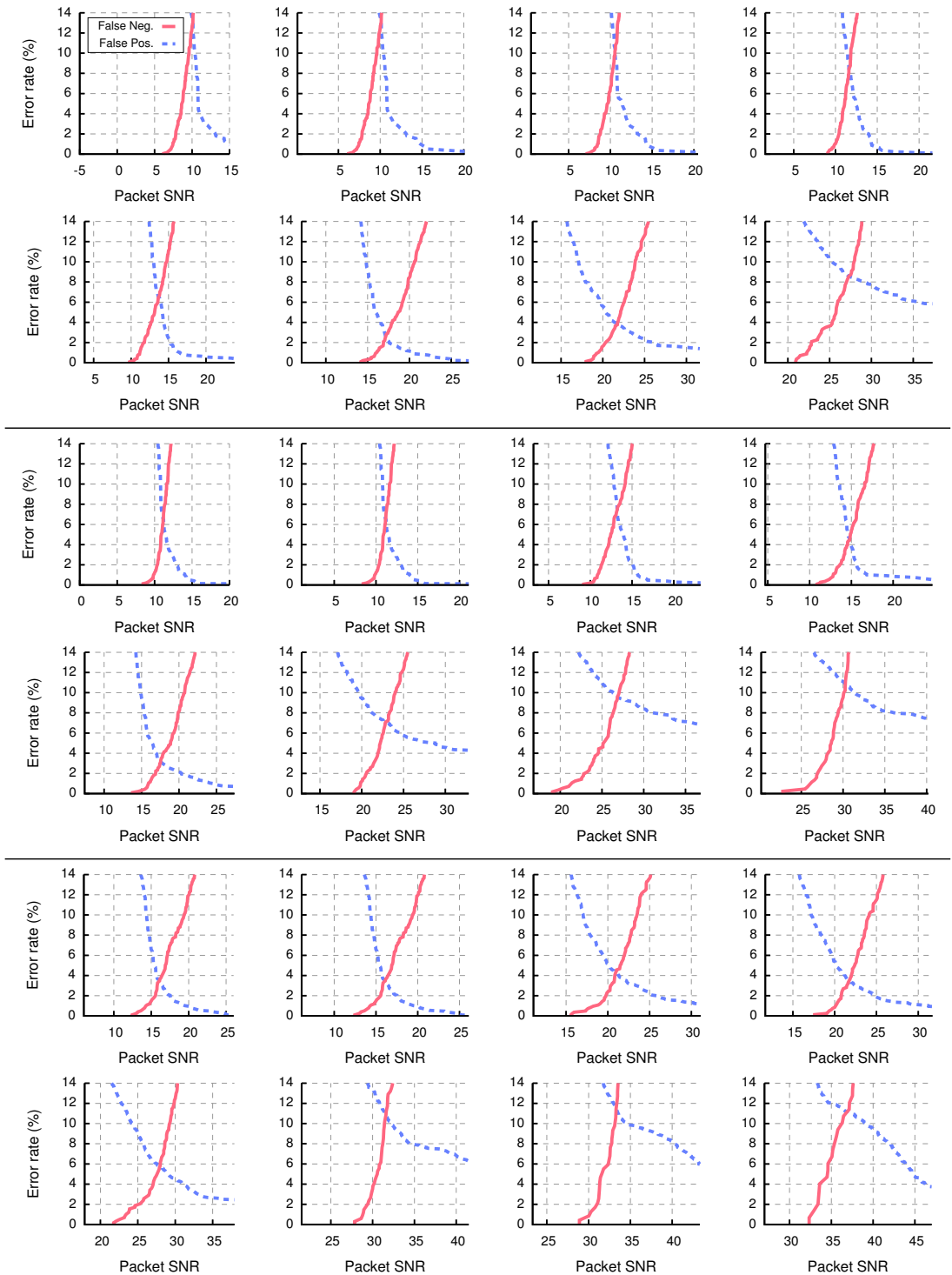

**Figure 6.4:** False negative and false positive error rates as a function of Packet SNR threshold. The figure shows results for wireless links that use one- (*top*), two- (*middle*), and three-stream (*bottom*) rates (MCS 0–MCS 23).

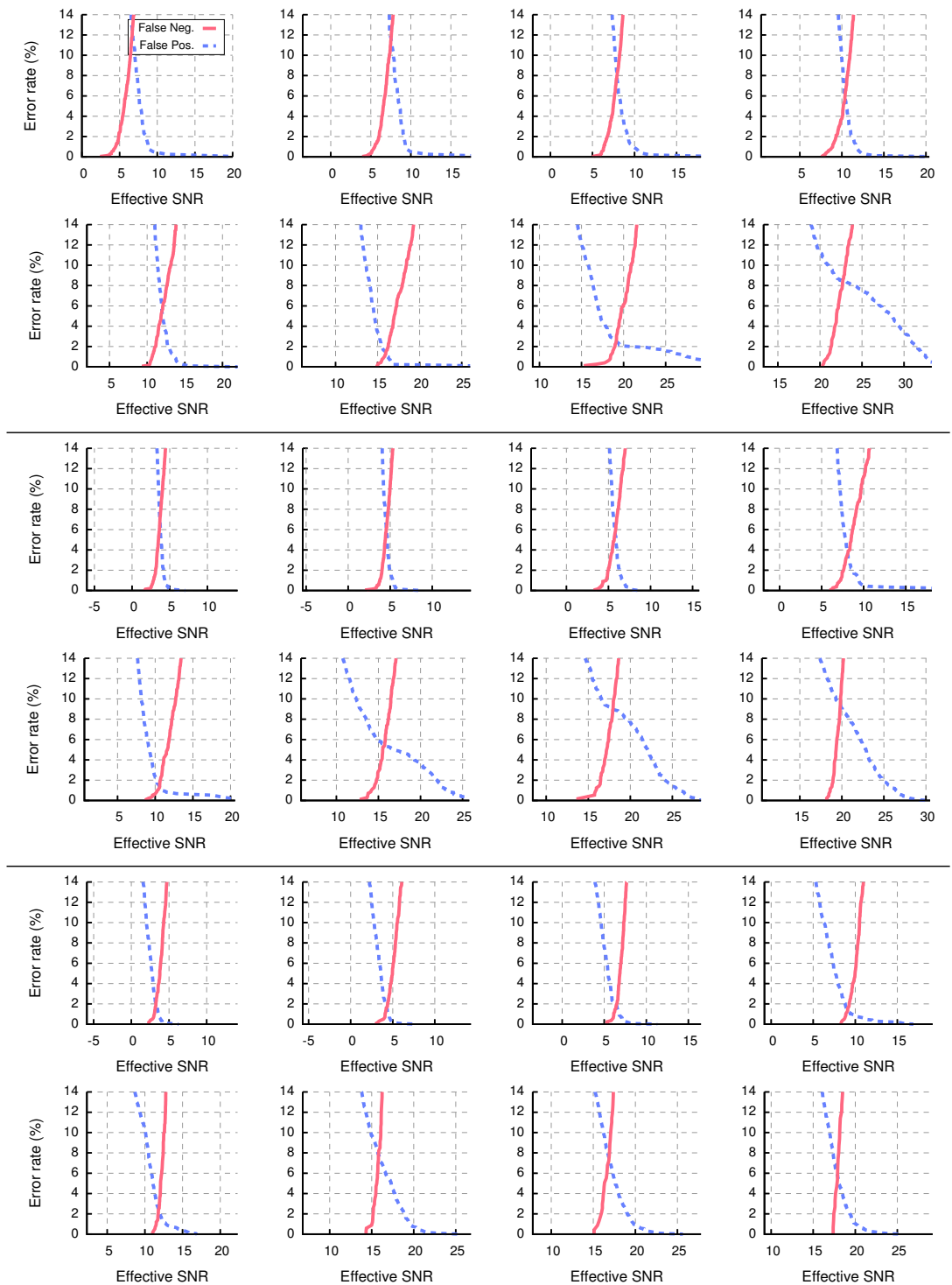

**Figure 6.5:** False negative and false positive error rates as a function of Effective SNR threshold. The figure shows results for wireless links that use one- (*top*), two- (*middle*), and three-stream (*bottom*) rates (MCS 0–MCS 23).

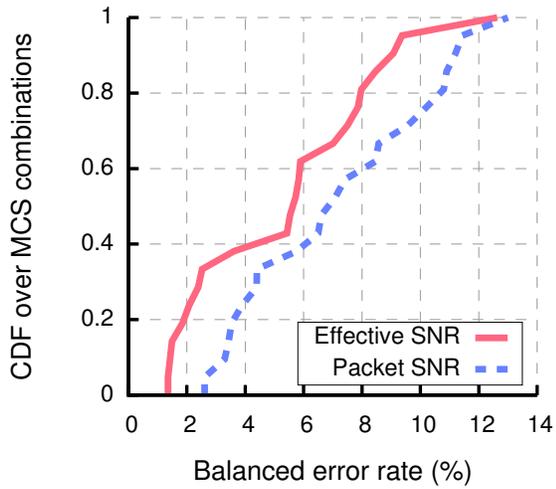

Figure 6.6: Balanced error rates for Effective SNR and Packet SNR in wireless testbed links.

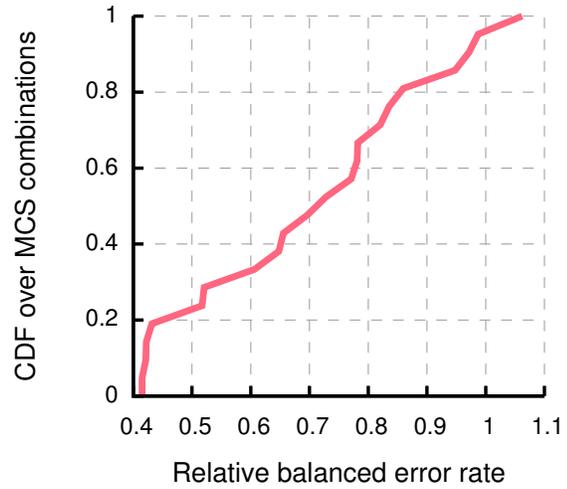

Figure 6.7: The balanced error rate for Effective SNR relative to Packet SNR in wireless testbed links.

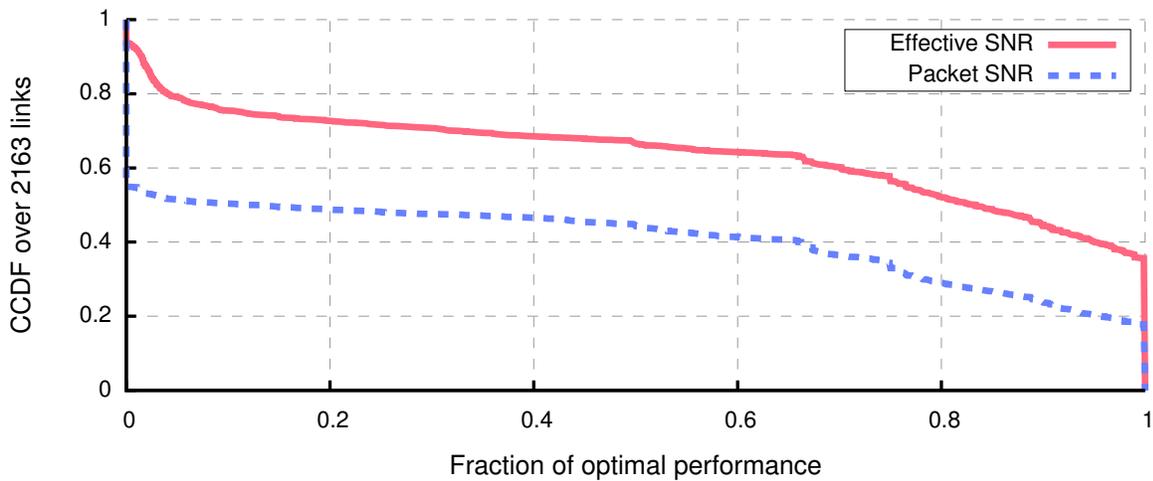

Figure 6.8: The performance of selecting rates based on Effective SNR and Packet SNR thresholds for 2163 wireless testbed links.

rates. (I explore rate adaptation in detail the next chapter.)

To understand how much the particular choice of threshold affects both algorithm, I experimented with raising or lowering the SNR thresholds uniformly across the different MCS choices. Raising the Effective SNR thresholds by 1 dB results in more conservative choices of rate and improved average performance: Fewer links achieve maximum performance, but the worst links that were over-selecting improve to a larger fraction of optimal. However, raising or lowering the Packet SNR had little effect on the curves in Figure 6.8, which suggests that the balanced error rate technique achieves locally-optimal thresholds.

Why does Effective SNR choose rates so much better than Packet SNR? The key reason is that Effective SNR is able to compute a different Effective CSI for each modulation and number of spatial streams based on the particular subchannels used. This means that Effective SNR can make decisions about each configuration independently, and hence can recognize when fading will make a strong link work poorly in one antenna mode but not another. In contrast, there is no way using RSSI measurements alone to predict Packet SNR in different configurations. Packet SNR must make a decision independent of fading, which often results in overly-aggressive choices.

This dramatic difference in performance highlights the key strength of Effective SNR: Its ability to predict performance in a wide space. Though this example only included selecting rates keeping all other metrics fixed, subsequent studies will consider additional dimensions of the wireless configuration space.

### 6.2.4   Summary

In this section, I analyzed whether Effective SNR accurately predicts packet delivery. I found that viewing wireless links through the lens of Effective SNR can lead to visual separation between rates and narrow transition regions within rates. The second part of this study showed that Effective SNR is a clear indicator of rate, usually narrowing down the possible set of best rates to one or two MCS within an antenna configuration. The third analysis showed that Effective SNR results in generally good choices for rate (median 83% of optimal), while Packet SNR often chose rates that simply did not work (median 10% of optimal).

In the rest of this chapter I demonstrate that Effective SNR has a few other useful properties that other channel metrics do not, namely that it can be used to solve joint optimization problems such as between transmit power and rate, and that the estimates it provides are robust to interference.

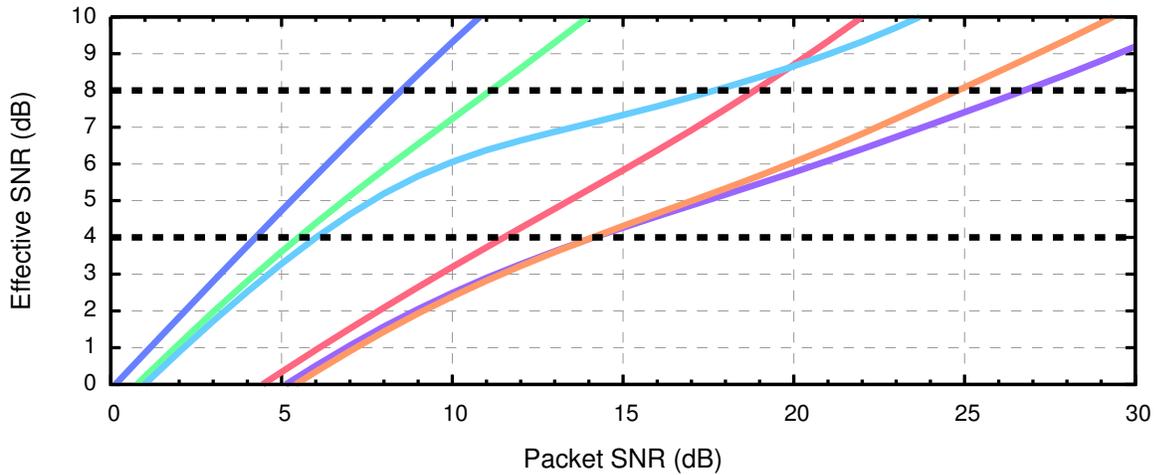

**Figure 6.9: Effective SNR (for QPSK) versus Packet SNR for flat (left) to faded (right) links.**

### 6.3 *Transmit Power Control*

The data above showed that my Effective SNR model can predict delivery for measurements taken over a range of transmit powers, among other choices of rate and spatial streams. I now show apply this model to the *joint interaction* of transmit power and rate, and I show that CSI measured at one transmit power level is useful to predict delivery at a *different power level*. This is valuable for power control applications, e.g., pruning excess power to reduce co-channel interference. Earlier work has shown that Packet SNR based on RSSI does not do this well [77, 91, 110].

#### 6.3.1 *Transmit Power Control in Faded Channels*

First, I analyze the effect of changing transmit power in faded channels. To do this, I simply scale the CSI measured at maximum transmit power for a link and compute the resulting Effective SNR over a range of power levels.

I found that changing transmit power has a different effect (in terms of delivery and highest rate) on different testbed links even if they start at exactly the same rate and SNR value. Figure 6.9 plots the Effective SNR versus Packet SNR relationship for six example SISO links from my 802.11n testbeds, chosen to represent a range of frequency-selective fading profiles similar to those in Figure 3.9. The links range from nearly flat to deeply faded. Correspondingly, they have different slopes.

On the left, Packet SNR matches Effective SNR for the nearly flat link. Since all subcarriers have the same strength, this link can be modeled as a single carrier with a single BER, and hence scaling transmit power has a linear effect on the Effective SNR. However, for the right-most, deeply faded links, the Packet SNR decreases from 25 dB to 15 dB ($10\times$ transmit

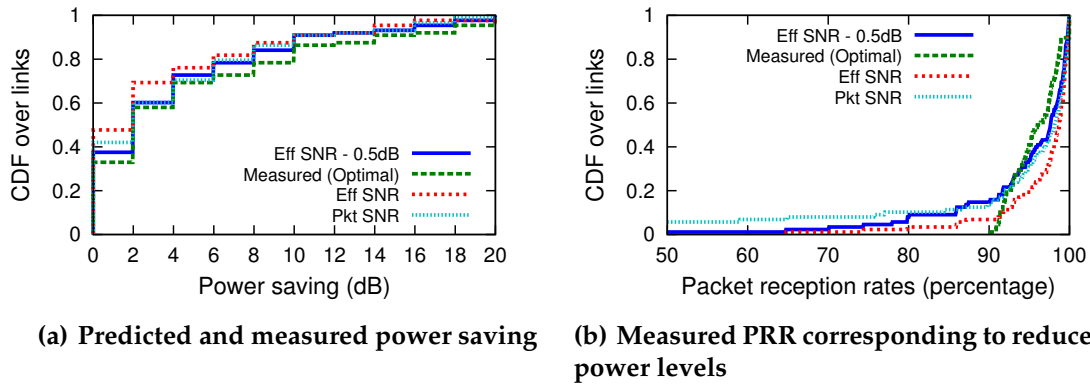

**(a) Predicted and measured power saving**

**(b) Measured PRR corresponding to reduced TX power levels**

**Figure 6.10: Power saving and performance impact of pruning excess transmit power. Pruning with Effective SNR is tight (within 0.5 dB) and does not degrade performance. Pruning with packet SNR degrades performance more without much extra savings.**

power reduction) between the dashed lines, while the Effective SNR only drops by 4 dB ($2.5\times$). This occurs because reducing the power has different effects on error rate at different SNR levels (see Figure 2.3)—so additional power has less impact on strong subcarriers than on weak subcarriers. This difference in how well devices harness power across links makes transmit power control non-trivial and explains why it has been hard for prior algorithms to use measurements at one SNR value to predict how well links will work at a different power level.

### 6.3.2 *Making Predictions with Effective SNR*

I test the ability of Effective SNR to make predictions across power levels by considering the goal of trimming excess transmit power. Excess transmit power is power that can be removed without causing the highest rate for the link to drop.

These experiments start with 88 SISO links from the Intel Labs testbed configured to radiate 10 mW of transmit power. I then take a single CSI sample per link. Considering transmit power reductions in increments of 2 dB, I use the threshold-based Effective SNR and Packet SNR algorithms to predict the best supported rate for each reduced power level, and choose the lowest power level with the same best rate. The measurements described earlier include the ground truth packet reception rate for each power level and thus they can be used to check the accuracy of the predictions.

Figure 6.10 shows the power savings and performance degradation of four different threshold schemes. A good result here is power savings without a loss of performance; the absolute amount of power savings is not meaningful as it depends on the testbed and the link. The Measured (Optimal) line shows the best that can be done. Measured PRRs at all power levels are used to guide power control decisions. Therefore, the final delivery probabilities are hardly decreased (all links have PRR ⩾90%). Most links save a little power, and some save a lot.

The graphs show that using Effective SNR to predict how much power to trim has a similarly good tradeoff. Impact on rate remains limited, yet power is saved, more than 10 dB for around 10% of the links. The gap between the Measured and the Eff SNR lines is because Eff SNR thresholds might be slightly conservative for some links.

To show that this trimming is tight, I considered the effect of using more aggressive—i.e. slightly lower (0.5 dB)—Effective SNR thresholds. This operation results in little additional power savings, but degrades performance for many more links.

Finally, I compared this algorithm to a scheme that uses Packet SNR to save power. The results show Packet SNR the savings are roughly similar, but more links have degraded performance and several stop working altogether.

### 6.4   Interference

I conclude this chapter by investigating how my Effective SNR-based model can cope with interference. This challenge is one of the largest potential weaknesses of this technique, because Effective SNR is based on measurements taken only during the packet preamble. There are three important components of dealing with interference, which I approach in turn.

#### 6.4.1   CSI During Interference

The first question is: If a weak interferer occasionally transmits while a packet is being received, does this weak transmission cause wild swings in the predicted link quality?

I studied the variation of CSI measurements during transient interference. I chose two nodes at UW that do not detect each other with carrier sense, and alternately designated one as the transmitter and the other as the interferer. The nodes were configured to send large packets designed to collide, while all other receiving nodes monitored the CSI to simulate a total of 20 links. The experiment also varied the transmit power of the node designated as the interferer from low to high to induce a large range of interfering channels, over which I evaluate the impact of interference on CSI measurements and my Effective SNR model.

For all but one of 20 links, the rate predicted by my model for the majority of correct packets was the same with and without interference; the remaining link was off by a single rate. In other words, the interference does not corrupt the CSI measurements, because the MIMO-OFDM training procedure can fairly accurately estimate CSI. Note that OFDM does not turn interference into inflated RSSI, unlike the spread spectrum modulations used in 802.11b. From these measurements, we can conclude that the mere presence of interference does not completely invalidate Effective SNR values, and thus transient interference will not necessarily cause wild swings in the recommend transmit configuration.

### 6.4.2    Recognizing Collisions

A second task is to recognize collisions when they occur, in order to distinguish between too-aggressive rate selection and interference loss. The solutions to these problems are very different. To work around persistent collisions from hidden terminals, a transmitter increases its MAC backoff counter and/or initiates the RTS/CTS procedure which is normally disabled for protocol efficiency. In contrast, when rates are over-selected the transmitter should reduce its rate but ideally need not add additional protocol delay. Accurately distinguishing between these two effects is known to improve performance in practice [50, 120, 123].

To recognize collisions, I propose to leverage a new MAC feature of the 802.11n packet aggregation mechanism. Block ACKs selectively acknowledge frames in a batch of packets transmitted as one continuous burst. Each packet in the burst has a separate checksum, and thus the Block ACK serves as block-based feedback of packet correctness just like the block-based checksums analyzed in PPR [50]. We can therefore use the error patterns in the Block ACK to recognize collisions: When the rate is over-selected, errors should be randomly distributed throughout the batch, and bursty when a collision clobbers a continuous part of the batch. The ability to recognize interference and hence decouple interference avoidance and rate selection has been shown to improve performance in many systems (e.g., SoftRate [120]).

### 6.4.3    Effective SINR

The final task is to dealing with persistent interference, i.e., if the RTS/CTS procedure does not successfully avoid collisions. For continuous interference the Effective SNR computed from CSI will likely provide an aggressive estimate, and the system will need another way to compensate. It may be possible to use an *Effective Signal-to-Interference-and-Noise Ratio (Effective SINR)* metric that incorporates CSI measurements from the interfering nodes to predict packet delivery taking into account the MIMO-OFDM fading properties of both the desired signal and any interfering transmissions. I leave the investigation of this technique to future work.

### 6.5    Summary

From the studies in this chapter, I conclude that Effective SNR consistently provides accurate estimates of packet delivery for nearly all links and all configurations without any per-link calibration. Viewing diverse wireless links in my testbeds through the lens of Effective SNR showed that there is a low degree of rate confusion and narrow transition region, contrary to the same results with Packet SNR. I presented my method to choose Effective SNR thresholds, and showed that Effective SNR can select rates that work well for most testbed links. The results in Section 6.3 demonstrate the flexibility of this approach by showing that CSI measurements are valid not just across different rates, but also across transmit power scaling. Finally, I discussed three ways in which my Effective SNR model, in conjunction with 802.11n protocol features, can work well in the presence of interference, though I leave detailed investigation of this problem for future research.

In the remainder of this thesis I evaluate the ability of Effective SNR to solve a variety of tasks. In the next chapter, I deploy my model as part of a rate adaptation system that selects the operating rate for a wireless link in mobile channels.

Chapter 7

# RATE SELECTION WITH EFFECTIVE SNR

In the last chapter, I showed that my Effective SNR model can accurately predict packet delivery for 802.11n. Here I close the loop, demonstrating how my model can be applied to solve an 802.11n configuration problem.

This chapter presents an in-depth study of the application of Effective SNR to the problem of rate selection for 802.11. This is a fundamental, well-studied problem because selecting a good rate at which to encode data is crucial for a link to work perform well, and if links do not perform well then no higher-level applications can be built. Though there is a wide variety of rate control algorithms, these probe- and channel-based schemes generally do not extend well to 802.11n and/or fast mobile channels.

I first compare an Effective SNR-based rate selection algorithm against state-of-the-art schemes for single-antenna 802.11a/g systems, which generally work well for SISO links. The goal is to show that Effective SNR performs as well as or better than these existing, well-studied probe- and channel metric-based schemes on their home ground, while my method has the advantages of simplicity, deployability, and generality.

I then show that my Effective SNR model extends well to 802.11n (MIMO), where other schemes do not work well. The results in this chapter will show that Effective SNR provides an accurate and response rate selection algorithm that provides good performance across SISO and MIMO configurations and a range of mobile channels.

In the next chapter, I will explain how to apply Effective SNR to a set of other configuration problems.

## 7.1  Experimental Methodology

I experiment with Effective SNR, an algorithm based on my model, plus SampleRate [14], the de facto rate selection algorithm in use today, and SoftRate [120], a research algorithm with the best published results.

I implemented a version of Effective SNR that randomly probes other antenna modes to collect CSI, and that sends Effective SNR estimates back to the transmitter. I ran it online against SampleRate in a human-scale mobility test. The results showed that the probing and feedback have little penalty: Effective SNR works better than SampleRate by a small (5%–10%) margin.

For a detailed comparison of Effective SNR with other algorithms, I turn to simulations. This is for two reasons: First, SoftRate runs on a software-defined radio, and cannot be implemented on a currently available commercial 802.11n NIC. Second, I want to compare the algorithms over varied channel conditions, from static to rapidly changing, to assess how consistently they perform. No algorithm will beat SampleRate by a significant margin on static channels, because it will eventually adapt to the channel. In contrast, SoftRate performs well even when the channel is changing rapidly. However, it is hard to generate controllable experiments in high-mobility settings. Traces let us perform these comparisons directly.

In this section, I first describe the rate selection (or adaptation) algorithms studied, and then present my trace-driven simulator that I use to perform the comparisons between the different strategies.

### 7.1.1   Rate Selection Algorithms

**SampleRate** [14] is an implicit feedback scheme that uses only information about packet reception or loss to guide rate selection. It maintains delivery statistics for different rates to compute the expected airtime to send a packet, including retries. It falls back to a lower rate when the airtime of the chosen rate exceeds (due to losses) the airtime of a lower rate. Standard implementations send a packet to probe 1 or 2 higher rates every 10 packets, to determine whether to switch to a higher rate.

The main weakness of SampleRate is its slow reaction to change. If the wireless channel quickly degenerates, SampleRate will incur multiple losses while it falls back through intermediate rates.[1] When the channel recovers, SampleRate's infrequent probing converges to the new highest rate slowly. Algorithms such as RRAA [123] aim to improve on SampleRate's weaknesses, but they are less widely used. The version of SampleRate I test is based on the minstrel [109] implementation in the Linux kernel. For 802.11n (MIMO) links, I use a version of SampleRate adapted for multiple streams, and based on the Linux minstrel_ht [28] algorithm.

**SoftRate** [120] is an explicit feedback scheme that uses information gathered during packet reception at a given rate to predict how well different rates will work. The input to these predictions is the bit error rate (BER) as estimated from side information provided by the convolutional decoder. SoftRate chooses rates based on the performance curves that relate the BERs for one rate (a combination of modulation and coding) to another. Each rate will be the best choice for some BER range. SoftRate has been shown to dominate trained SNR-based algorithms such as CHARM [55], so I do not evaluate against those approaches.

---

[1] The original SampleRate [14] did not reduce rate for retries, but some implementations [55] and the version used in modern kernels [109] do. This turns out to be important for good performance.

SoftRate is defined for SISO channels, like SampleRate, and its predictions hold only for fixed transmit power and antenna modes. It does not extend to MIMO systems; instead, separate SoftRate probes would be needed for each separate antenna configuration. I only use it for 802.11a/g experiments. To cover the full SISO range, I extended the MIT implementation of SoftRate to the 64-QAM modulation and added support for 2/3 and 5/6 rate codes.

**Effective SNR** uses my model in a very simple way, based on Algorithm 4.2. Given recent channel state information and per-MCS Effective SNR thresholds, it computes the highest rate configuration that is predicted to successfully deliver packets. It runs at the receiver, measuring CSI on received packets and returning rate changes to the sender along with the ACK like SoftRate. Finally, to protect against poor choices near a rate boundary in our model, I fall back one rate if consecutive packets must be retried and the Effective SNR level has not changed. This is a fixed rule.

Like SoftRate, this algorithm obviates the search phase. There is no calibration of dynamic thresholds. This is not rate *adaptation* so much as rate *selection* that changes only because it tracks the channel's evolution. And unlike SoftRate, the predictions of the model hold over different antenna modes. This lets it run over 802.11n rates as easily and in the same way that it runs over 802.11a/g rates. Thus, I report results from both 802.11a/g and 802.11n runs for this algorithm.

**Optimal.** Finally, I take advantage of simulation to add upper bounds on achievable performance. This lets me assess how well the algorithms perform on an absolute scale. The Optimal scheme has an oracle that knows the true highest rate that can be successfully delivered at any given time. This is of course impractical, but the simulator can provide it. The Delayed Optimal scheme knows the optimal rate that worked on the channel for the previous packet and uses it for the next transmission; unlike Optimal, it does not know the future. Since SoftRate and Effective SNR use an estimate of this previous channel state, and SampleRate infers the recent channel state, they are unlikely to beat Delayed Optimal. The gap between Delayed Optimal and Optimal is also likely to be large because of inherent wireless channel variability—the Optimal algorithm gets the benefit of instantaneous transient improvements in the channel.

### 7.1.2 Trace-driven Simulator

The simulator I built uses a trace from a real mobile channel and implements all algorithms described above.

### Channel Trace

I collected real channel information for the simulations. I walked around UW CSE while carrying a laptop configured to send back-to-back small packets to stationary testbed nodes that record the CSI. The CSI measures frequency-selective fading over real, varying 20 MHz MIMO channels. This is typically not observed with more narrowband experimentation, e.g., on the USRP. Recall that CSI is estimated during the preamble of the packet transmission, independent of the modulation and coding of the payload. Therefore, the mobile transmitter can quickly cycle through all antenna configurations (SIMO, MIMO2 and MIMO3) by sending a single short UDP packet at the lowest rate for each configuration.[2] This enables fine grained sampling of the channel, approximately every 650 µs. The following results are derived from a trace with approximately 85,000 channel measurements taken over 55 seconds, spanning varying RF channels that range from the best 3-stream rates to SISO speeds.

The measured CSI from the trace is interpolated to 56 carriers and serves as the ground truth for the channel in the packet simulator I describe next.

*Simulator*

To simulate rate selection algorithms operating over a mobile channel, I built a simulator from three interacting parts: (1) An 802.11n packet simulator, which determines whether packets will be delivered successfully given a particular instance of the wireless channel. This simulator also computes the physical-layer feedback used by the SoftRate and Effective SNR algorithms. (2) Implementations of the SampleRate, SoftRate, and Effective SNR algorithms. (3) An 802.11n MAC simulator, which maintains the state of the 802.11n protocol, including the current time, the length of transmissions, and MAC backoff parameters. The MAC simulator acts as a bridge between the packet simulator and the rate selection algorithms, taking rate choices from the rate selection algorithm and returning to it information about packet success or failure and the appropriate physical-layer feedback at that point in time.

**802.11n Packet Simulator.** I wrote a custom 802.11a/g/n packet simulator in a combination of MATLAB and the MIT C++ GNU Radio code for SoftRate. Given the CSI describing a particular wireless channel, the simulator implements packet reception as shown in Figure 7.1. This includes demodulation for BPSK through 64-QAM, deinterleaving, and convolutional decoding with soft inputs and soft outputs. Packets are correctly received when there are no uncorrected bit errors, or else they are lost. Thus given a measured CSI trace, the packet simulator computes which MCS combinations would successfully deliver packets for each CSI record in the trace.

The physical-layer feedback values are computed by the packet simulator. While applying error correction decoding to the received bitstream, the simulator also computes

---

[2] Cycling through antenna modes to measure the full MIMO CSI could be avoided using the 802.11n extension spatial streams (Section 4.6.2), but the IWL5300 does not yet support this technique.

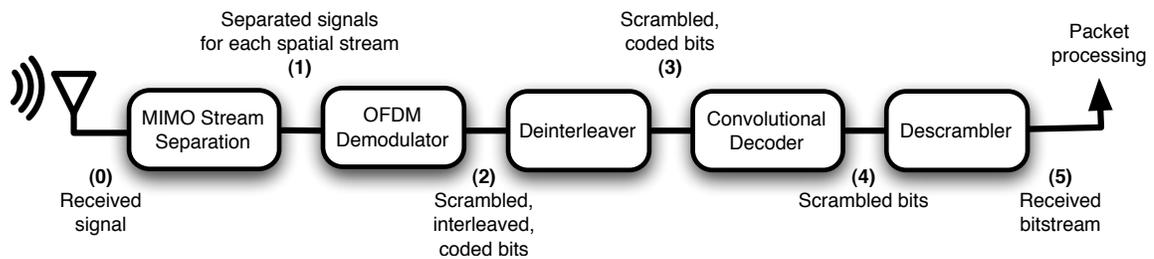

**Figure 7.1:** **The 802.11n MIMO-OFDM decoding process. The MIMO receiver separates the RF signal (0) for each spatial stream (1). Demodulation converts the separated signals into bits (2). Bits from the multiple streams are deinterleaved and combined (3) followed by convolutional decoding (4) to correct errors. Finally, scrambling that randomizes bit patterns is removed and the packet is processed (5).**

SoftRate's soft-output BER estimate for each CSI record and for each SISO rate. The simulator also computes the Effective SNR values for each CSI, but to reflect real-world NIC limits, Effective SNR is not given the measured CSI. Instead, I corrupt the CSI at the level of ADC quantization. This typically induces an error of $\pm 1.5$ dB in the output Effective SNRs.

**802.11n MAC Simulator.** A second component of the simulator implements the 802.11n MAC. This includes randomized backoff and link-layer packet aggregation. Initially, the simulator is reset to time 0 with default parameters for the MAC protocols, and then a rate selection algorithm will choose a rate for the next transmission. The simulator calculates whether that transmission is received, updates its internal MAC state, and then returns information about packet delivery to the rate control algorithm, as well as the appropriate physical-layer feedback. After every transmission, the simulator executes 802.11n's randomized backoff process and computes the new time for the start of the next transmission.

In 802.11n, transmitters send aggregated packet batches up to 65,000 bytes long. Batches are shortened for slower rates, when the transmission time is instead limited by the 802.11n standard's 4 ms restriction on transmission duration. Given an MCS, the simulator computes how long the transmission will last, then whether the batch is received successfully. As a single packet batch transmission can last up to 4 ms, this may overlap multiple channel probes in the trace. As senders use a fixed MCS for the entire transmission, the MAC simulator requires that $\geqslant 80\%$ of simulated packets for those channel records be received correctly in order for the batch to be received (this allows for the effects of coding).

SoftRate operates using the 80th percentile soft estimate from the range of packets. The Effective SNR feedback for a particular packet batch is given from the CSI of the first measurement overlapping that transmission. This models a varying channel that samples for CSI periodically, as happens when CSI is measured during the packet preamble.

**Rate Selection Algorithms.** The SampleRate, SoftRate, and Effective SNR algorithms are implemented as described previously, and get their inputs and outputs from the packet and MAC simulator components. Each algorithm chooses a rate based on its internal state, and the simulator returns whether the rate succeeds as well as any physical layer feedback, namely SoftRate's BER estimates and the Effective SNR values, measured during the transmission.

**Metrics.** The goal is to evaluate the ability of these algorithms to respond to changing channel conditions. Thus, the primary metric is the delivered MAC layer rate per time, modeling a UDP application. Higher-layer factors, such as TCP reactions to loss, will affect how this rate translates to throughput. These effects are ignored in the results presented here.

The results reported are the median of five trials where the simulator is initialized with different random seeds. To vary mobility, I scale the trace at different speeds; for example, $4\times$ mobility means that the records are assumed to arrive $4\times$ faster than they were measured.

### 7.2 SISO Rate Adaptation Results

I first examine the performance of Effective SNR for selecting SISO rates. The goal of this experiment is to establish a reasonable baseline, showing that Effective SNR performs as well or better than existing well-studied SISO rate adaptation algorithms, while using a simpler algorithm. If so, this will provide initial validation that the accurate packet delivery predictions provided by my Effective SNR model are useful in practice.

#### 7.2.1 *Effective SNR vs. Optimal*

I begin by comparing Effective SNR performance against the Optimal algorithm. Figure 7.2 shows the rate over time for Effective SNR and Optimal over the SISO trace. The rate is averaged over a window of 200 ms to smooth the data for readability. Effective SNR performs excellently. It is below Optimal but consistently tracks the Delayed Optimal algorithm, a realistic upper bound. Overall, Effective SNR delivers 90% of packets, with about 10% over-selection of rates.

Note that Packet SNR was observed to fare quite poorly in mobile channels [120]. Since Effective SNR reflects link fading, its estimates are more accurate (Chapter 6) and are stable ($2\times$–$3\times$ less variance).

#### 7.2.2 *Effective SNR vs. 802.11a/g algorithms*

Next, I compare Effective SNR with SampleRate and SoftRate, in order to see how it performs against current state-of-the-art SISO rate adaptation algorithms.

Figure 7.3 shows the delivered rate of each algorithm versus time. While it is hard to separate the lines on the graph, at $1\times$ speed, Effective SNR slightly outperforms SoftRate, which slightly outperforms SampleRate. These results were surprising, because the SoftRate work [120] indicated a large gap between SoftRate and SampleRate. In deeper analysis, I discovered that reducing the rate for retries is an important factor that gives SampleRate short-term adaptability. Without this rate fallback (the "SampleRate without fallback" line), SampleRate loses 25%–50% of its performance in mobile channels (Figure 7.4). This variant without fallback is the SampleRate algorithm that was the basis for earlier comparisons.[3]

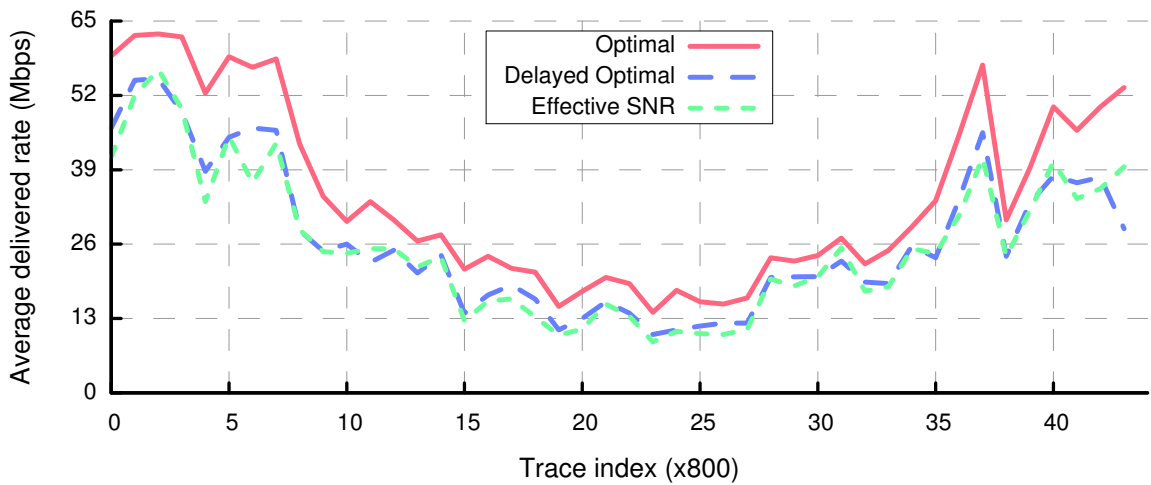

**Figure 7.2: Effective SNR SISO performance versus Optimal in human-speed mobility.**

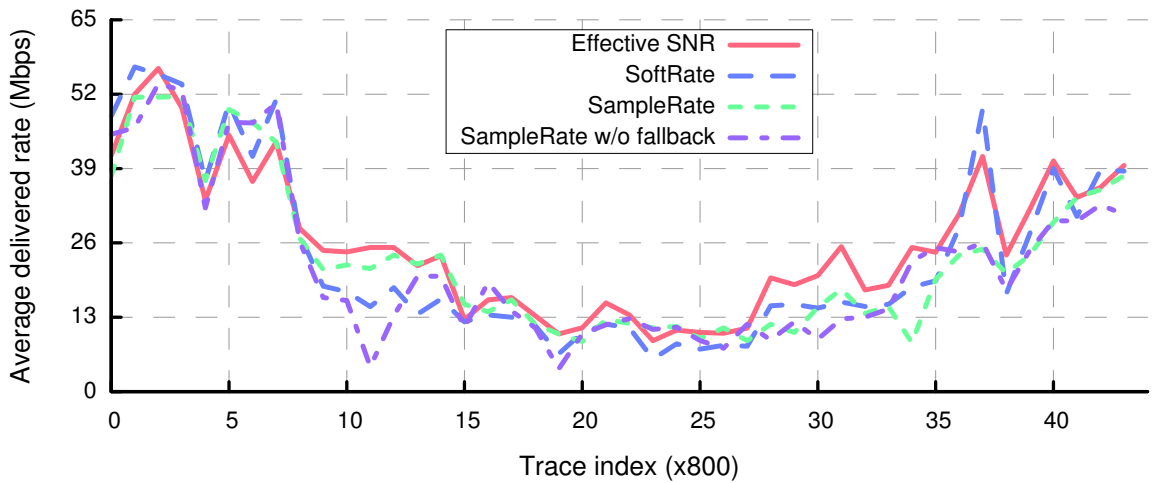

**Figure 7.3: Effective SNR, SoftRate, and SampleRate SISO performance in human-speed mobility.**

Figure 7.4 shows the effects of mobility on SISO channels. Each line plots the total amount of data delivered during the trace as a function of the speed at which the trace is played. The speeds range from 1× to 16×, corresponding to movement speeds of between walking speeds ≈3 mph when using the normal simulation, to about 50 mph for the fastest playback. The y-axis value is normalized by the Optimal algorithm's performance to better illustrate how the relative performance changes at different speeds.

---

[3] M. Vutukuru, personal communication, and code inspection.

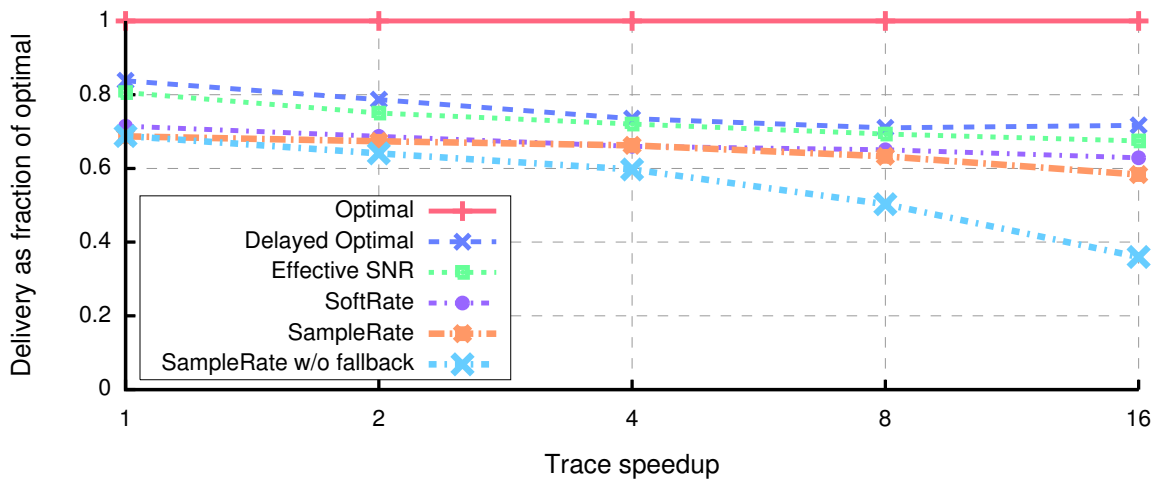

**Figure 7.4:** **OPT, Effective SNR, SampleRate, and SoftRate SISO performance in fast mobile channels.**

This plot shows that all schemes fall off with increased speed; the gap between Optimal and the Delayed Optimal algorithm increases from about 20% to about 30% at the fastest speeds. However, even in these mobile channels, Effective SNR holds up quite well and tracks the Delayed Optimal algorithm within 5%.

SoftRate performs slightly worse at the normal human speeds, but maintains a nearly constant performance, about 70% of Optimal, as the trace speed increases. SampleRate degrades the fastest with increasing mobility, and the version that does not reduce rate on retries finally less than 40% of the Optimal performance, while the other algorithms maintain better than 60% of Optimal performance at all speeds.

Finally, I note that while the performance differences between the schemes can be significant, other evaluations have reported larger differences. Note that they studied throughput based on TCP traffic, which will magnify performance gaps by reacting to packet loss. The UDP-like results I generated using my simulator capture the underlying accuracy of the individual schemes instead.

### 7.3 MIMO Rate Adaptation

Now I extend the evaluation to MIMO channels. This will show the generality of my model, which can flexibly support multiple spatial streams. I compare Effective SNR to an 802.11n-enabled version of SampleRate, to understand whether the larger search space will increase

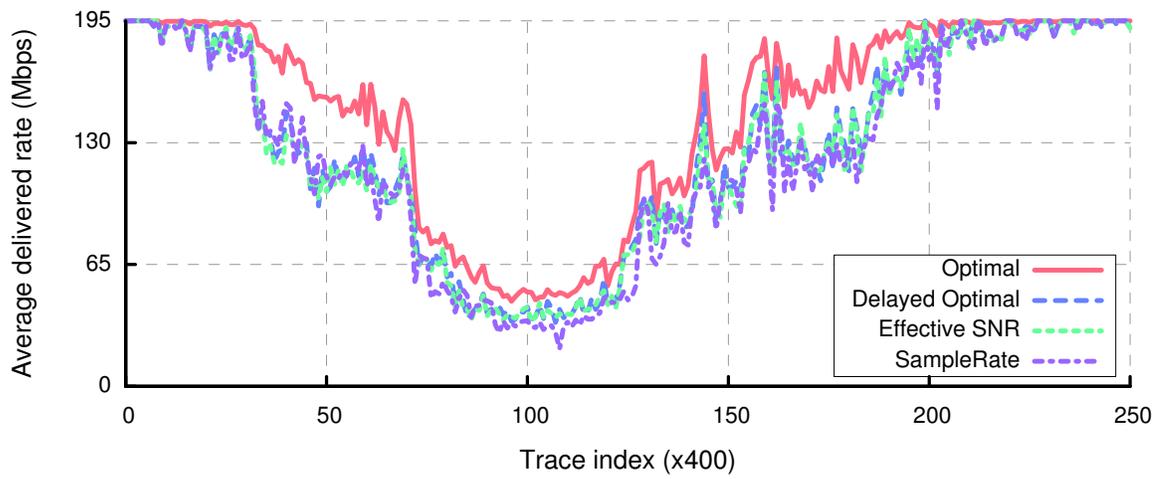

**Figure 7.5: Optimal, Effective SNR, and SampleRate MIMO performance in human-speed mobility.**

the performance gap between the two schemes.

Figure 7.5 and Figure 7.6 show the performance of an unmodified Effective SNR algorithm selecting among 802.11n MIMO rates, as well as (modified) SampleRate and the two Optimal algorithms. This graph does not include a line for SoftRate, as it is not defined for multiple streams. These figures are in the same form as for SISO, except the range of rates has grown by a factor of 3 to support up to 195 Mbps. The MIMO trace is longer and has more packets-per-second, and thus includes enough data to speed up the trace by a factor of up to $256\times$.

Overall, the Effective SNR trends in these graphs are similar to those in the SISO graphs. At human mobility speeds, Effective SNR tracks the Delayed Optimal algorithm and delivers excellent performance, with 75% accuracy and 10% over-selection. In faster mobile channels, Effective SNR tracks the Delayed Optimal algorithm until the speeds increase to about $128\times$, after which there is a slightly larger gap with for MIMO than for SISO. This arises likely because Effective SNR must now choose between 24 rates instead of 8. It is slightly more likely to choose rates lower than the highest rate that would have worked.

These graphs also show that—as with SISO links—SampleRate performs well in human speed mobility, only slightly worse than Effective SNR. But for SampleRate, this trend *does not hold*, and as the speed of the trace gets faster the performance degrades rapidly to around one-third of Optimal, and less than half of Effective SNR. The difference between SampleRate and Effective SNR highlights that Effective SNR is able to handle the large MIMO rate space even in rapidly-changing channels, while the probe-based SampleRate algorithm cannot.

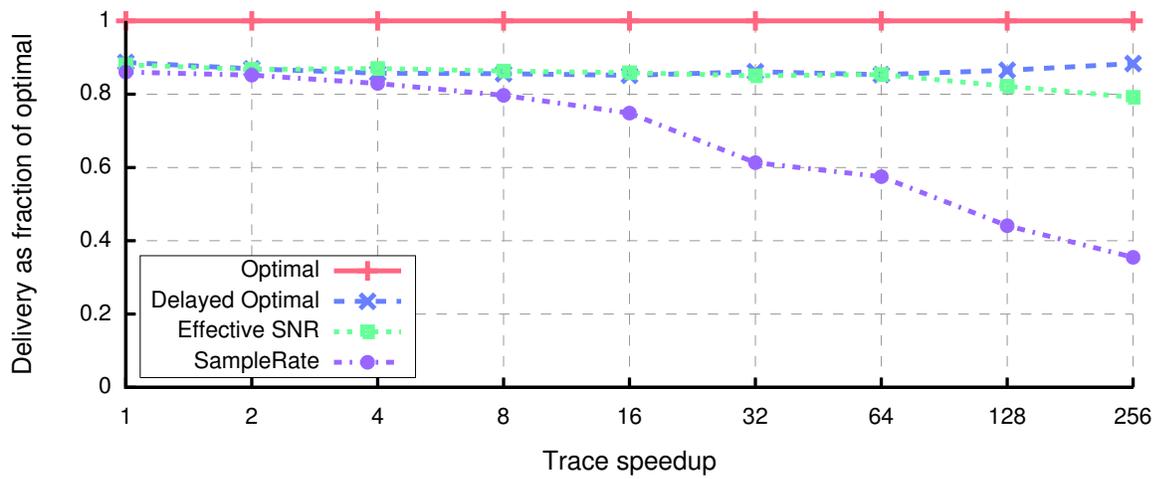

**Figure 7.6:** **Optimal, Effective SNR, and SampleRate MIMO performance in faster mobile channels.**

### 7.4  Enhancements: Transmit Antenna Selection

I conclude this chapter with an example that highlights the strength of my Effective SNR in that it accommodates choices other than just rate. In particular, I amended the implemented Effective SNR algorithm to select the best transmit antenna set.

Transmit antenna selection can be useful in practice, for instance in an 802.11n AP that selects which antennas to use to send packets to a legacy 802.11a/g client (plus uses all antennas to receive packets). With three antennas to choose from, the expected theoretical gain in SNR is a little over 2.5 dB [30]. For a SISO link, this gain is likely enough to advance to a higher rate.

I ran my antenna selection-enabled Effective SNR algorithm on a 3x1 MISO channel, extracted from the 3x3 MIMO trace. These measurements correspond to those that would be measured if the transmitter exploited the "extension spatial streams" CSI probe to measure the 3x1 MISO channel as I described in Section 4.6.2.

Using the MISO CSI feedback, the Effective SNR algorithm chose the antenna with the highest Effective SNR for the next transmission. This gave a gain in the total packet delivery of 5%. For comparison, the Optimal antenna selection algorithm achieved a 10% increase by always knowing which antenna was best.

While transmit antenna selection presents a relatively small gain for this trace, it comes at no cost and does not complicate the Effective SNR algorithm. In contrast, no other rate adaptation schemes as directly support these enhancements. They would instead require customized, multi-dimensional probing algorithms and coarse adaptation of antennas to implement antenna selection.

Antenna selection is one of many ways that my Effective SNR model can be applied beyond simple rate selection. In the next chapter, I present a study of four different network configuration applications.

### 7.5 Summary

In this chapter, I presented a detailed evaluation of Effective SNR in the context of configuring the rate for 802.11a/g and 802.11n links in a mobile setting. The results for the single-antenna 802.11a/g systems showed that Effective SNR performs as well as or better than existing state-of-the-art algorithms for rate adaptation, with a much simpler rate selection algorithm.

A key result of this evaluation is that this good performance extends to 802.11n, while the probe-based SampleRate algorithm suffered in the larger search space. This validates my fundamental claim that probe-based algorithms will suffer in large search spaces with dynamic channels, and it motivates the need for and benefits of my Effective SNR approach.

Finally, I also showed that Effective SNR easily supports additional enhancements such as antenna selection. In the next chapter, I will flesh out Effective SNR applications by applying my model to a set of other configuration problems along different dimensions of the configuration space.

Chapter 8

# FURTHER APPLICATIONS OF EFFECTIVE SNR

The detailed evaluation presented in the previous chapter showed that my Effective SNR model provides good performance when selecting rates for 802.11a/g and 802.11n links, one important application in wireless link configuration. In this chapter, I present an evaluation of the Effective SNR for four wireless configuration problems to further illustrate the broad applicability of my work.

I chose four problems that represent the key applications needed to configure a future wireless network. The first three are selection algorithms for access points, operating channels, and multi-hop paths through networks. These represent the main decisions (beyond rate) that need to be made in flexible topologies such as Wi-Fi Direct networks. The last application is to classify whether a device is mobile, in order to switch between algorithms designed for static vs. mobile links, or to trigger a search for a new, better operating point when the device begins to move. Together these are the building blocks of an automatically-reconfiguring network that provides good performance.

## 8.1   Experimental Methodology

To understand how well different techniques perform at the applications studied in this chapter, I took comprehensive measurements of both packet delivery and the RF channel in the University of Washington testbed. These data provide the ground truth performance of links in different configurations and also a record of the channel state during the experiments. Using this data, I perform offline simulations of the different algorithms.

### 8.1.1   Dataset

I measured the links between 24 static devices in my testbed at the University of Washington. The measurements in this chapter were taken 3 years after the measurements used in Chapter 6 and Chapter 7, and they include a different set of devices. In addition to evaluating the applications in question, the results in this chapter will also tell us whether these NICs in practice experience physical degradation that invalidates their in-factory calibration.

I took measurements on all 35 channels, each 20 MHz wide, that the IWL5300 devices support. Of these 35 channels, 11 overlap in the 83 MHz-wide unlicensed 2.4 GHz band, and the remaining 24 non-overlapping channels are spread across three non-contiguous bands between 5.170 GHz and 5.835 GHz. In an experiment, one sender transmits packets with random payloads to 23 receivers. The transmitter sends a total of 2400 packets by interleaving 100 packets from each of the 24 MCS that correspond to 1-, 2-, or 3-stream 802.11n rates. Each receiver uses 3 antennas for spatial diversity and/or spatial multiplexing. I cycled through all transmitters and all channels over the course of a few hours, and I took data at night to attempt to minimize the impact of interference.

The above measurements included 100 packets for each link, MCS, and channel. From these data, it is straightforward to compute the ground-truth packet reception rate (PRR) for each of these configurations. In this chapter, I decouple the ability to make accurate choices at the application-level from the rate selection algorithm needed to realize it. I thus assume an ideal rate selection algorithm that always correctly chooses the fastest MCS. This provides an Optimal baseline against which I compare Packet SNR- and Effective SNR-based application configuration algorithms.

### 8.1.2 Simulating RF Measurement-based Algorithms

While measuring the performance of the link, I sent exhaustive packet probes at all rates, and collected the RSSI and CSI for each correctly received packet. Of course, this collected information would not be available to an algorithm attempting to quickly configure the network. Instead, the algorithms are given only the RF measurements from the first packet received. I use this to compute Packet SNR ($\rho$) and Effective SNR ($\rho_{\text{eff}}$). By using only the first measurement, I emulate the performance that a configuration algorithm would obtain in practice, where it would use only a single probe.

In the next three sections, I evaluate how well my Effective SNR model can make access point, channel, and path selections on this dataset. I use a different mobile dataset for the last section, and describe it therein.

## 8.2  Access Point Selection

The first protocol operation a wireless client device performs is to join an existing network. This operation typically consists of scanning for a known access point (AP) by sending broadcast packets called *probe requests* on different frequencies until a recognized AP is found. In a typical home today, there will likely be one valid *probe response*: The single home AP. However, in a dense Wi-Fi network, such as an enterprise Wireless Distribution System or a Wi-Fi Direct peer-to-peer network, the client needs to choose from among many available responding devices that connect it to the same network. This is the *access point selection* problem.

The general access point selection algorithm examines the probe responses to predict the link performance for each access point, and then it chooses the fastest one (Algorithm 8.1). The standard approach used today is for the client to measure the Packet SNR from each

---

**Algorithm 8.1** AccessPointSelection(AP Set A, Sender s)

---

1: **return** $\arg\max_{a \in A}$ GetMetric$(a, s)$   // *choose the AP with the best downlink metric*

---

---

**Algorithm 8.2** GetMetric-PacketSNR(Sender s, Receiver r)

---

1: Measure the Packet SNR $\rho$ at r from s
2: **return** $\rho$

---

---

**Algorithm 8.3** GetMetric-EffectiveSNR(Sender s, Receiver r)

---

1: Measure the CSI at r from s   // *a full CSI including all TX and RX antennas*
2: Compute the Effective SNR estimates $\rho_{\text{eff},m}$ for each MCS m
3: Determine whether each MCS works by comparing $\rho_{\text{eff},m} \geqslant \tau_m$
4: **return** the bitrate $B(m)$ for the fastest working MCS m

---

probe response and connect to the strongest AP (Algorithm 8.2). Here, SNR measurements are used as a proxy for an estimate of the downlink throughput, and the AP with the highest SNR is considered the best. (Other factors than link quality, such as interference and contention with other clients, can affect throughput as well; some systems can take these factors into account, but they are not relevant here. I discuss these and other related issues in Chapter 9.) In this section, I evaluate whether using Effective SNR to predict the downlink rate can improve this decision process (Algorithm 8.3).

### 8.2.1 *Characterization of Access Points*

I begin by characterizing whether access point selection matters in my testbed. How much does a good or bad choice impact performance?

To generate data for this evaluation, I first filtered the data set to the 11 channels in the 5 GHz band for which there is no overlapping UW Wi-Fi network. Then I considered the access point selection problem by considering each client and channel in turn. In particular, for node c playing the role of a client, I defined A to be the set of access points that responded to a probe from that client on a particular channel. By considering each channel independently each client generates 11 data points, for a total of $11 * 24 = 264$ simulated client association attempts. To ensure that AP choice can matter, I excluded 17 clients that had fewer than 3 responding APs on a particular channel, leaving 247 total.

Figure 8.1 and Figure 8.2 show the results, framed as the difference in throughput (relative or absolute) from the best choice AP for the worst, median, and average choices. These graphs show that for this testbed, choosing the wrong AP can hurt: The worst responding AP offers less than half of the best AP's throughput in 95% of cases, and in most cases the random (average) and median APs are better but not close to optimal. In absolute terms, the downlink from the median AP is 50 Mbps to 100 Mbps worse than from the best AP in most cases. A poor choice of access point can dramatically hurt performance in practice.

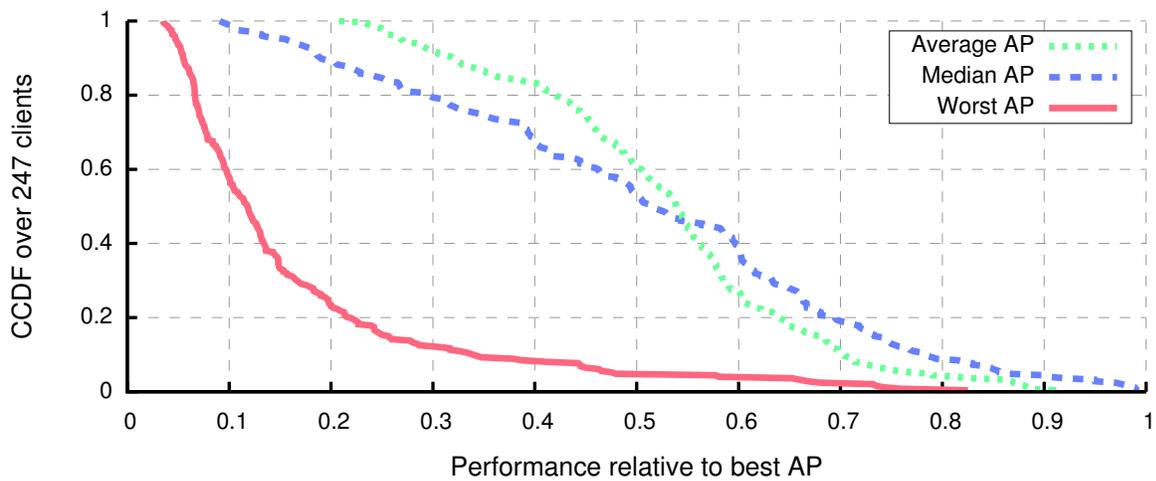

**Figure 8.1:** For each client, the relative difference in throughput over access points compared to the best choice.

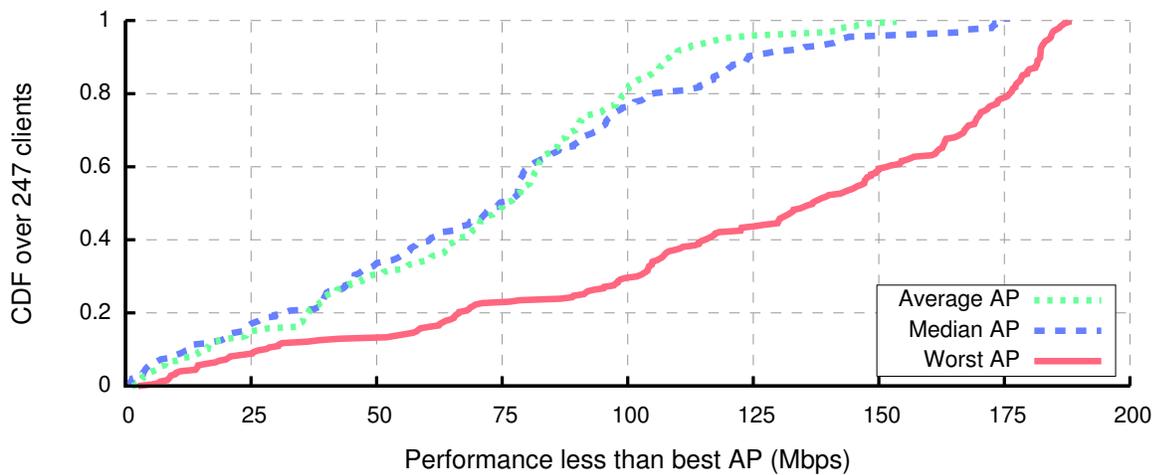

**Figure 8.2:** For each client, the absolute difference in throughput over access points compared to the best choice.

### 8.2.2 *Access Point Selection Performance*

As presented above, Algorithm 8.1 shows the framework for selecting access points typically used today, with Packet SNR (Algorithm 8.2) usually used as the metric of comparison between access points. I calculate the Effective SNR link metric as in Algorithm 8.3—a simple instantiation of the procedure described in Chapter 4.

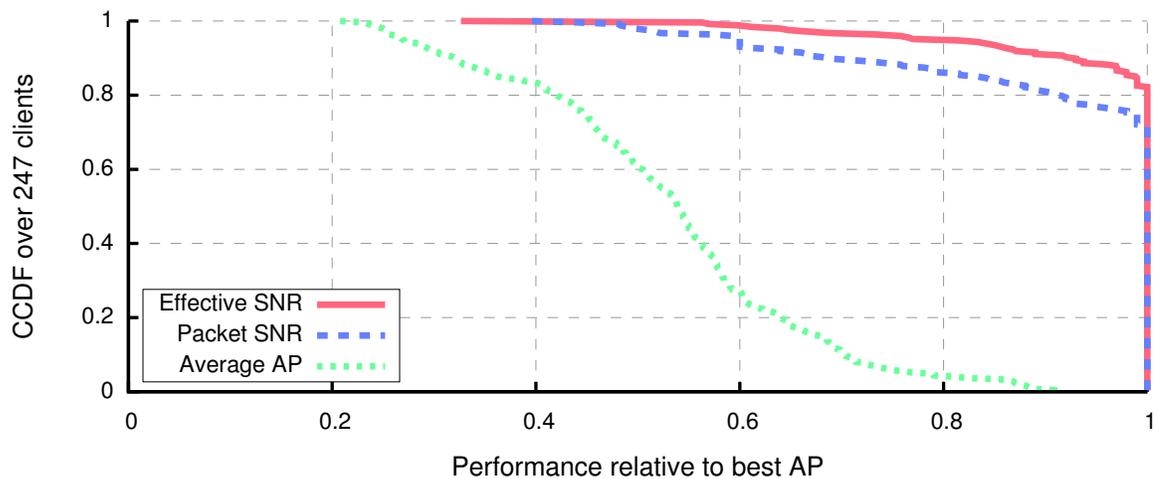

**Figure 8.3: CCDF of relative performance of AP selection using Packet SNR and Effective SNR, compared to Optimal.**

Figure 8.3 shows the performance of Packet SNR and Effective SNR-based access point selection algorithms relative to the optimal algorithm. I plot the complementary CDF of the results, so that each $(x, y)$ point in the graph shows the fraction of clients $y$ that achieved performance at least within the fraction $x$ of the optimal algorithm. The line for a perfectly accurate algorithm would be located in the upper right corner of the graph: All clients would achieve the best possible throughput. This graph shows that both algorithms perform well, choosing a near optimal access point in almost all cases. I attribute this overall good performance to the fact that the potential access points are spread across a large testbed, exhibiting a wide range of SNRs. This large geographic spread means that there may be one or a few nearby access points that offer a clear best choice, and Packet SNR—which is correlated with distance—can correctly identify good choices.

Although both algorithms perform well, Effective SNR offers improved performance. The Effective SNR algorithm finds the best access point for 83% (204) of clients, versus 72% (177) for Packet SNR. Considering only the suboptimal choices, those made by Effective SNR are better: 3/4 (31/43) of incorrect choices are within 80% of optimal, versus only half (35/70) when using Packet SNR.

Next, Figure 8.4 shows the absolute performance loss of the two algorithms in choosing the access point. Here, each $(x, y)$ point represents the fraction of links $y$ that choose an access point within $x$ Mbps of the best access point. The area over the curves represents the performance lost by each algorithm: An accurate algorithm would be in the top left corner, losing little throughput for only a few links.

Examining this graph, we can see that these benefits translate to raw bitrate as well. Ac-cess points selected by Effective SNR links are within 10 Mbps of optimal for 90% (223/247) of cases, but only 78% of selections meet this criterion when using Packet SNR. For Packet SNR, the 90th percentile performance loss is 42 Mbps, versus 9 Mbps with Effective SNR. These results show that the Effective SNR-based access point selection algorithm works well and makes better choices than one based on Packet SNR. Using the area over the curves as a measure of the inefficiency, this area is 2.7× larger when using Packet SNR.

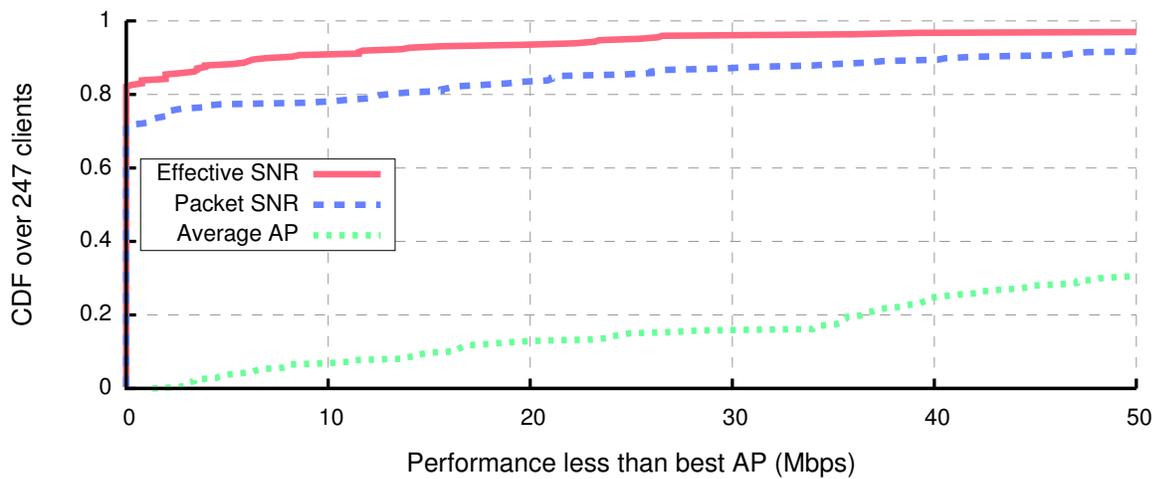

**Figure 8.4: CCDF of the throughput lost using APs selected by Packet SNR and Effective SNR, compared to Optimal.**

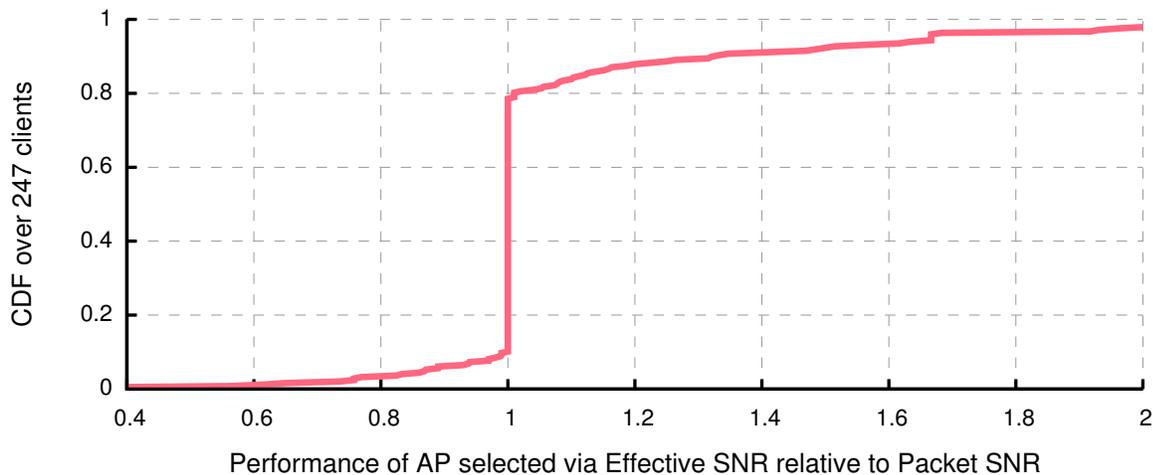

**Figure 8.5: The relative throughput selecting APs by Packet SNR or by Effective SNR.**

That Effective SNR is statistically better than Packet SNR over the testbed does not mean it works better in all cases. I examine the head-to-head performance of selecting access points via Effective SNR or Packet SNR in Figure 8.5. For each link, I plot the ratio of the performance of the access point picked with the Effective SNR strategy to that of the access point chosen by maximizing Packet SNR. If the two APs perform equally, the ratio will be 1; if the Effective SNR strategy chose a better AP the ratio will be larger. The algorithms chose APs with equal performance for 69% (171/247) of the clients, Effective SNR makes a better choice in 21% (53) of cases, and Packet SNR is better for the remaining 10% (23). Though Effective SNR does not always lead to a better choice, it does so twice as often as Packet SNR. The graph also shows that when Packet SNR does make a better choice, Effective SNR comes close—within 2/3 for 5 in 6 of these cases—while it often improves on the choice from Packet SNR by a larger margin.

**Effective SNR occasionally worse.** It may be surprising that Effective SNR, with its ability to better understand subchannel effects, can ever result in a worse choice of access point than using the Packet SNR computed from RSSI. I present one likely explanation.

Recall that these algorithms compute channel metrics using a single RSSI or CSI measurement. Generally, a single CSI probe is not as accurate as multiple measurements, because of error in the estimates of individual subcarriers. On the other hand, a single RSSI probe measures only the total power across carriers and is consistent across packets.

**Summary.** When selecting access points, both Packet SNR and Effective SNR make good choices; each algorithm selects the fastest access point in most cases. However, the ability of Effective SNR to capture channel effects leads to better choices more often and generally closer-to-optimal performance when it makes a choice incorrectly.

### 8.3   Channel Selection

Using the new Wi-Fi Direct standard, 802.11 devices that wish to send data directly (instead of through the access point as in 802.11 infrastructure mode) can create a peer-to-peer link. Depending on the amount of interference in the network (e.g., from other clients of the access point) and the quality of the link between the two devices, they may wish to move the link to a different operating channel in order to improve performance. This is one example of the *channel selection* problem: To quickly choose the best operating frequency for a pair of nodes to communicate. In this section, I define the "best" channel to be the channel that provides the highest throughput in the absence of interferers.

The channel selection problem is similar to the access point selection problem, and has a similar algorithmic solution (Algorithm 8.4). It can use the same GETMETRIC functions for Packet SNR (Algorithm 8.2) and Effective SNR (Algorithm 8.3). The primary difference is a reordering of the parameters: Rather than a fixed receiver with fixed channel choosing between multiple senders, a fixed sender and receiver must choose between multiple channels.

---

**Algorithm 8.4** CHANNELSELECTION(CHANNEL SET C, SENDER s, RECEIVER r)

---
1: **for all** c ∈ C **do**
2:    Both s and r switch to channel c
3:    Compute the channel metric $m_c$ using GETMETRIC(s, r)
4: **end for**
5: **return** $\arg\max_{c \in C} m_c$                    *// choose the channel with the best metric*

---

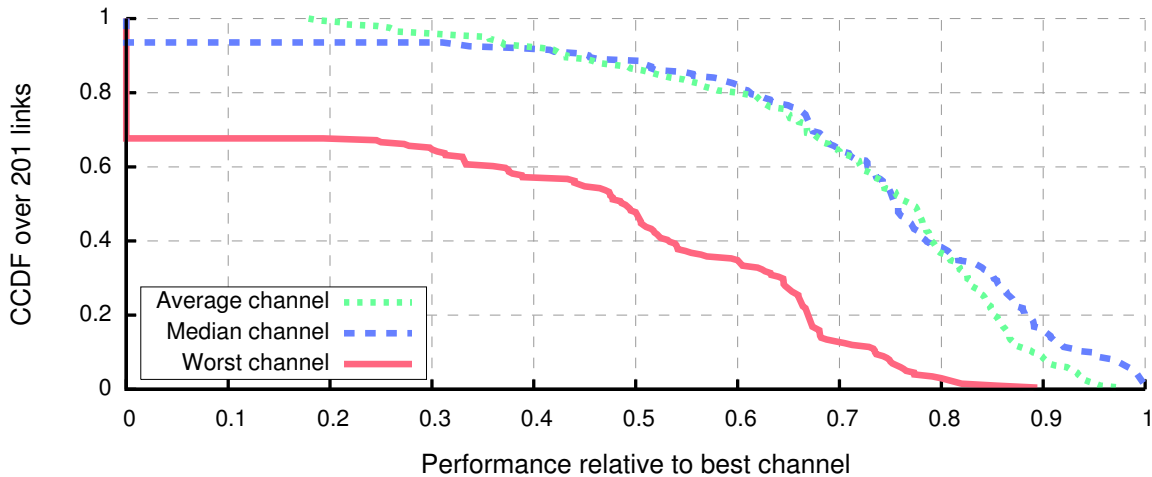

**Figure 8.6: The relative difference in throughput over 802.11n channels.**

### 8.3.1 Characterization of 802.11 Channels

To start my investigation of channel selection algorithms, I first measured how the operating frequency affects 802.11n links in practice.

As in Section 8.2, I filtered the data to the 11 channels in the 5 GHz band for which there are no co-channel university APs. (Note that channel selection generally makes sense *within* one frequency band. Selection across bands is trivial: Because of better antenna gain and Friis' Law effects, a 2.4 GHz channel has typically 10 dB–15 dB stronger SNR than a 5 GHz channel for the same nodes.) I further eliminated from consideration any pairs of devices that did not obtain at least 6.5 Mbps throughput on at least 3 of the 11 channels. This left 201 unidirectional links, approximately a third of the $24 * 23 = 552$ links in the testbed.

*Results*

Figure 8.6 and Figure 8.7 show how the throughput of the worst, median, and average channels compares to the best channel for these links. Note that because the channel set is fixed and independent of connectivity, the worst channel might deliver no throughput at all—unlike AP selection, in which the client was choosing from only those nodes that responded to its probe. About a third of the links had at least one such channel.

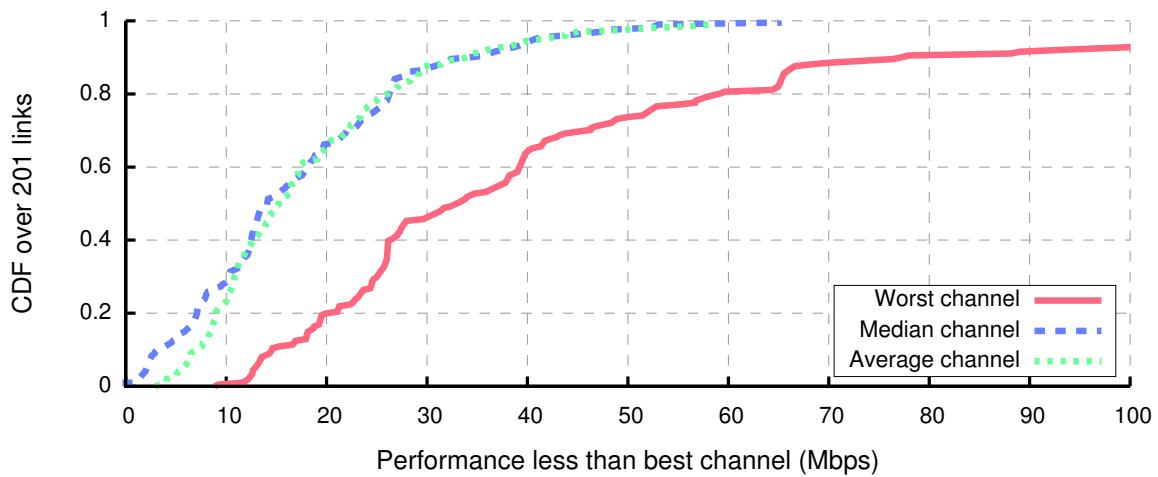

**Figure 8.7: The absolute difference in throughput over 802.11n channels.**

These figures demonstrate that the choice of channel can dramatically impact performance. Figure 8.6 shows that the worst channel offers less than half of the best throughput for more than half the links. In absolute terms, this difference can be quite large: The worst channel is a median 33 Mbps worse than the best channel, and for a few links (about 7%) the difference is more than 100 Mbps. For these links, some channel will deliver no packets at all, while another delivers packets at nearly the maximum bitrate. An unlucky choice of channel can cripple performance and result in little to no throughput.

The median and average channels perform about equivalently in our testbed, but they are significantly worse than optimal. These channels yield less than half the optimal throughput for 15% of the links and for only one third of links do these come within 80% of the best throughput. These figures show that for very few links do most channels perform optimally. The median or average channel is 15 Mbps worse than the best channel for most links, and the gap is larger than 25 Mbps for 20% of links.

### 8.3.2 Channel Selection Accuracy

I evaluate Packet SNR and Effective SNR-based channel selection algorithms based on Algorithm 8.4 and using the link metric functions described in Algorithm 8.2 (Packet SNR) and Algorithm 8.3 (Effective SNR).

Figure 8.8 shows the performance relative to the optimal algorithm when using Effective SNR or Packet SNR to choose between channels, using the same format as I used for access point selection in Figure 8.3. Again, we see that both algorithms perform well, though Effective SNR is a better predictor of application performance. Effective SNR chooses an optimal channel for 121 links (60%), whereas Packet SNR is optimal for only 71 links (35%). The Effective SNR-based algorithm is within 90% of optimal for 168 links (84%), 80% for 181 links (90%), and 70% for 191 links (95%). In contrast, maximizing the Packet SNR only gets within 90% of optimal for 112 links (56%).

Next, I compare the absolute performance loss of the channel selection algorithms in Figure 8.9. Again, the area over the curves represents the performance lost by each algorithm. This area is 3.3× larger for the Packet SNR-based selection algorithm, showing that Effective SNR is significantly more accurate. This difference translates to about 9 Mbps faster links when selecting channels via the Effective SNR.

Finally, Figure 8.10 shows the ratio of the performance of the channels selected by Effective SNR and Packet SNR. Recall that a ratio larger than 1 means that Effective SNR chose a faster operating channel. The algorithms choose channels with equal performance for 82 (41%) of the 201 links, while the Effective SNR-based algorithm chooses a better channel for 94 (47%) links and the Packet SNR-based algorithm for the remaining 25 (12%). Additionally, the gains from Effective SNR are much larger than its losses: The Packet SNR strategy chooses a channel that performs 20% better than the Effective SNR-selected channel for only 4/25 (16%) links when its channel is better, while Effective SNR chooses a 20% better channel for 50/94 (53%) of cases. In other words, the Effective SNR channel selection algorithm is more likely to pick a better channel by a factor of about 4 (94/25), and the difference is more likely to be significant by a factor of about 3 (53%/16%).

*Summary*

These results shows that both Effective SNR- and Packet SNR-based channel selection strategies perform well in my testbed. However, the Effective SNR channel selection strategy is significantly more accurate: It chooses an optimal channel for 70% more links, it offers about 9 Mbps more bandwidth per link when selecting suboptimal channels, and it is more likely to choose a better channel than a worse channel by a factor of 4.

Note that the benefits of Effective SNR relative to Packet SNR are larger for channel selection than they were for AP selection. I attribute this to the fact that Packet SNR is very similar for fixed links across channels, while in contrast Packet SNR varies widely across geographically distributed APs. With fewer outliers, Packet SNR measures small SNR differences across similar channels. In contrast, Effective SNR can accurately assess the impact of subchannel fading effects, leading to improved performance.

### 8.4 Path Selection

While today's access point and wireless distribution system (WDS) infrastructure networks use tree-structured topologies and have only a single path between any two nodes, a future device-to-device wireless network such as Wi-Fi Direct may offer many paths along which packets can be routed. Choosing which path in a multi-hop wireless network will provide the best throughput is the *path selection* problem, and it can be thought of as a generalization of the access point selection problem I described in Section 8.2. Indeed, a client implicitly makes a routing decision when joining a WDS network—which access point it chooses can make a large difference in its connection quality to the root access point that serves Internet access. Research in multi-hop routing for wireless mesh networks [10, 96] has shown that the choice of path can effect a large difference in connection quality.

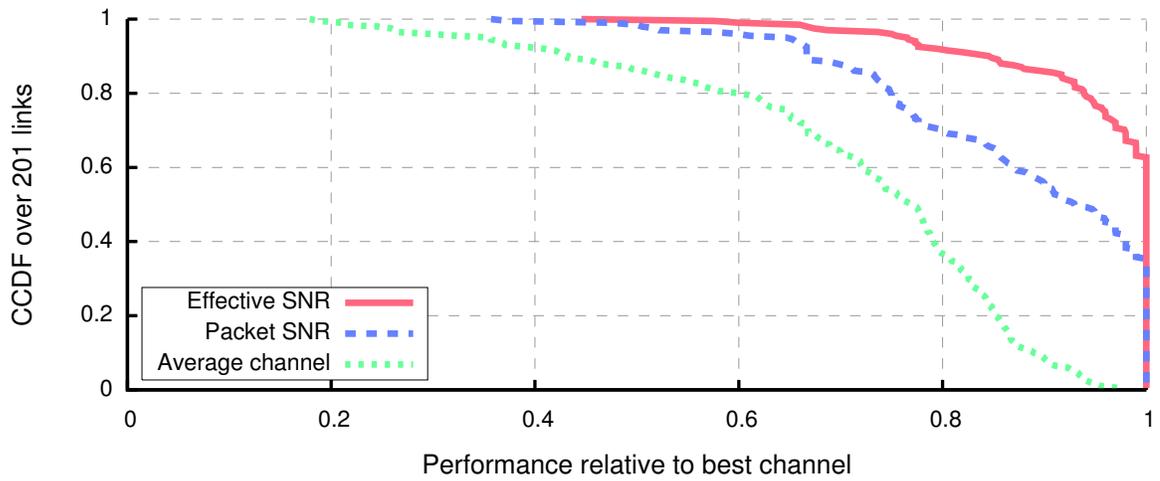

**Figure 8.8: Channel selection algorithm performance relative to an optimal algorithm.**

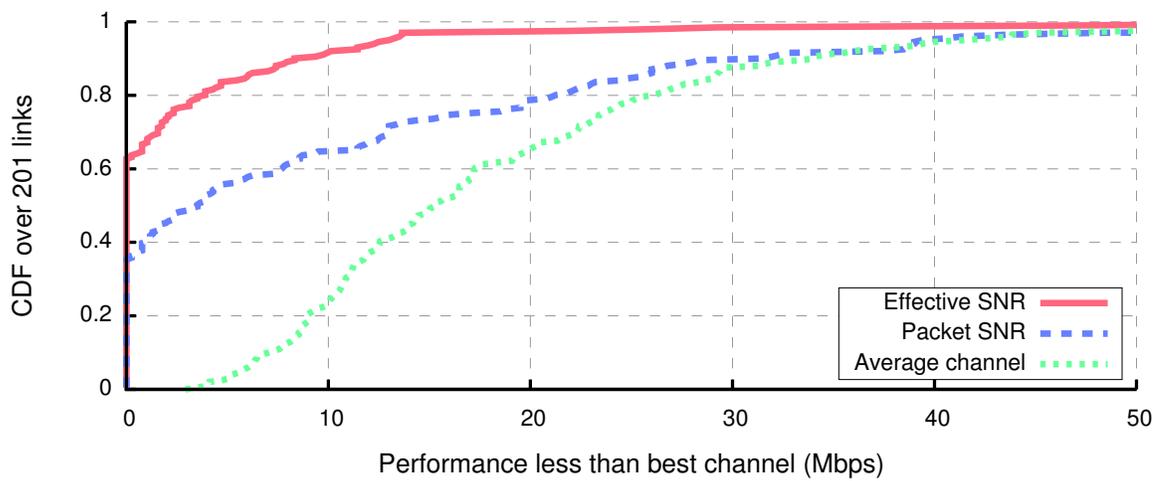

**Figure 8.9: Channel selection algorithm performance loss from optimal algorithm.**

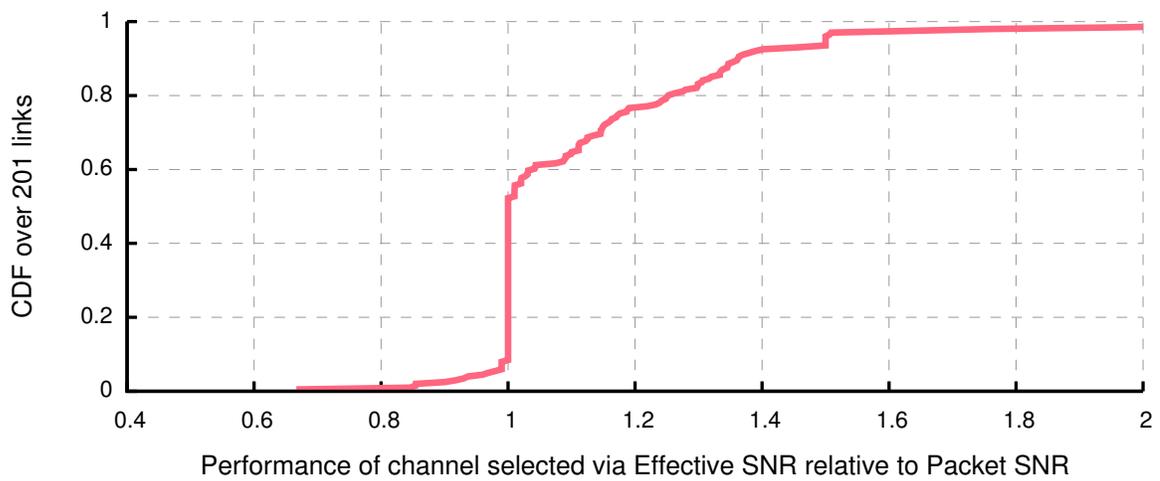

**Figure 8.10: Channel selection performance with Effective SNR relative to Packet SNR.**

The practical state of the art in this area is the recent work by Bahl et al. [10] on an opportunistic repeater scheme for 802.11a. In this design, when a client with a strong link detects *rate anomaly* [42]—that is, that its throughput is hurt by a client with a weak link monopolizing airtime—the strong client evaluates whether relaying that client's packets would improve throughout for both. In certain scenarios, they showed that this could improve aggregate performance of the network by 50%–200%.

While this solution is practical and works well, the use of MIMO and other hardware techniques in 802.11n significantly complicates the picture. First, the scheme of Bahl et al. uses a link's RSSI (as a measure of Packet SNR) to select between the 8 available 802.11a rates. In contrast, as I have shown in Chapter 6, Packet SNR does not accurately predict the rate for 802.11n links, nor does it enable devices to choose between different MIMO modes. Second, Bahl et al. used a homogeneous network of single-antenna 802.11a chipsets; but the set of devices in most 802.11n networks will be heterogeneous, with differing numbers of antennas and asymmetric transmit/receive capabilities. While it is not clear how to handle these challenges via Packet SNR, Effective SNR offers the ability to overcome them. In this section, I evaluate the ability of Effective SNR to deliver the benefits of opportunistic repeaters in 802.11n networks.

Note that the problem of path selection is similar to AP selection, except that when choosing between repeaters (or a direct link) the entire path must be considered rather than merely the last hop. For simplicity, I assume that the network diameter is small such that pipelining [96] is of limited benefit, and do not consider schemes that forward along multiple unreliable paths such as ExOR [15]. I next describe the basic path selection algorithm I evaluate and characterize the multi-hop paths in my testbed.

### 8.4.1 Path Selection Algorithm

I describe a simplified path selection algorithm in Algorithm 8.5. This *relay selection* algorithm only considers paths with a single intermediate node (called a *relay*). In step 4, this algorithm computes the *expected transmission time (ETT)* [26] of the multi-hop path as the sum of the time to transfer the packet along each hop. This metric makes the optimistic assumption that there is no protocol overhead, and hence provides an overestimate of actual performance. However, it lets us compare the ability of Packet SNR and Effective SNR

---

**Algorithm 8.5** RELAYSELECTION(RELAY SET R, SOURCE s, DESTINATION d)

---

1: $t_{\text{direct}} \leftarrow 1/\text{PREDICTBITRATE}(s, d)$        *// time to send a bit directly from s to d, or $\infty$*
2: **for all** $r \in R$ **do**
3:     *// time to hop through r is the sum of the times of the two hops*
4:     $t_{\text{relay},r} \leftarrow 1/\text{PREDICTBITRATE}(s, r) + 1/\text{PREDICTBITRATE}(r, d)$
5: **end for**
6: $r_{\text{opt}} \leftarrow \arg\min_{r \in R} t_{\text{relay},r}$        *// find the optimal relay $r_{opt}$ with the shortest path*
7: **if** $t_{\text{relay},r_{\text{opt}}} < t_{\text{direct}}$ **then**
8:     **return** $r_{\text{opt}}$        *// if optimal relay offers a shorter path*
9: **else**
10:     **return** $\emptyset$        *// if the direct link is best*
11: **end if**

---

algorithms to predict the performance of one-hop paths. To balance this optimism, I only consider paths that provide at least 20% throughput improvement over a direct link.

Relay selection is only a subset of the full path selection problem, but it is simple and recoups much of the potential gain. I considered all $24 * 23 * 11 = 6072$ source-destination-channel tuples in the UW testbed, where I consider each $5\,\text{GHz}$ channel as a different instantiation of the network. Of these 6072 node pairs, 2037 (34%) have a direct link. Adding in optimal one-hop relays enables a further 2317 (38%) of node pairs to connect, for a total of 4354 (72%) connected links. The remaining 1713 (28%) of node pairs require two relays to connect, and would probably be best helped by switching to the $2.4\,\text{GHz}$ band in order to obtain higher SNR and longer links.

How much can relaying help in this testbed? To evaluate this, I calculated the bitrate of the optimal one-hop relay choice. I also computed the bitrate of the optimal path using a modification of Algorithm 8.5 to handle multiple hops. (Note that, because it ignores overheads that increase roughly linearly in the number of hops, the path bitrate is even more optimistic than the relay bitrate.)

Figure 8.11 shows the net bitrate of the direct link, optimal relay, or best multi-hop path between the 6072 node pairs. This figure demonstrates that the use of a single relay captures most of the improvement for node pairs that can communicate faster than 20 Mbps. Though the relay strategy leaves a quarter of the node pairs disconnected, the optimal multi-hop paths would only enable those nodes to communicate at an optimistic estimate of 10–20 Mbps at best.

Figure 8.12 shows the achievable improvement in bitrate between the node pairs. Only 25% (1621/6072) of node pairs gain 20 Mbps or more using these strategies, and 60% (964) of these also gain 20 Mbps using a single relay.

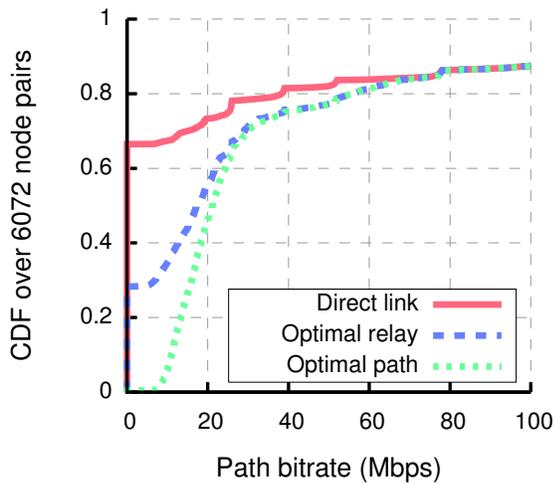

**Figure 8.11: Performance of the direct link and the optimal relay or multi-hop path.**

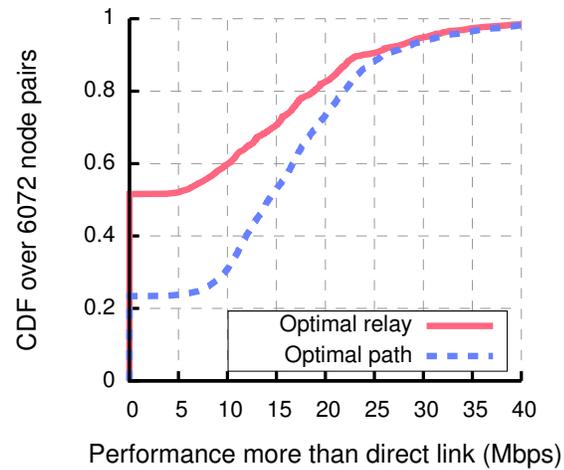

**Figure 8.12: Performance gain from relay or multi-hop path.**

These results show that relay selection recoups most of the gains of full path selection in this testbed. Thus, I restrict the evaluation to selecting a good relay. To complete the description of the algorithms, I now explain how to choose relays using Effective SNR and Packet SNR. To complete Algorithm 8.5, we need only define the function PREDICTBITRATE, which we use to predict the bitrate of the link studied.

*Relay Selection with Effective SNR*

To define the function PREDICTBITRATE for Effective SNR, we can simply use GETMETRIC-EFFECTIVESNR (Algorithm 8.3)—because the link metric of interest for Effective SNR is indeed simply the predicted bitrate of the link.

*Relay Selection with Packet SNR*

Unfortunately, defining a PREDICTBITRATE-EFFECTIVESNR function for Packet SNR is more complicated. We have already seen in Chapter 6 that Packet SNR is not a good indicator of packet delivery for a specific MCS, so the approach we used for Effective SNR in Algorithm 8.3 would offer poor performance.

Instead, I observe that there should still be a positively correlated relationship between Packet SNR and expected bitrate, which I use instead to predict the bitrate. (Note that this will not tell *which* specific modulation and coding schemes to use, only what the expected bitrate will be given ideal rate selection among the many choices.) This approach will provide a better way to predict throughput from Packet SNR.

To generate this function, I measured the optimal throughput and Packet SNR for all the links in my testbed. I then divided this data into 1 dB-wide SNR bins, and took the median throughput of each bin as the throughput prediction for that Packet SNR value. By connecting these medians (and dropping some sparse bins so that the line monotonically increases), I derived a monotonic function that uses the Packet SNR measurement to predict the throughput of a link. I plot the original data as a scatterplot in Figure 8.13, with the interpolated throughput prediction also shown as a solid line. To define PREDICTBITRATE-PACKETSNR in order to define a Packet SNR-based relay selection algorithm, I simply look up the Packet SNR of the link and return the throughput predicted by this line.

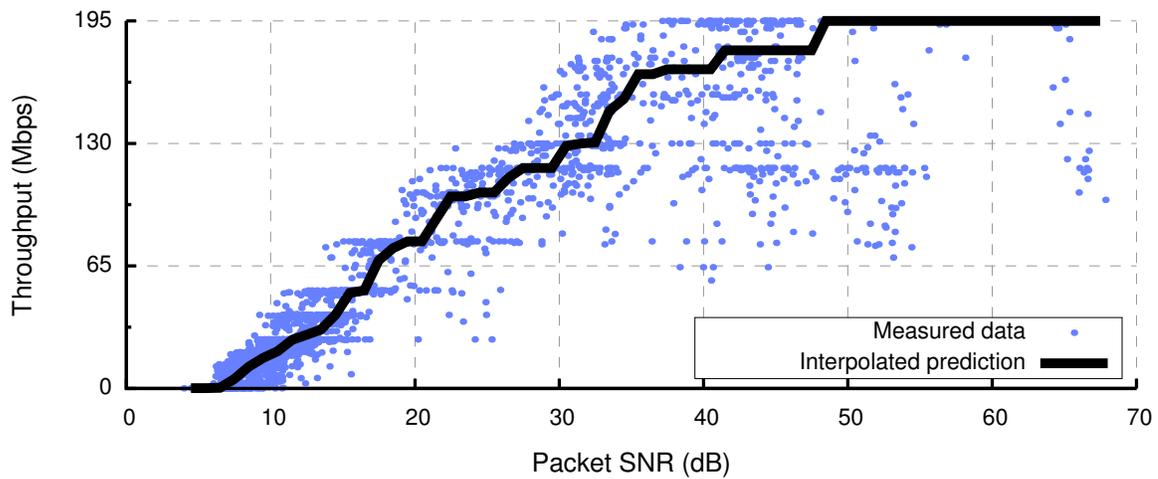

**Figure 8.13: Deriving a prediction for throughput as a function of Packet SNR.**

*Practical Coordination Algorithm*

Algorithm 8.5 describes the calculation of the rate, but it does not describe directly how the two bitrate predictions are obtained. Recall the access point selection algorithm, in which the client sent a probe to all potential APs and evaluated the downlink performance of each AP from the probe responses. For relay selection, the sender instead sends a probe to all potential *relays*, estimating the link performance from their response. If the relays include in their response their predicted bitrate to the destination, the sender has all the information needed to estimate the performance of the relay path. This induces little added coordination on the network, provided the relays have already calculated the bandwidth to the destination. If not, the destination can send a single probe response which can be measured by all relays.

Having defined both algorithms for relay selection, I next evaluate their performance.

*8.4.2   Relay Selection Performance*

I use Algorithm 8.5 with Packet SNR or Effective SNR to select relays on the same dataset.

Figure 8.14 plots the relative performance of each algorithm to the optimal relay algo-rithm; it has a truncated y-axis. Note that the algorithms, which choose to use a relay or not based on predictions of performance, do not choose to use a relay for the same set of links. Taking the union of the links for which either algorithm chooses a relay, Figure 8.15 shows the ratio of the path chosen using Effective SNR to the path chosen with Packet SNR. Both algorithms perform very well, choosing an optimal relay in more than 80% of cases.

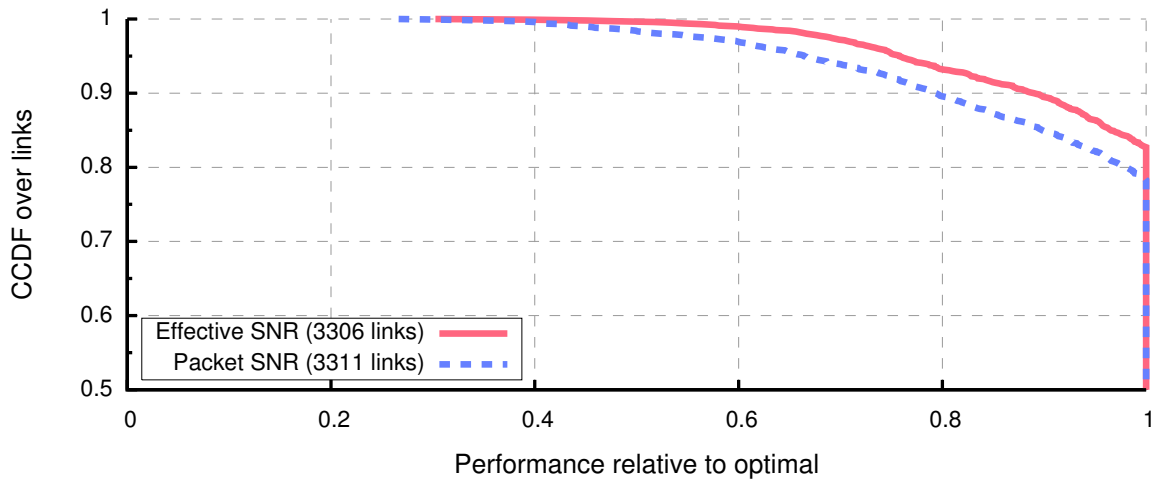

**Figure 8.14: Relay selection algorithm performance relative to an optimal algorithm.**

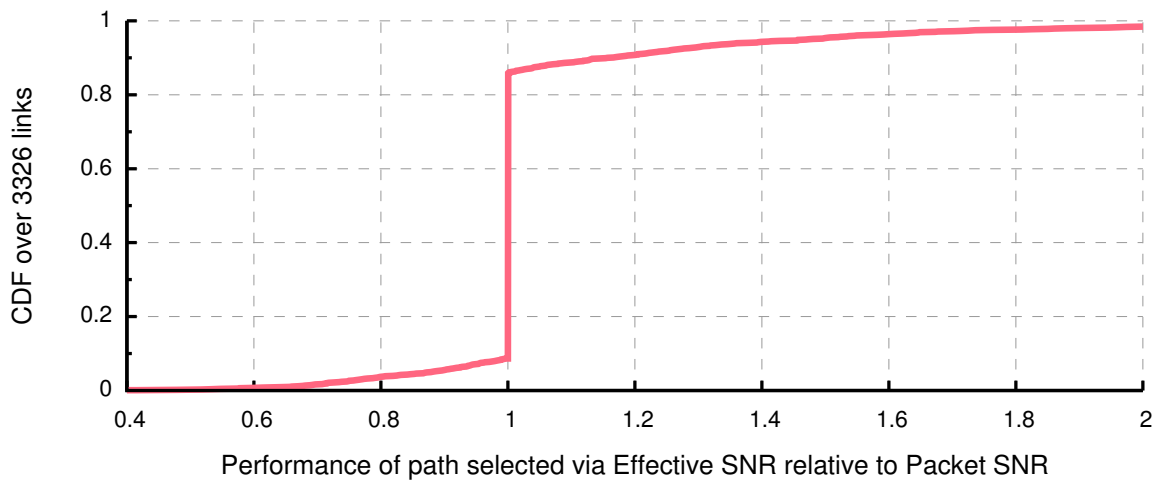

**Figure 8.15: The relative throughput selecting relays by Packet SNR or by Effective SNR.**

The head-to-head comparison shows a potentially surprising result: The Packet SNR and Effective SNR-based algorithms have nearly indistinguishable performance (and thus I truncated the y-axis in Figure 8.14). Effective SNR chooses a better path for 471/3326 (14%) of links and a worse path for 289 (9%). The Effective SNR and Packet SNR lines in Figure 8.14 are tight. Unlike for access point or channel selection, the ability of Effective SNR to account for subchannel effects does not improve performance much over simply using Packet SNR for relay selection.

To explain this phenomenon, recall from Figure 8.11 that the links for which performance improves using a relay are those links with low speeds (somewhat less than 100 Mbps). This is because the best possible relay path—a path with two maximum-rate hops (195 Mbps)—would have an idealized rate of 97.5 Mbps. Consequently, direct links that are faster will not benefit from a one-hop relay. These results indicate that the interpolated throughput values in Figure 8.13 are fairly accurate for slower links.

Why would Packet SNR accurately predict the expected performance for slower links? I offer two explanations. First, note that the throughput prediction task that I use to select paths is easier than the packet delivery (Chapter 6) or rate selection problems (Chapter 7). Rather than knowing which particular configuration work provides the best performance, the algorithm need simply predict the expected performance value—and only well enough to eventually select the right relay. This statistical property is easier to get right, and need not be exact in order to find a good path: A slight over- or under-estimate of one or both hops might not change the final path selection. For example, if a hop delivers about 78 Mbps, it does not matter much whether this is achieved as MCS 12 (two streams at 39 Mbps each) or MCS 19 (three streams at 26 Mbps each). Similarly, for a path using two 78 Mbps hops (total idealized throughput 39 Mbps), an erroneous prediction of 65 Mbps for one hop results in an estimate of 35.5 Mbps which is close to the truth and might still lead to a correct relay selection.

Secondly, slower links are likely to be using single-stream SIMO or dual-stream MIMO2 configurations rather than the three streams used to achieve the absolute fastest rates. The packet delivery results in Chapter 6, particularly as demonstrated by Figure 6.3(b) and Figure 6.3(c), showed that Packet SNR can be fairly accurate for these configurations, which exploit spatial diversity using the excess receive antennas.

These two factors combined explain why Packet SNR performs well for relay selection in my testbed. This property should generalize, as relaying is generally most useful for slow links. Effective SNR simply offers little gain relative to SNR in this application, though it does not require training the Packet SNR/throughput relationship. Still, Effective SNR performs as well as Packet SNR (if not slightly better) and given its other benefits it is a natural fit for this application.

## 8.5  Mobility Classification

The previous three applications in this chapter used the Effective SNR in algorithms that configure various network parameters. In this section, I use the Channel State Information (CSI) underlying the Effective SNR model to determine whether a wireless device is moving. Though this application does not directly use the Effective SNR, this primitive provides an important complement to the network-level configuration problems—in wireless systems, simply knowing whether a device is mobile can improve performance and reliability.

Detecting mobility can be used to enhance reliability in networks that support dynamic topology, such as today's cellular phone networks, enterprise Wi-Fi wireless distribution systems (WDSes), and networks that support relaying mechanisms such as described above. By proactively looking for a better AP or relay when the device starts moving, service quality can be improved and downtime reduced.

A recent implementation by Ravindranath et al. [92] detected mobility using the accelerometer in a mobile phone. While this technique is accurate and responsive, it has a few disadvantages. The use of an on-board sensor means that detection can only be performed by the mobile client, and thus requires protocol changes to communicate a device's mobile state to the other endpoint of the link. This solution is not backwards-compatible. Also, this technique can only be implemented on devices that have accelerometers, and it requires that this sensor be powered on.

In this section, I explore whether it is possible to classify whether a device is mobile based solely on passively measured RF information. If successful, such an implementation would eliminate all of these drawbacks by requiring no extra hardware and supporting unilateral adoption by either endpoint of the link, including the static device. Ravindranath et al. made a preliminary attempt to classify mobility using RSSI, but they were not successful. They list three challenges: (1) that RSSI is unstable even for static links in a quiet environment; (2) that RSSI varies by different amounts at different absolute signal strengths and thus needs to be calibrated; and (3) that RSSI was extremely sensitive to movement in the environment and triggered many false hints. Here, I show that the CSI can overcome these challenges and provide a robust solution.

### 8.5.1 Experimental Setup

I configured a SIMO experiment using a single-antenna laptop as the client device and eight of the testbed nodes as three-antenna monitors. The client sent 100,000 back-to-back short packets using MCS 0 (1 stream, 6.5 Mbps), approximately one packet every 200 μs for 20 s. This sampling rate is higher than practical for Wi-Fi. In particular, packets sent at slow rates or in batches can last as long as 4 ms. I thus downsampled the trace from 100,000 packets every 200 μs to about 5,000 packets every 4 ms before any of the processing described in the rest of this section.

I ran four experiments to analyze a variety of static and mobile channels. For two experiments, I placed a *static* client in the UW CSE Networking Lab, with students present, but not moving in the room. I next took a trace with *environmental mobility* in which I left the client static, but waved my hand within a few centimeters of the antenna and then walked around the room and opened doors. Finally, I took a *mobile device* trace in which I picked up the laptop and moved it around within a meter of its original location. Chronologically, the traces were taken in the order described within a 10-minute window, with the second static trace taken last.

My goal in this section is to develop a simple classification scheme that can distinguish between these scenarios. In particular, I would like to be able to tell, from RF channel effects alone, whether the device is in a static or mobile channel (with device or environmental mobility). A secondary goal is to distinguish between the two types of mobile channels. Ravindranath et al. argued that both of these two goals are difficult, if not impossible, with simple RSSI. I believe that the fine-grained information conveyed by the CSI can overcome these deficiencies.

### 8.5.2 Classifying Mobility with RSSI

I begin by analyzing RSSI variation over time to confirm that it is indeed difficult to distinguish between these four traces. Figure 8.16 shows the RSSI in dBm measured by one receiver for each scenario. In each plot, the three lines each show the RSSI for one of the three receive antennas. With the typical Wi-Fi noise floor around $-92\,$dBm, these graphs show the three antennas for a strong receiver with a Packet SNR around 45 dB.

I note several interesting effects visible in these measurements. First, the RSSI is actually extremely stable in static scenarios, varying roughly 2 dB across 20 seconds. This stability deviates from the observations by Ravindranath et al., likely because the newer 802.11n hardware I used is well-calibrated, compared with older 802.11g hardware they used to run experiments with the MadWiFi driver.

Second, though RSSI does vary with environmental mobility, the variation is fairly small and mostly limited to the periods of activity directly next to the client. Later in the trace, when I moved across the room, the RSSI variation decreased to match the static scenario. It also appears that the variation is not completely correlated across antennas; in several parts of this trace (e.g., at the beginning and around $10\,$s–$12\,$s) one or two antennas see variation in RSSI while the others do not. These periods of partial variation may be indicative of a static device with environmental movement.

Finally, the mobile trace exhibits the RSSI variation with the largest magnitude, and shows consistent variation throughout the trace and across all antennas. This is a dramatic outlier compared to the other traces, and strongly reflects the effects of movement.

Based on this visual evidence, which mirrors the results for the other 7 receivers, I believe it likely that the static scenario actually *can* be identified using RSSI, and hypothesize that it may also be possible to distinguish between environmental and device mobility. However, I deferred exploring these possibility further because, as I will show next, the CSI can

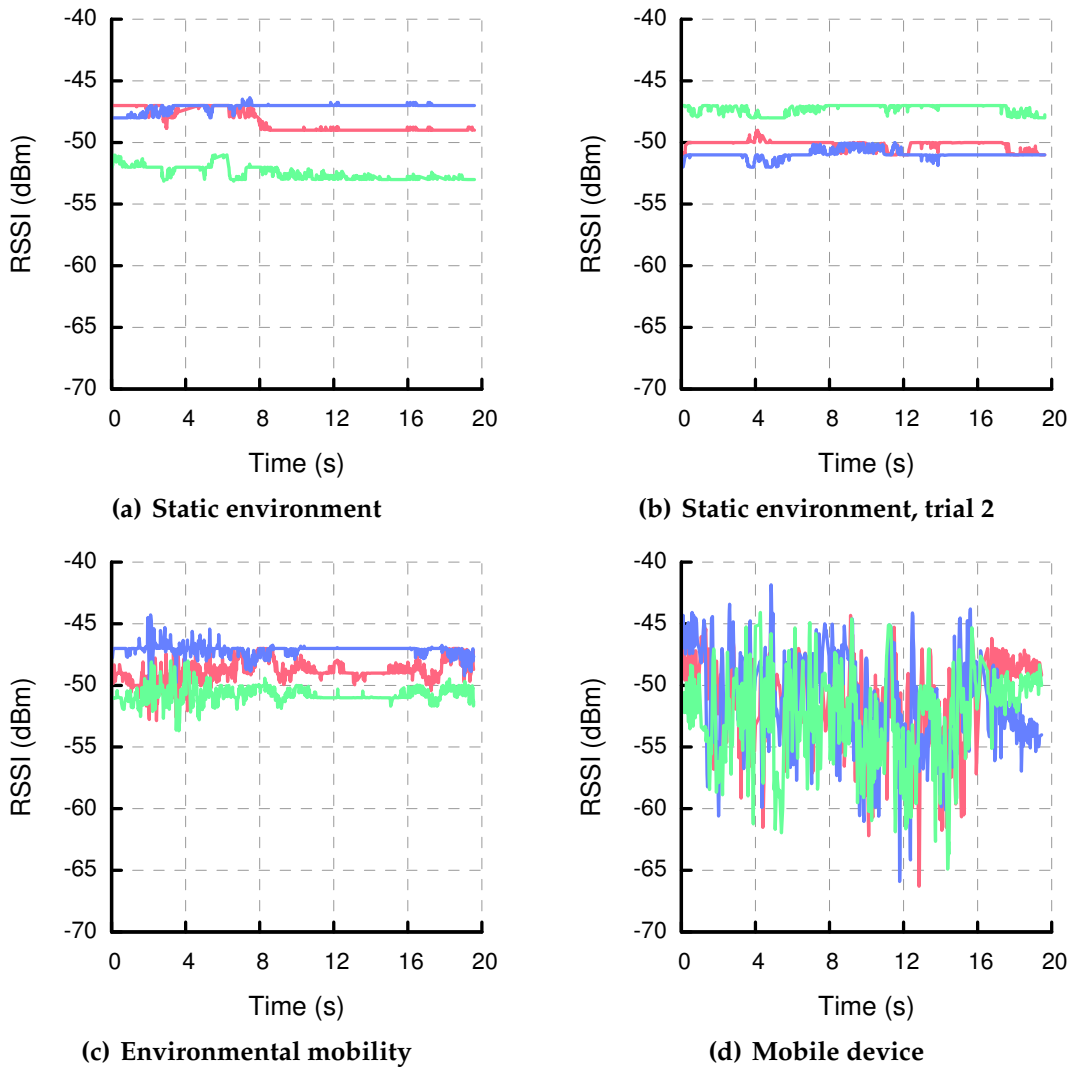

**(a) Static environment**

**(b) Static environment, trial 2**

**(c) Environmental mobility**

**(d) Mobile device**

**Figure 8.16: RSSI variation in different mobility scenarios.**

conclusively classify a device's activity into these three states.

### 8.5.3 Classifying Mobility with CSI

The last subsection examined the four traces I took—two static traces, one with a fixed device and environmental mobility, and the last with a mobile device—and showed that RSSI variation differs visually across in these four scenarios. Here, I examine the same four traces through the lens of the CSI.

*Quantifying CSI Variation: Pearson Correlation*

Recall that the RSSI yields a single power measurement for each antenna for every packet, whereas the CSI gives a set of matrices of complex numbers that represent magnitude and phase on different spatial paths and frequencies. To measure the deviation in RSSI, we could simply look at its variation—e.g., absolute difference between samples, or windowed variance—over time, as I showed visually in Figure 8.16. In contrast, it is not obvious how to quantify the variation in CSI over time.

I chose a simple method to quantify the variation of CSI, by using the *Pearson correlation function* for each spatial path between a transmit-receive antenna pair. The Pearson correlation is the "standard" correlation for two $n$-element vectors $\vec{x}$ and $\vec{y}$:

$$corr(\vec{x}, \vec{y}) = \frac{\sum_{i=1}^{n}(x_i - \overline{x})(y_i - \overline{y})}{\sqrt{\sum_{i=1}^{n}(x_i - \overline{x})^2 \sum_{i=1}^{n}(y_i - \overline{y})^2}}. \tag{8.1}$$

Here $\vec{x}, \vec{y}$ are indexed by $i$ and have respective means $\overline{x}$ and $\overline{y}$.

To apply this to CSI, let $\vec{r}_{p,t}$ represent the magnitudes of the CSI coefficients across subcarriers for spatial path $p$ at time sample $t$. Then we can quantify the change between sample $t$ and sample $t + 1$ by $corr(\vec{r}_{p,t}, \vec{r}_{p,(t+1)})$. The correlation will be close to 1 if the CSI matches across time, i.e., the channel is not changing, and closer to zero if the channel varies quickly so that the CSI samples are close to being independent.

*CSI Variation Examples*

Using the same link as in Figure 8.16, I present the temporal correlation for the four experiments in Figure 8.17. Again, each plot shows one line for each of the 3 receive antennas. These plots show that the static traces have near-perfect correlation, close to 1.0 for the full 20 s trace. In contrast, both the environmental and device mobility traces vary significantly, as low as 0.9 in environmental mobility and 0.3 when the device itself moves. These results thus indicate that the Pearson correlation can classify whether the channel is static, and in changing channel can differentiate whether just the environment is changing or the device itself is moving.

That the environmental and device mobility scenarios exhibit dramatically different correlations might be surprising. To explain this, recall that the frequency-selective channel effects measured by CSI usually result from indoor multipath effects, in which multiple copies of the transmitted signal arrive at the receive antenna after propagating along different (possibly reflected) rays through the RF environment. In many cases of environmental mobility only some of these rays will be affected. If instead the device is itself moving, all paths will be affected. This provides an intuitive explanation for why the temporal correlation is significantly stronger for environmental mobility, though both correlations

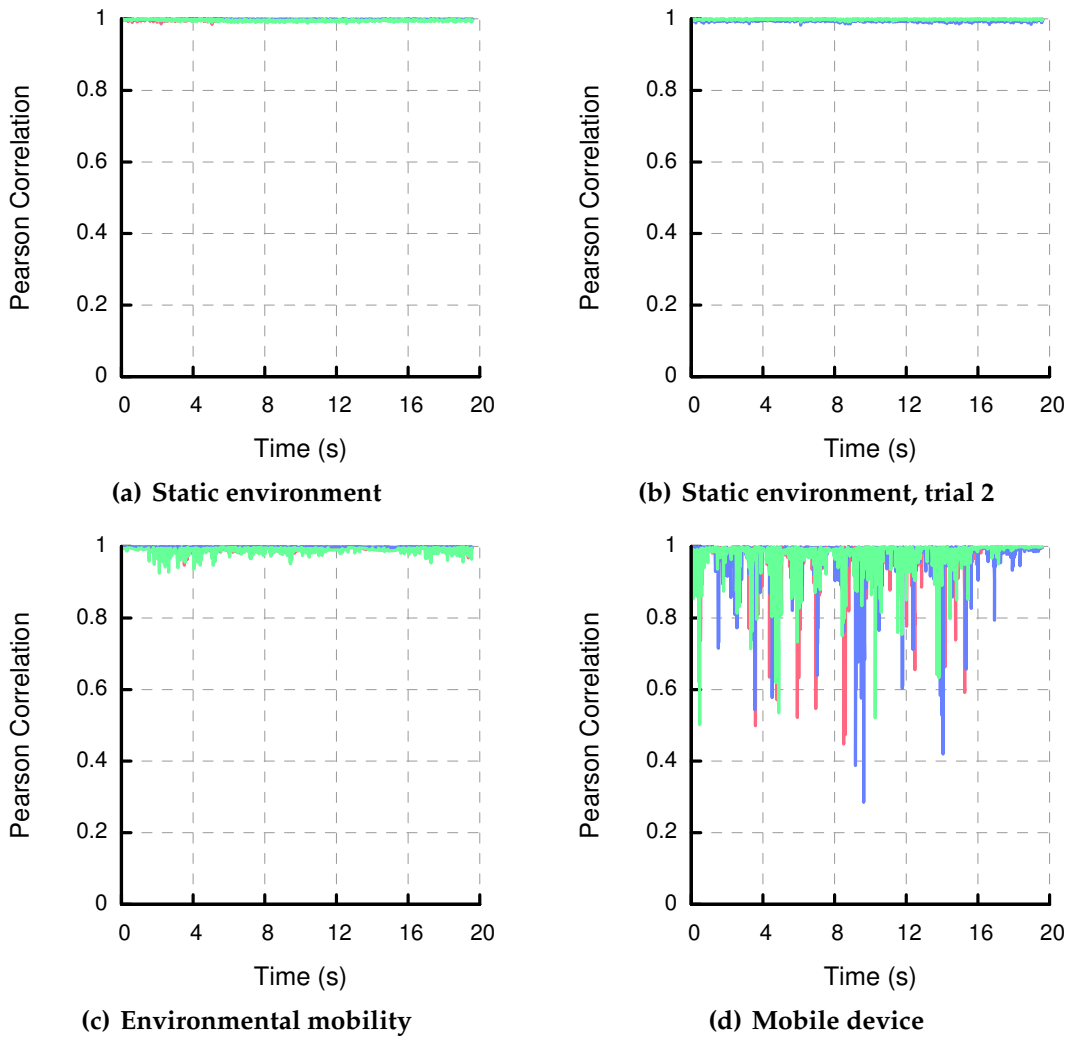

**(a) Static environment**

**(b) Static environment, trial 2**

**(c) Environmental mobility**

**(d) Mobile device**

**Figure 8.17: CSI variation as measured by correlation in different mobility scenarios.**

deviate significantly from the static correlation of 1.

**CSI correlation for weak links.** The Pearson correlation over CSI has thus overcome two of the three drawbacks that Ravindranath et al. outlined for classifying mobility using RSSI. First, the correlation of CSI across time is stable for static links. Second, the Pearson correlation can distinguish between environmental mobility and a moving device. The third drawback is that RSSI-based classifiers do not work as well for weaker links: How does my CSI-based classifier perform in this scenario?

Figure 8.18 analyzes the RSSI and CSI variations during a static trace when the weakest antenna has a Packet SNR of only 5 dB. I found that, indeed, Pearson correlation does not work as well for this weak link. In particular, the correlation against CSI is as low as 0.5,

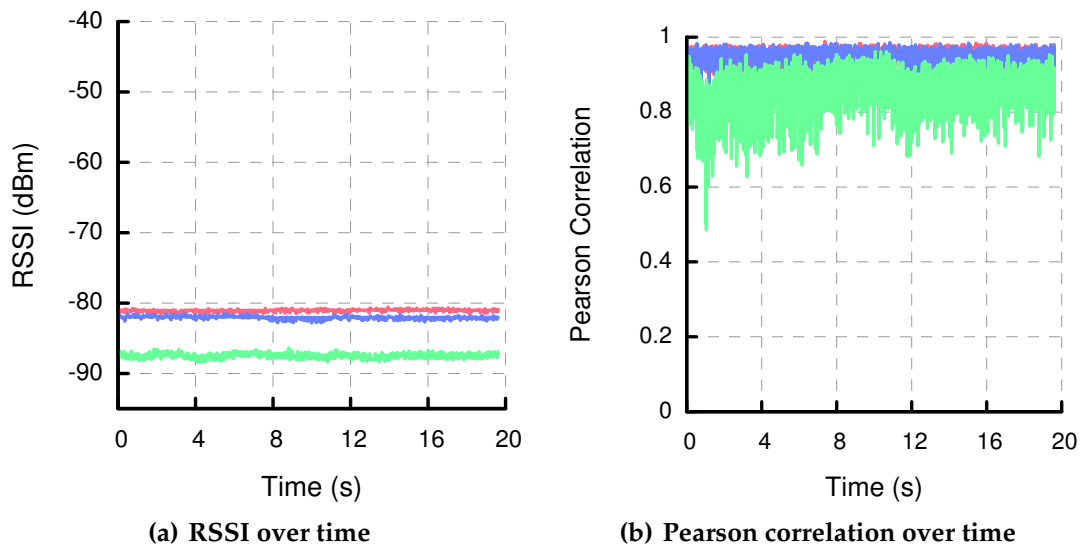

**(a) RSSI over time**

**(b) Pearson correlation over time**

**Figure 8.18: RSSI and CSI variation for a weak link in a static environment.**

on par with the mobile device for the stronger link shown in Figure 8.17. The RSSI shown in the left graph also exhibits similarly larger variation for this weak link compared to the strong link analyzed above.

To fix this, I introduced a *windowed correlation* scheme. Before computing the Pearson correlation values, I average together 10 consecutive CSI samples. This has the effect of smoothing out the noise corrupting the CSI estimates and restoring the high correlation to the static traces, while only slightly affecting the low correlations for mobile traces.

Figure 8.19 shows the CDF of the windowed correlation values (denoted *corr*, and combined across antennas) over time for these same two links. To better illustrate the difference across experiments, I capture the deviation from a perfect correlation of 1.0 by plotting $1-corr$ on a logarithmic scale. Thus a value of 0.001 on the x-axis means a correlation of 0.999, while a value of 0.1 means a correlation of 0.9. Recall that lower correlations (right on the graph) imply that the channels are changing faster.

For both the weak and the strong link, the mobile trace stands out. Its line is to the right, meaning its correlation is lower, than the other measurements for both links. Focusing on the lowest correlations (the points in the bottom 20% of each CDF), we can see that only the mobile traces achieve a correlation below 0.9 (x-axis value 0.1 or higher). Though this only happens for a small portion of the trace, these points are spread throughout the mobile portion of the trace (see Figure 8.17(d)). For the stronger link, there is clear separation (at least an order of magnitude) between the mobile trace, the environmental mobility trace, and the static traces. For the weak link, it is difficult to distinguish environmental mobility

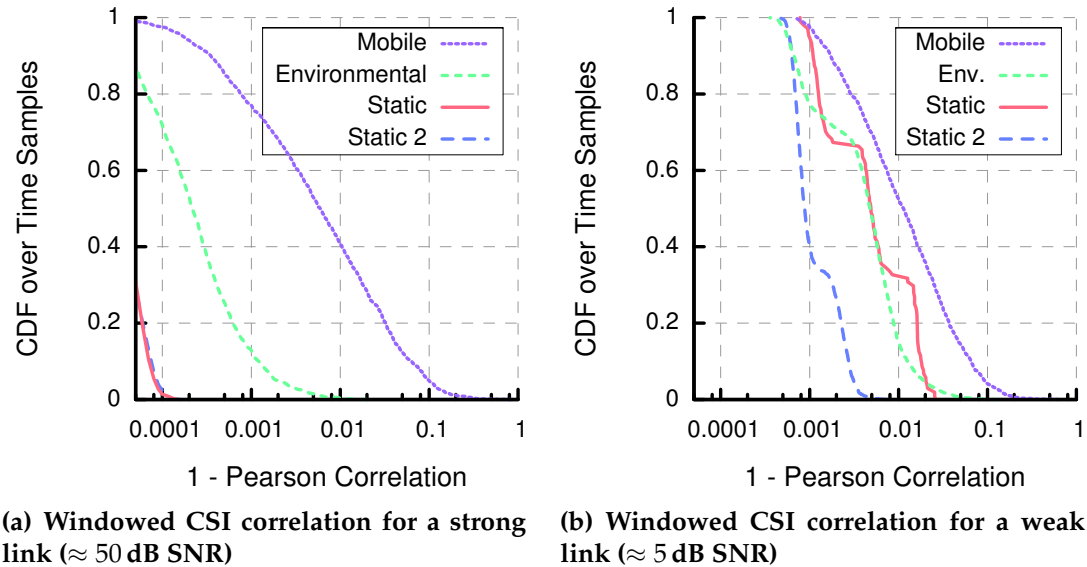

**(a)** Windowed CSI correlation for a strong link ($\approx 50\,\text{dB}$ SNR)

**(b)** Windowed CSI correlation for a weak link ($\approx 5\,\text{dB}$ SNR)

**Figure 8.19:** Windowed CSI variation for a strong and a weak link.

from the static trace. This indicates that the secondary goal of distinguishing these scenarios may only be achievable for strong links.

*CSI Correlation Results*

Thus far, I have presented CSI correlation results for one strong and one weak link.

The basic mobility classification is as follows. Continually run windowed temporal correlations across CSI measurements for active links, and record the smallest correlation values found. When the smallest correlation is below 0.9 (or over 0.1 in the complementary scale I plot), the classifier will output that the device is mobile. When the largest correlation is below 0.99 (or above 0.01 in the complementary scale), the classifier will output that there may be environmental mobility.

Looking at Figure 8.19, we can see that this algorithm will accurately classify mobility, but can result in some static scenarios falsely classified as experiencing environmental mobility. Recall that the traces studied here are 20 second traces of CSI activity; though only approximately 5%–20% of points fall below (above) these thresholds, the peaks are scattered throughout the 20 s trace (e.g., see Figure 8.17). Over the period at which mobility state can change—at least a few seconds—the classifier should see multiple events.

In Figure 8.20, I present the CDFs of windowed CSI over all eight links, separated into the three mobility conditions scenarios. Note that the mobility and environmental mobility conditions have 8 lines, one for each receiver, while I have combined both static traces into 16 lines on a single graph. The black vertical lines show the thresholds for environmental

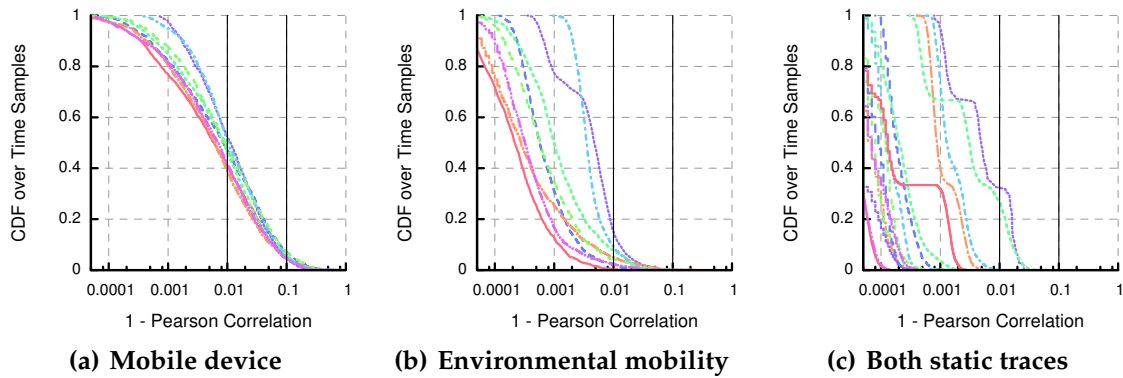

**(a) Mobile device**  **(b) Environmental mobility**  **(c) Both static traces**

**Figure 8.20: Windowed CSI variation for all eight receivers.**

mobility (0.01) and device mobility (0.1).

These results show that the features I identified looking at the first two links extend to the entire testbed. Notably, all 8 links are properly classified when in device mobility (right of 0.1 on the x-axis), while none of the 24 traces in the other two conditions are false positives. For this experiment, my windowed CSI correlation can identify a mobile device across all 8 receivers, some in different rooms and on different floors.

Next, all 8 links are properly classified as being in environmental mobility, with correlations below 0.99 (right of 0.01 on the x-axis). Only two static traces reach this level, and those have low Packet SNRs that are near the noise floor. For the combined 16 static traces across the two experiments, most of the links have correlations above 0.999 (left of 0.001 on the x-axis) for the entire 20 s period. (Again, the exceptions are the links with very low SNR.) These latter two figures indicate that windowed CSI correlation can distinguish between static channels and channels with environmental mobility except at very low SNR values.

### 8.5.4 Summary

The results in this section offer a promising indicator that RF information can indeed provide an accurate classification of device mobility, distinguishing between three different types of RF channels. Though the evaluation is insufficient to prove that these techniques work in all cases, my windowed CSI correlation technique was able to accurately classify device mobility from a static channel for eight different receivers spread across my 802.11n testbed. Additionally, for the 6 of 8 receivers that had strong links, my method could also distinguish between fully static environments and those environments in which people and objects moved near a fixed transmitter. The overall results suggest that the results could be much better, either by parameterizing by the observed SNR level or by using better recognition algorithms.

These results offer evidence that RF measurements can overcome the three drawbacks identified by Ravindranath et al. [92] and can thus provide useful information for mobile system behavior. The advantages of an RF-based approach include lower power consumption, compatibility with today's protocols, and the ability for these techniques to be implemented at either end of the link—to work with unmodified legacy devices and devices that do not include sensors.

### 8.6   Summary

In this chapter, I evaluated three algorithms using Effective SNR and one algorithm using Channel State Information to illustrate the flexibility of my model and also flesh out the space of Wi-Fi network configurations. My Effective SNR-based algorithms offer good performance for these applications and make them practical.

Effective SNR provides strong performance improvements over the Packet SNR for the access point and channel selection applications. Interestingly, I found that although Effective SNR can inform multi-hop path selection in the network, it does not perform significantly better than Packet SNR because of the 802.11n spatial diversity techniques that improve the accuracy of Packet SNR.

Finally, inspired by recent work that uses information from on-board sensors to improve link- and network-level performance, I designed and evaluated a pure CSI-based mobility classifier. This classifier can be used to complement the prior applications in a variety of ways, such as informing the nodes in the network when it might be worth looking for a new operating point.

Chapter 9

**RELATED WORK**

In this chapter, I place my thesis in the context of existing work on wireless technology and wireless systems. I view my research as lying at the intersection of three bodies of work: (1) understanding real 802.11 wireless channels, (2) theoretical analysis of wireless link performance, and (3) practical algorithms for configuring wireless systems. I use this framework to guide the discussion of related work.

### 9.1  Understanding Real 802.11 Wireless Channels

Since the advent of Wi-Fi technology in the late 1990s, a number of studies have investigated the performance characteristics of 802.11, and uncovered several issues.

Initial studies of 802.11b in Roofnet [3] and at the University of Washington [95] found Packet SNR calculated from RSSI to be a weak predictor of packet delivery. Similar results were observed for sensor networks [127]. One reason for this disconnect was poor calibration of NICs, which has since improved. Today's NICs have calibration procedures that include in-factory measurements of the raw silicon manufacturing variability and on-line measurements conducted when in active use (e.g., as described in Chen and Hsieh [23]) to compensate for thermal conditions and power supply effects. As an example, Intel's open-source drivers [49] perform seven calibration steps whenever the device powers on, changes transmit power level, or switches channels to ensure accurate operation of components such as baseband and RF oscillators, and linear transmit amplifiers. Broadcom advertises that its Wi-Fi solutions "are capable of self-calibrating based on usage temperature and other environmental conditions" [16]. These modern calibration solutions have largely overcome the inaccuracies that plagued early Wi-Fi chipsets.

Another reason that Packet SNR was observed not to predict performance well for 802.11b was the corruption of RSSI estimates by interference [95, 119]. This effect is caused by the spread-spectrum technologies of 802.11b, and the OFDM and MIMO techniques used in 802.11a/g/n today reduce this effect greatly.

Despite these two improvements, a fundamental reason for variation across links with 802.11a/g/n OFDM comes from frequency-selective fading, which does not affect spread-spectrum modulations in 802.11b. These effects have been noticed in several studies of real hardware from a variety of manufacturers [38, 76] and system analyses [67, 115]. I present experimental measurements confirming these effects in Chapter 3 and Chapter 6.

An early 802.11n study by Shrivastava et al. [107] found that the use of multiple antennas can improve physical layer performance for real hardware. My measurements and model provide a better understanding of the channel that can explain the underlying cause of these gains as well as quantify their potential benefits.

Finally, understanding and developing models for real 802.11 channels is a large part of the 802.11 standards working group. Its members have studied raw channel performance in a variety of environments and mobility conditions and have developed models that capture these results, now included as part of the IEEE 802.11-2007 [44] and IEEE 802.11n-2009 standards [45]. My channel state information measurement tool has been used by 802.11 working group members to inform these models, as well as to provide corrective amendments [85].

## 9.2   Theoretical Analysis of Channel Metrics

In the face of fading effects that affect real wireless channels, there has been a large body of theoretical work on the performance of systems in these environments.

The 2008 study by Vlavianos et al. [119] of metrics for devices that operate in real wireless channels uncovered problems with all metrics accessible in Wi-Fi cards at the time. My work fills this gap by both building a practical tool that exposes better physical layer information and developing a practical methodology to compute an accurate channel metric.

Nanda and Rege proposed the Effective $E_b/N_0$ [79] in 1998 as a way of capturing performance over generic faded channels, of which OFDM in 802.11a/g and MIMO-OFDM in 802.11n are instances. Other estimates for faded channels, such as the subcarrier variance proposed by Lampe et al. [67], may be slightly simpler to compute but are less accurate. As such, the Effective SNR notion has been adopted by many communities and is the basis of the theoretical model in my thesis.

Most work on OFDM with convolutional coding (as in 802.11a/g) begins with the Effective SNR or Effective BER and adds simulated faded channels to build closed-form expressions for error rates under coding [9, 82, 114]. My model is related, but simpler: I avoid simulating complex, implementation-dependent coding effects in favor of using fixed, per-rate thresholds. In prior models, dealing with a different implementation or a different code meant changing the internals of the computation. My model can naturally extend to handle different implementations via an adjustment of thresholds.

Effective SNR and related metrics such as the Mutual Information Effective SNR Metric (MIESM) [41, 51, 57, 72, 75] have also been extended to MIMO-OFDM. These extensions have been evaluated using simulated channel models for technologies like 802.11n or LTE/WiMAX, and focused on designing a metric that is tuned to closely predict simulated delivery, independent of complexity. My model is related to these, but more practical. My model uses simpler internals, and I convert CSI to Effective SNR in a way that better matches the equal modulation and power allocation used by 802.11n and offers a better API for practical use.

Most importantly, I experimentally evaluate my model at the application level for real 802.11 NICs and RF channels; I am not aware of other work that uses Effective SNR measures for Wi-Fi outside of simulation or analysis. My studies of Effective SNR in real channels led me to design my model to account for important artifacts such as quantization error and to include a better understanding of protocol and implementation concerns.

### 9.3 Wireless Network Configuration Algorithms

I evaluated my Effective SNR model for packet delivery in the context of a variety of 802.11n configuration problems. Each of these problems has a history of task-specific algorithmic implementations, which I discuss here. The key contribution of my thesis is to replace each of these task-specific algorithms with a single unified algorithm that is simple and accurate.

#### 9.3.1 Rate Adaptation

The problem of efficiently finding a good rate configuration for wireless networks is a well-studied one, since a good rate selection algorithm is necessary to do anything else with wireless technology.

The rate control algorithms in use today use a form of guided search to adapt rates based on packet delivery statistics. Lucent's ARF algorithm [56], OAR [99], and Sample-Rate [14] were early rate adaptation algorithms of this type. RRAA [123] enhances these by dynamically enabling or disabling the RTS/CTS mechanism depending on whether hidden terminals are a problem.

The state of the art in rate adaptation based on packet delivery is the Linux kernel's minstrel [109], a version of SampleRate/RRAA adapted for modern Wi-Fi hardware that can use lower rates for packet retransmissions. The minstrel algorithm has also been adapted for 802.11n [28], performing parallel searches between the multiple MIMO modes with different numbers of spatial streams and channel bandwidths. MiRA [84] is a research algorithm that takes a similar approach. These guided search approaches work well for slowly varying channels and simple configurations (e.g., a few rates with fixed transmit power and channel).

For rapidly varying channels, these algorithms become less effective. Camp et al. [17] demonstrated the importance of varying the time constants used to generate summary statistics for minstrel-like algorithms. Recently, RapidSample [92] used hints from smart-phone sensors to detect mobility and switch to a simplified SampleRate-like algorithm that walks up and down the rates in an agile manner. This provides better performance when devices are moving, but it is not obvious how to extend RapidSample's logic to 802.11n where there is not a single linear set of rates. Still, I believe this is a promising direction.

Some research algorithms propose to use Packet SNR based on RSSI to adapt to rapidly varying channels. RBAR [43] was an early algorithm that aimed to select transmission rate by measuring SNR at the receiver on the RTS packet, and feeding back a choice of rate based on precomputed SNR thresholds in the CTS response. This type of exchange forms the basis of the feedback algorithms now implemented in the 802.11n standard. In the above mentioned work, Camp et al. used knowledge of device speed to more accurately adapt these SNR thresholds. The hybrid rate control work by Hartcherev et al. [39], the SGRA [126] algorithm, and the CHARM [55] algorithms all attempted to train the SNR thresholds of a link on-line, and use these learned thresholds to inform delivery-based adaptation when SNR changed suddenly. My Effective SNR can fit into the frameworks of these algorithm types, but it provides a more accurate indicator of performance and does not need on-line training to work for a specific wireless channel.

HYDRA [64] used software radios to experimentally evaluate various ARF- and RBAR-inspired variants of rate adaptation for 802.11n over emulated and real wireless channels. However, they used narrow 2 MHz channels and only two transmit antennas, so their results fail to capture the frequency- and spatially-selective nature of real wireless channels that my experimental data and evaluation have shown are important in practice.

Recent work has returned to the theoretical approach and made headway by measuring symbol-level details of packet reception. In particular, SoftRate uses the output of soft-Viterbi decoding for each symbol to estimate the Effective BER [120]. This allows it to predict the effects on packet delivery of changing the rate. AccuRate uses symbol error vectors and a full channel simulator for the same purpose [103]. Error Estimating Coding [22] accomplishes the same goal by changing or supplementing the link-layer coding scheme. Though they obtain accurate Effective BER estimates, these methods are not defined for selecting other useful parameters, such as transmit power, and they do not extend from 802.11a/g to 802.11n, e.g., when selecting antennas or numbers of spatial streams.

Compared to all of these approaches, my Effective SNR metric is simple, accurate and quick, providing competitive or better performance in static and mobile channels. It is also more general: With a single CSI measurement, I can extrapolate performance in a wide space of rates, spatial streams, antenna selections, channel widths, and transmit power levels. I have also shown that Effective SNR can be implemented on commodity NICs, and I evaluated it over real wireless channels with mobile and fixed clients. My deeper understanding of fading should also aid attempts to use the faster OFDM rates in challenging outdoor mobile environments [27] that have previously been hampered by an inability to explain or predict performance in a reasonable way.

Finally, some proposals obtain better performance by changing the physical layer. FARA [89] drastically changes the fundamentals of the communication to modulate and code data differently to adapt to each subchannel's best performance. This is not compatible with practical distributed schemes like 802.11 that require that each packet can be demodulated in isolation, and this approach currently requires custom hardware [90]. Other proposals designed to better integrate with 802.11 combine transmission with more efficient channel-dependent coding [71], frequency-aware interleaving [13] or partially-correct ARQ schemes [50]. I believe the better estimate of overall error given by the Effective SNR and the understanding of where errors come from given by the CSI can enhance all these schemes.

### 9.3.2 Transmit Power Control

Work on transmit power control falls into two main categories: Saving energy and increasing spatial reuse (or both). PCMA [77] increased network capacity by a factor of two in simulated dense networks using a distributed transmit power reduction protocol. MiSer [88] focused on maximizing data-per-Joule and was able to increase this metric by 20% in simulated 802.11a networks. Son et al. [110] performed practical power control work in single-rate sensor networks. Symphony [91] is a recent, more practical work that was experimentally evaluated in an indoor 802.11a network using multiple rates and multiple channels. It used a synchronous, two-phase rate adaptation and power adaptation protocol to reduce transmit power by 3 dB and increase network-wide throughput by 50%. These are a few representative samples of a wide body of work.

Finally, all these proposals for transmit power control require complex probing and adaptation mechanisms. Several of these studies noted that they had to measure performance at each different power level because it was hard to predict the impact of a power change, even knowing the best rate in the current state. To yield practical protocols, they also assume symmetric channels. This does not match reality in 802.11n networks which have fundamentally asymmetric transmit and receive behavior, such as receiver spatial diversity. My Effective SNR model can overcome these deficiencies by being able to extrapolate the effects of power control with lightweight measurements that can capture the effects of asymmetric channels. The example evaluation in Chapter 6 suggests that, because of a good predictive model, we can use the Effective SNR to directly and confidently select a reduced transmit power without degrading link performance.

### 9.3.3 Antenna Selection

Antenna selection algorithms have long been available and are well-studied [115]. Some 802.11a/g devices such as the Intel IPW3945 and similar chipsets from Atheros and Broadcom include multiple antennas but only a single transmit/receive chain. In these forms, the NICs would simply choose to receive from the antenna with the strongest RSSI, and would also use this antenna to transmit packets [24, 113]. More advanced techniques took subchannel fading into account [125].

However, in indoor 802.11a/g systems, these techniques usually provided little gain because switching antennas did not alleviate the primary problem of frequency-selective fading—my evaluation also confirms this result. In contrast, another experiment used RSSI-based antenna selection between multiple differently-polarized antennas for an 802.11 ground link to an unmanned aerial vehicle [118]. In this highly dynamic scenario, the authors found that antenna selection provided almost 70% throughput improvement, suggesting that these techniques may have increased benefit as wireless is used in more dynamic environments.

Modern multi-antenna techniques make these algorithms more interesting. For instance, a battery-operated client may want to disable excess receive antennas to save power; the 802.11n standard [45] introduces a new Spatial Multiplexing Power Save mode for this purpose. It also includes an antenna selection feedback protocol to help link endpoints negotiate antenna use. However, these optional protocols are generally unimplemented; instead, the algorithms in use today remain based on extensive probing. For example, the rate control algorithm implemented in Intel's 802.11n wireless drivers [49] contains two concurrent adaptation loops. One is the standard rate adaptation loop that probes different numbers of streams: SIMO, MIMO2, and MIMO3. The second loop switches between antenna sets within a mode, for instance using antenna pairs AB, AC, and BC to alternately send the two-stream MIMO2 packets. Antenna selection thus enlarges the state space, exacerbating the configuration problem and slowing down the convergence time of these adaptation algorithms. The Effective SNR model I present can cut through this joint configuration space and directly choose a good operating point.

### 9.3.4 Access Point Selection

Most existing work on access point selection focuses on other aspects than raw link bitrate, because other factors such as AP load and wired network performance can matter more in practice [11, 53]. Proposed systems modify clients, access points, or both [11, 12, 81, 117] to probe and/or estimate these factors and to better balance load. Recent state-of-the-art enterprise systems such as DenseAP [78] achieve load balancing by (1) centrally calculating the best potential access point for the client based on triangulation and measurements from the client's probe requests, and then (2) forcing the client to associate there by only letting that AP send a probe response.

My work with Effective SNR is complementary to these techniques. Though these algorithms focus on load balancing, they often include a component that requires estimation of the channel between the client and the AP; my evaluation showed that the use of Effective SNR will improve this step. Secondly, these procedures have not been updated for the multi-antenna techniques used in 802.11n, in which these predictions will need to be made across heterogeneous APs and for asymmetric links. Effective SNR can transparently adapt to these scenarios, while the existing Packet SNR-based procedures do not handle these steps.

### 9.3.5  Channel Selection

Like AP selection, most channel selection work focuses on load balancing between contending links and networks. Rather than estimating the achievable rate between two devices, these algorithms (e.g., [4, 5, 62, 97]) probe the free airtime on the available channels because that is a large contributor to actual throughput. They then organize, in a distributed or centralized manner, either the local or global set of access points across channels to make efficient use of available mechanisms.

Again, Effective SNR complements this work by providing a simple, accurate, quickly measurable link performance estimator that can be plugged as a subroutine into algorithms of this type. My model can also be used in new emerging peer-to-peer wireless scenarios as envisioned in MultiNet [20] and SampleWidth [21], and instantiated today in Wi-Fi Direct [122].

### 9.3.6  Multi-hop Networks

Research on multi-hop routing is typically framed in the context of mesh networks, and it focuses on maintaining an efficient distributed routing infrastructure (e.g., [6, 26, 86, 98]). The IEEE standardized the 802.11s amendment in 2011 [46] to support mesh networking, using a protocol called hybrid wireless mesh protocol (HWMP) to select paths. The default link metric in HWMP is close to the idealized ETT [26] metric I used but includes more accounting for overhead. As with ETT, computing this metric requires knowing the wireless bitrate and the packet reception rate, both of which the Effective SNR can predict accurately and quickly.

Other research on multi-hop networks focuses on pipelining transfers along long mesh paths [68, 69, 96], using network coding to improve performance of crossing flows [59, 61, 80], or propagating data more effectively across the network by making use of many unreliable links [15].

As a result of the complex nature of these solutions, work on mesh networks tends to simplify other aspects of network design, for instance by using homogeneous single-antenna nodes and fixing the entire network to a single bitrate and uniform transmit power, so that they only have to probe packet delivery at a single rate. For the example case of network coding, one recent extension added the ability to choose between only two 802.11b rates [80], handling only the state of Wi-Fi in 1997 without many rates, OFDM, or MIMO. Another proposed a distributed probing framework to handle concurrent rate adaptation, evaluated it using simulation only [63]. Incorporating my Effective SNR model would simplify the channel estimation components of these schemes and enable them to handle the broad configuration space posed in heterogeneous MIMO and OFDM networks.

### 9.4 Follow-on Research

Other authors have used my CSI measurement tool and Effective SNR models for applications that go beyond the ones described in this thesis. I describe these projects here because they are related to the work in my thesis. They highlight the value to the community of refining and releasing my research prototype [37].

My CSI measurement tool has been used independently of my Effective SNR model in several projects. Researchers at Duke University and the Hong Kong University of Science and Technology have built three systems [100, 102, 124] that use CSI information for indoor localization. Researchers at Intel Corporation and at Carnegie Mellon University have incorporated measurements taken with my prototype into their work on building accurate models of wireless channels [85] and better emulating them (in the emulator by Judd et al. [18, 54]). In the context of improving rates, a group at the University of Texas at Austin used CSI measurements to design techniques that better take advantage of frequency diversity [13].

My Effective SNR model has been used in two ways. First, it has been adopted as the rate selection algorithm in recent projects that use software radios to conduct research on 802.11n [70], as these platforms enable the necessary CSI to be exposed. Second, in the area of better handling the block-oriented nature of 802.11e/802.11n protocols with packet batching [112], researchers have used my Effective SNR model to forecast better predictions of future bitrates in highly mobile wireless channels [101].

Chapter 10

## CONCLUSIONS AND FUTURE WORK

Modern Wi-Fi (IEEE 802.11n) devices can provide flexible, portable, high-performance connectivity at low cost. This unprecedented functionality is poised to enable a new class of rich applications built by combining functionality from many devices. The key missing component is a network connectivity layer that "just works", providing good performance overall and quickly adapting to changing application demands and mobile wireless environments.

There are two components to such a network layer. The first is a protocol to connect the devices logically. For 802.11n, this is readily available in the form of Wi-Fi Direct [122], a recently standardized specification for building wireless peer-to-peer networks targeted at these applications. Support for Wi-Fi Direct is actively being developed in major consumer operating systems including Linux, Mac OS X, Windows, and Android.

The focus of my thesis is on the second key component: A way to configure the physical layer parameters and network topology to best meet application needs. The nature of modern wireless technology and heterogeneity of wireless devices combined with the inherent multi-device coordination makes this a hard problem. There is a large number of possible configurations from which the network must choose a good operating point.

In this thesis, I have demonstrated that my Effective SNR model provides a practical mechanism to quickly evaluate how well configurations work, which can be used to efficiently configure the physical layer. I have also shown that it can flexibly handle a wide variety of configuration problems and parameters that no prior approaches considered in tandem.

In this chapter I summarize my thesis and its contributions and present next steps for this work.

### 10.1  Thesis and Contributions

My thesis is that *it is possible to rapidly and accurately predict how well different configurations of MIMO and OFDM wireless links will perform in practice, using a small set of wireless channel measurements*. I demonstrated this thesis by building an Effective SNR-based model for wireless networks and evaluating it in the context of IEEE 802.11n.

My Effective SNR model can evaluate a particular physical-layer configuration using a simple interface. The model takes as input a single MIMO and OFDM channel measurement, a target set of transmitter and receiver device configurations, and produces an estimate of how well that combination will deliver packets. This flexible API can express a wide class of configuration tasks, as illustrated by the algorithms I presented in Chapter 6, Chapter 7, and Chapter 8. In particular, I applied my Effective SNR model to rate selection, antenna selection, joint transmit power and rate control, access point selection, channel selection, and relay selection. I also showed how to use CSI to determine whether a wireless device is mobile.

I demonstrated that this model is practical; I presented data on its low computational overhead and implemented it on commodity wireless hardware. My model uses measurements already taken by devices in order to receive packets, and can compute its output in much less time (4 μs) than it takes to transmit a packet (at least 48 μs). In most cases, only a few bytes that represent configuration decisions need to be exchanged, such as a receiver feeding back a particular requested rate to a transmitter. My detailed prototype evaluation of the model in a wide variety of configuration problems provides experimental proof that the model is practical and accurate for real devices operating in real wireless channels.

My specific contributions are as follows:

- I developed a model that accurately predicts the error performance of different MIMO and OFDM configurations on wireless channels. This model is flexible to support a wide variety of transmitter and receiver device capabilities, device implementations, and configuration problems. I presented an implementation of my model using a commodity 802.11n wireless device that demonstrates its feasibility in practice and handles the practical considerations of operation over real links using real, non-ideal hardware. This includes a detailed experimental evaluation of my system that shows that this model accurately predicts packet delivery over real 802.11n wireless links in practice.

- I detailed how to use this model in a system that can solve a large number and variety of configuration problems similar to those described in Section 1.1. I evaluated this system in the context of a wide variety of 802.11n configuration problems. These evaluations show that the predictions output by my model lead to good performance in practice, and exceed the performance of prior probe-based and RSSI-based approaches.

- As part of my thesis I have produced an 802.11n research platform based on open-source Linux kernel drivers, open-source application code, and commodity Intel 802.11n devices using closed-source firmware that I customized. I have released this tool publicly, and at the time of writing it is in use at 23 universities, research labs, and corporations.

### 10.2 Future Work

In this section, I consider paths for future research. First, I think Effective SNR provides a path to building a high-performance networking layer that works well, and would like to pursue this. Second, I present three further configuration problems that I believe Effective SNR can help solve, but which I have not yet validated.

#### 10.2.1 Using Effective SNR in Practical Systems and Protocols

My thesis suggests that it will be possible to build a high-performance networking layer that "just works". While my practical Effective SNR model is a step in that direction, much work remains to be done. I proposed several different ways that my model could be used, e.g., via computation at the receiver side or transmitter side of a link, but there is still a large step to reducing it to practice.

One place to start is by integrating my model into control algorithms for Wi-Fi Direct networks. Open source Wi-Fi Direct implementations are close to being refined enough to support experiments. Building a working combined system would provide invaluable practical experience and research lessons. One key problem includes identifying the cases where Effective SNR results in poor choices (e.g., the few cases where Effective SNR chooses poorly performing access points) and working around them to provide good performance in practice. I discuss a few more important tasks below.

#### 10.2.2 Spatial Reuse

Coping with interference is an important problem in wireless networks, and one that I have not yet solved. An important problem for further research is understanding and managing spatial reuse, that is, understanding when multiple transmitters can send concurrently on the same spectrum. This problem is the subject of great importance in today's AP wireless networks, as systems become increasingly dense. This problem will likely be ameliorated to some extent in future wireless networks that can take advantage of multiple channels, but interference will always be a primary limiting factor in scaling networks.

Existing work has shown large gains from spatial reuse, especially in the area of eliminating corner cases of hidden terminals that can degrade link performance to almost nothing. Today's solutions use expensive distributed coordination mechanisms (i.e., RTS/CTS [58]) with large overheads that are often disabled in practice, and today's research proposals on spatial reuse for Wi-Fi [106, 121] have simply fixed the entire network to a single rate during experiments because of the large search space.

Work in communications theory has defined an Effective SINR notion that extends the Effective SNR to take interference into account. These models typically assume that interference, like noise, is modeled by the AWGN model. However, I believe that it will not be this simple in practice. One reason is that interference is not really equivalent to noise, but is frequency- and spatially-selective as well. Second, how to extract good performance from real Wi-Fi devices in persistent interference is not well-understood. Extending my model to support a practical version of this notion would likely be useful.

### 10.2.3   Saving Energy with Effective SNR

Effective SNR could be highly integrated into the development of better methods to manage the power consumption of battery-operated devices. In particular, clients could select access points or relays with the express aim of minimizing wake time. By choosing a close relay that uses fast rates, a client can spend less time awake. By disabling receive antennas on the mobile device and using advanced mechanisms such as beamforming on the transmitter, the client can make further power savings. I highlighted the importance of these 802.11n parameters in an earlier measurement study [34], but have not shown how to optimize them. I believe that my Effective SNR model can be used in conjunction with power-aware metric functions such as that described in Section 4.3.2, but this solution has not yet been shown to work well in practice.

### 10.2.4   Practical Benefits of Beamforming

Transmit beamforming is a well-studied area of research in communications theory. Most theoretical systems aim to optimize Shannon capacity assuming ideal hardware. In this case, they use the well known water-filling algorithm [115: p. 183] to allocate power across subchannels proportional to SNR. However, this approach may be ineffective in practice because real hardware does not support these idealized requirements.

Real transmit hardware can only support signals with a particular dynamic range, and so cannot perfectly support water-filling. Secondly, in practical systems like 802.11, different subchannels are modulated identically and thus cannot make good use of the asymmetric power across subchannels. Work is needed to understand the practical constraints of real hardware, and then to design a beamforming algorithm for 802.11n that limits the space of beamforming matrices to account for these factors. I believe this problem is interesting in its own right, and that my Effective SNR model can be used to evaluate allocations of power with the goal of minimizing bit errors and finding the best working modulation and coding scheme across all subchannels.

### 10.3   Summary

Wireless systems such as 802.11 are configured today using probing for rate adaptation and for a host of other applications. This probing is the standard strategy because communications-theoretic approaches to configuring network parameters are considered too inaccurate to work well. However, in my thesis I have shown that it is indeed possible to connect theory back to real wireless systems operating over real wireless channels. I have presented an Effective SNR model for wireless systems that use modern physical layer techniques like MIMO and OFDM, and I have shown that it works well for IEEE 802.11n. Going forward, I hope and expect that Effective SNR will be integrated into the control plane for future wireless networks and help enable the next generation of device-to-device wireless applications.

# VITA

Daniel Halperin obtained Bachelor of Science degrees in Computer Science and Mathematics from Harvey Mudd College in 2006. At the University of Washington, he obtained his Master of Science degree in Computer Science and Engineering in 2008, and his Doctor of Philosophy in Computer Science and Engineering in 2012.